\documentclass[prc,aps,nofootinbib,superscriptaddress,showkeys,showpacs,twocolumn,10pt,floatfix]{revtex4-1}
\usepackage[colorlinks,urlcolor={blue}]{hyperref}
\usepackage{eurosym}
\usepackage[usenames]{color}
\usepackage{epsfig}
\usepackage{graphicx}
\usepackage[small]{subfigure}
\usepackage{setspace}
\usepackage[T1]{fontenc}
\usepackage[utf8]{inputenc}
\usepackage[english]{babel}
\usepackage{mathrsfs}
\usepackage{amssymb}
\usepackage{amsmath}
\usepackage{amsfonts}
\usepackage{siunitx}
\usepackage{amsopn}
\usepackage{mathtools}
\usepackage{float}
\usepackage{chemformula}
\usepackage{comment} 
\usepackage{tikz}
\usepackage{bm}
\usepackage{enumitem}   
\usepackage{booktabs}
\usepackage{multirow}
\usepackage{verbatim}
\usepackage{ulem}
\usetikzlibrary{arrows}
\newcommand{\red}[1]{\textcolor[rgb]{1.0,0.0,0.0}{\bf \sout{#1}}}

\newcommand{\cya}[1]{\textcolor[rgb]{0.0,0.6,1.0}{\bf #1}}
\newcommand{\blu}[1]{\textcolor[rgb]{0.0,0.0,1.0}{\bf (comment: #1)}}

\begin{document}
\title{Embedding short-range correlations in relativistic density functionals %for nuclear matter
%mean-field models 
through quasi-deuterons}
%energy shifts}
\author{S. Burrello } 
\email{burrello@lns.infn.it, ORCID: 0000-0002-1132-4073}
\affiliation{Technische Universit\"{a}t Darmstadt, Fachbereich Physik, Institut f\"{u}r Kernphysik, Schlossgartenstra\ss{}e 9, D-64289 Darmstadt, Germany}
\author{S. Typel} 
\email{stypel@ikp.tu-darmstadt.de, ORCID: 0000-0003-3238-9973}
%s.typel@gsi.de}
\affiliation{Technische Universit\"{a}t Darmstadt, Fachbereich Physik, Institut f\"{u}r Kernphysik, Schlossgartenstra\ss{}e 9, D-64289 Darmstadt, Germany}
\affiliation{GSI Helmholtzzentrum f\"{u}r Schwerionenforschung GmbH, Theorie, Planckstra\ss{}e 1, D-64291 Darmstadt, Germany}

\date{\today}

\begin{abstract}
\begin{description}
\item[Background] The formation of clusters at sub-saturation densities, as a result of many-body correlations, constitutes an essential feature for a reliable modelization of the nuclear matter equation of state (EoS).
Phenomenological models that make use of energy density functionals (EDFs) offer a convenient approach to account for the presence of these bound states of nucleons when introduced as additional degrees of freedom. However, in these models clusters dissolve, by construction, when the nuclear saturation density is approached from below,
revealing inconsistencies with recent findings that evidence
%evidencing 
the existence of short-range correlations (SRCs) even at 
%\red{large} \blu{maybe too generic?}
larger densities.
%a larger density. 
 
\item[Purpose] The idea of this work is to incorporate SRCs in established models for the EoS, in light of the importance of these features for the description of heavy-ion collisions,
%reactions, 
nuclear structure and in the astrophysical context. Our aim is to describe SRCs at supra-saturation densities by using effective quasi-clusters immersed in dense matter as a surrogate for correlations, in a regime where cluster dissolution is usually predicted in phenomenological models. %It would be indeed advisable to include these features to produce more reliable modelization of the nuclear equation of state.

\item[Method] Within the EDF framework, we explore a novel approach to embed SRCs within a relativistic mean-field model with density dependent couplings through the introduction of suitable in-medium modifications of the cluster properties, in particular their binding energy shifts, 
which are responsible for describing the cluster dissolution. As a first exploratory step, the example of a quasi-deuteron within the generalized relativistic density functional approach is investigated. The zero temperature case is examined, where the deuteron fraction is given by the density of a boson condensate. 

\item[Results] For the first time, suitable parameterizations of the cluster mass shift at zero temperature are derived for all baryon densities. They are constrained by experimental results for the effective deuteron fraction in nuclear matter near saturation and by microscopic many-body calculations in the low-density limit. A proper description of well-constrained nuclear matter quantities at saturation is kept through a refit of the nucleon meson coupling strengths. The proposed parameterizations allow to also determine the density dependence of the quasi-deuteron mass fraction at arbitrary isospin asymmetries. The strength of the deuteron-meson couplings is assessed to be of crucial importance.
Novel effects on some thermodynamic quantities, such as the matter incompressibility, the symmetry energy and its slope, are finally discerned and discussed.

\item[Conclusions]The findings of the present study represent a first step to improve the description of nuclear matter and its EoS at supra-saturation densities in EDFs by considering correlations in an effective way. In a next step, the single-particle momentum distributions in nuclear matter can be explored using proper wave functions of the quasi-deuteron in the medium. %\red{and the obtained mass fractions}. 
The momentum distributions is expected to exhibit a high-momentum tail, as observed in the experimental study of SRCs by nucleon knockout with high-energy electrons.
%with the aim to explore implications in the widest scope of astrophysical applications as well as for general aspects of reactions dynamics, such as the clustering processes emerging in heavy-ion collisions.

\end{description}
\end{abstract}
\date{\today}
\maketitle

%%%%%%%%%%%%%%%%%%%%%%%%%%%%%%%%%%%%%%%%%%%%%%%%%%%%%%%%%%%%%%%%%%%%%
\section{Introduction}
\label{sec:intro}
The Equation of State (EoS) of strongly interacting matter is a fundamental ingredient in the theoretical description of compact stars. %\red{. It plays} \gre{, playing also} 
It plays also a crucial role in many astrophysical simulations of, e.g., core-collapse supernovae and neutron star mergers \cite{shenPRC2011, hempelASJ2012, lattimerPREP2016, abbottPRL2018}. Tight links have been also established between the properties of the EoS and those of finite nuclei, concerning both their structure and reaction dynamics \cite{klahnPRC2006, zhengPLB2017, zhengPRC2018}. A large variety of approaches has been employed therefore in the last decades to develop reliable models for the EoS, which are constrained by both nuclear physics experiments and astronomical observations. For astrophysical simulations, tables of global, multi-purpose EoSs are in particular needed,  
%\red{employing the most} \gre{being provided by} 
being provided by sophisticated theoretical methods available in literature \cite{oertelRMP2017}.

One class is given by microscopic ab-initio models, which try to solve the nuclear many-body problem using advanced methods and realistic interactions that are constrained by scattering data and properties of few-nucleon systems, see, e.g., Refs.\
\cite{Akmal:1998cf,Carlson:2014vla}
%\cite{Epelbaum:2008ga,Bogner:2009bt,Gezerlis:2013ipa,Furnstahl:2021rfk}. 
%\blu{Let us add some references here.} 
Other approaches based on the effective-field theory exploit systematic developments from quantum chromodynamics and symmetry concepts, e.g. in chiral perturbation theory, providing also uncertainty estimates
\cite{epelbaumRMP2009,hammerRMP2020,Furnstahl:2021rfk}. 
However, these models fail in properly describing the formation of clusters at densities below nuclear saturation, where these bound states of nucleons emerge as many-body correlations generated by the short-range nucleon-nucleon interaction. Such low-density conditions are encountered in various systems: the debris of heavy-ion collisions (HIC), the post-bounce evolution of core-collapse supernovae, and the surface of nuclei
\cite{Typel:2014tqa,Tanaka:2021oll} where the formation of clusters is a prerequisite for cluster radioactivity and, in particular, $\alpha$-decay of heavy nuclei. The emergence of clusters is definitely an essential feature for the modelization of a realistic EoS \cite{radutaPRC2010}.

Phenomenological models, whose parameters are constrained from experimental results of HIC and astronomical observations and/or by directly fitting properties of finite nuclei and nuclear matter near saturation, offer a convenient alternative to microscopic models to approach this problem. A widely used class of phenomenological approaches is based on energy density functionals (EDFs), which are usually derived in the self-consistent mean-field approximation with an effective in-medium interaction without a direct connection to the nucleon-nucleon interaction in free space \cite{benderRMP2003}. There are various versions of this approach, e.g., non-relativistic models using Skyrme or Gogny type interactions \cite{dutraPRC2012, sellahewaPRC2014} or relativistic models based on the exchange of mesons \cite{dutraPRC2014}. In recent years, several attempts have been made to also directly link EDFs to microscopic ingredients \cite{furnstahlEPJA2020, grassoPPNP2019, marinoPRC2021}. For example, a special class of functionals inspired by effective field theories (EFTs) and bench-marked on ab-initio predictions have been designed \cite{yangPRC2016, grassoPRC2017}. They were applied to finite nuclei \cite{burrelloPRC2021} and finite temperature EoS calculations of pure neutron matter (PNM) \cite{burrelloEPJA2022}. Moreover, few steps were made towards the construction of a power counting in EDF \cite{yangPRC2017, burrelloPLB2020}.

However, EDFs derived from phenomenological mean-field models fail as well, when only nucleons are considered as basic constituents. Further progress is only achieved if clusters are introduced as additional explicit degrees of freedom at low densities \cite{typelPRC2010}. The dilute matter is then depicted as an ideal mixture of nucleons and all nuclei from the table of isotopes in thermodynamic equilibrium. Such a model is called nuclear statistical equilibrium (NSE) and is widely used in the astrophysical context, 
%\red{is considered valid only as long as the interaction between the %constituents can be neglected. It} \gre{where it} 
where it leads to a reliable description for the chemical composition of stellar matter at sub-saturation densities \cite{radutaPRC2010, burrelloPRC2015}. 
%\red{Models} \gre{However, models} 
However, models like NSE are considered valid only as long as the interaction between the constituents can be neglected.  Thus they fail at higher densities where in-medium effects become important, leading to the dissolution of clusters and the transition to cluster-free nuclear matter.
The dissolution of a cluster, i.e., the so-called Mott effect, which is expected when approaching nuclear saturation density from below, can be produced through the introduction of an excluded-volume mechanism, which is just a simple geometric concept \cite{radutaPRC2010, hempelNPA2010, hempelPRC2011, sagunNPA2014}. 
More microscopically, the formation and dissolution of light clusters in nuclear matter can be treated using a quantum statistical approach with thermodynamic Green's functions, see, e.g., Refs.\ \cite{Ropke:1982vzx,Ropke:1982ino}.

An alternative scenario was proposed in the last decade, when interacting clusters were introduced as explicit degrees in relativistic density functionals, firstly concentrating on light hydrogen and helium isotopes, whose properties are modified in the medium \cite{typelPRC2010}. Contrary to the chemical picture traditionally adopted in NSE-like models, where the properties of the correlated states of nucleons are assumed to be independent of the medium, in such a physical picture, the in-medium effects are %\red{more microscopically} 
addressed by introducing a proper modification of the masses of the clusters, as inspired by the quantum statistical approach.
%\red{, whose effective binding energies 
%\red{steeply} increase with density \cite{typelEPJA2020}.}
Following this idea, the effective binding energies are expected to increase with density \cite{typelEPJA2020}.

Many-nucleon correlations in the continuum, which still survive above the Mott density, are then described in terms of effective resonances or quasi-clusters. In current phenomenological approaches with the excluded-volume mechanism or with medium-dependent mass shifts, these states are statistically suppressed by construction beyond saturation, so that only nucleons should remain as independent quasi-particles \cite{Pais:2017}. Mean-field type descriptions of nuclear matter above saturation consider thus the system as a free Fermi gas, with the usual step function in the single-particle momentum distribution at zero temperature. Such a picture is however inconsistent with recent experimental results from nucleon knock-out reactions on nuclei using inelastic electron scattering \cite{fominPRL2012, henRMP2017}. These studies clearly evidence the smearing of the nucleon Fermi surface and the emergence a high-momentum tail (HMT) in the single-nucleon momentum distribution of cold nucleonic matter, ascribable to the existence of sizeable nucleon-nucleon short-range correlations (SRCs) even at saturation density \cite{arringtonPPNP2012, ciofidegliattiPREP2015, henRMP2017}. 

Experimental investigations assessed that SRCs pairs are formed by approximately 20\% of nucleons in various measured nuclei \cite{egiyanPRL2006, subediSCI2008, henSCI2014}. They are characterized by large relative %\red{momenta} 
and small center-of-mass (c.m.) momenta. Moreover, some results concluded that their magnitude is spin- and isospin-dependent, with a clear dominance in the neutron-proton channel \cite{subediSCI2008, henSCI2014, duerNAT2018, duerPRL2019}, that affects the ratio of minority and majority species in asymmetric nuclear matter, in the bulk part and tail of the single-particle momentum distributions \cite{duerNAT2018, schmooklerNAT2019}. 
Extrapolating these experimental results from finite nuclei to infinite nuclear matter, useful information around saturation density was then deduced.
At higher densities, only numerical investigations exist to predict the density dependence of these SRC pairs. From the analysis, the probability for nucleons to form SRC pairs seems to have a minimum in the neighbourhood of the saturation density, owing to the interplay between the tensor component and the repulsive core of the nuclear force \cite{riosPRC2014, yangPRC2019, liPRC2016}.

SRCs may %\cya{also} 
also change the balance between kinetic and potential contributions to the energy of the system. Thus their introduction is expected to significantly affect the controversial density dependence, in particular at supra-saturation densities, of the nuclear symmetry energy \cite{caiPRC2016, liPPNP2018}, which quantifies the difference between the total energy of PNM and symmetric nuclear matter (SNM).  
Many theoretical and experimental investigations are currently investigating the density dependence of this quantity \cite{liPREP2008, russottoPRC2016, zhengPRC2016, burrelloPRC2019, burrelloFRO2019}. 

The purpose of the present work is to incorporate SRCs in established models for the EoS, in light of the key-importance of these features for the description of HIC, nuclear structure and in the astrophysical context \cite{weinsteinPRL2011, vancuykPRC2016, guoPRC2021, luNPA2021}. Our aim is to explicitly treat SRCs at supra-saturation densities by using effective quasi-clusters immersed in dense matter as a surrogate for correlations, through the introduction of proper in-medium modifications of the cluster mass shifts. In this regime cluster dissolution is usually predicted in actual realizations of phenomenological models.  Within the EDF framework, we propose thus a novel approach to embed the SRCs within a relativistic mean-field model (RMF) with density dependent coupling \cite{typelNPA1999} through a substantial modification of the cluster mass shift at high densities. 

Given its phenomenological nature, the adopted approach does not allow to investigate the origin of the SRCs. Several effects coexist in the high-density regime that can lead to a smearing of the single-particle distribution functions and the appearance of HMTs as known from the description of heavy-ion collisions in transport theory \cite{bussPR2012}. They are due to different (repulsive) components of the nucleon-nucleon interaction and can be hardly disentangled. A deeper insight on these features would indeed require a more microscopic treatment of these SRCs, going beyond the scope of the present work.

As a first exploratory step, the example of a quasi-deuteron within this generalized relativistic density functional (GRDF) is currently explored, since two-body SRCs in the neutron-proton $^3S_1$ channel are much more important than other many-body correlations. The zero temperature case is examined, where the deuteron is represented by a boson condensate that determines the mass fraction and leads to specific conditions to the parameterization of the mass shift.
The purpose of this work is then to propose possible mass shift parameterizations that will be employed to determine the density dependence of the quasi-deuteron mass fraction at arbitrary isospin asymmetries. The final ambitious goal is to investigate the effect of accounting for the SRCs in an effective way on the EoS, and some related thermodynamic quantities, at supra-saturation densities. In this context, it is worthwhile to mention that a recent study employed the concept of a mass shift in a similar approach to model the effective interaction of a possible heavy particle with baryon number $B=2$, the so-called {\it sexaquark}, in nuclear matter.
It is treated as a boson condensate like the deuteron and affects the EoS of compact-star matter and thus the properties of neutron stars \cite{Shahrbaf:2022upc}.

The manuscript is structured as follows. In Section \ref{sec:grdf} the theoretical formalism is illustrated and the fundamental principles and basic formulas of the GRDF are given. The case of zero temperature is studied, where quasi-deuteron condensation is expected, and the role of the deuteron-meson coupling strength is discussed. Section \ref{sec:mass_shift_constraints} explores the different constraints for the deuteron mass shifts. Section \ref{sec:mass_shift_param} concentrates on the mass shift parameterization, as suitably derived for nuclear matter at zero temperature. 
%\red{, under the assumption of deuteron condensation at high-density and compared to the traditional treatment. Results concerning the} 
Then, the corresponding density dependence of the quasi-deuteron mass fraction is obtained. 
%\red{as well as for} 
The impact on the EoS and on some general %\red{nuclear matter} 
properties of nuclear matter at arbitrary neutron-proton asymmetries is shown in Section \ref{sec:results}. Conclusions and %\red{an outlook} \cya{outlooks}
an outlook are finally given in Section \ref{sec:conclusions}.
Details on the formal derivation of some quantities, on the conversion of parameters and analytical expressions for the mass shift parameters are %\red{finally} 
furthermore collected in four appendices.

%%%%%%%%%%%%%%%%%%%%%%%%%%%%%%%%%%%%%%%%%%%%%%%%%%%%%%%%%%%%%%%%%%%%%
\section{Theoretical formalism}
\label{sec:grdf}

%%%%%%%%%%%%%%%%%%%%%%%%%%%%%%%%%%%%%%%%%%%%%%%%%%%%%%%%%%%%%%%%%%%%%
\subsection{Generalized Relativistic Density Functional}

The GRDF is a density functional derived from a RMF with nucleons and further degrees of freedom. Their effective interaction in the medium is described by the exchange of mesons with density dependent couplings \cite{typelNPA1999, typelPRC2010}. In such a model, light clusters are explicitly introduced, allowing for a unified treatment from matter with bound nucleons at low densities to matter possibly made of only neutrons ($n$) and protons ($p$) at high densities. All degrees of freedom are represented by quasi-particles with self-energies that incorporate the effects of the interaction. For sake of simplicity, only $^2$H nuclei (labeled as $d$ in the following) are added to nucleons as degrees of freedom in light of their expected prevalent importance discussed in Section \ref{sec:intro}.

Let us thus consider the general case of asymmetric nuclear matter (ANM) composed of
neutrons, protons and deuterons with particle number densities %\red{$n_{n}$, $n_{p}$, and $n_{d}$} 
$n_i$ ($i = n, p, d$), baryon numbers $A_i$ and charge numbers $Z_{i}$. The system is usually characterized by specifying the
baryon density $n_b=n_{n}+n_{p}+2n_{d}$, the isospin asymmetry $\beta=(n_{n}-n_{p})/n_{b}$ and the temperature $T$.  
In this section, the basic formulas of the theoretical formalism are given for the most general case of finite temperature. However, the zero temperature case will be considered in the analysis performed in the following sections, leaving the analysis at finite $T$ for future work. In this work, the masses of neutrons and protons will be taken as equal to the average
nucleon mass $m_{\mathrm{nuc}}$ so that $m_{n}=m_{p}= m_{\mathrm{nuc}}$. The same values as given in Ref. \cite{typelPRC2010} are considered. %\gre{Formulas are expressed in a form using $\hbar = c = 1$.}

%\textcolor{blue}{(later)
%\begin{eqnarray}
%\label{eq:vec_dens}
%n_{n}^{(v)} &=& \frac{1+\alpha-X_{d}}{2}n_{b} = \frac{k_{n}^{3}}{3\pi^{2}} \nonumber \\
%n_{p}^{(v)} &=& \frac{1-\alpha-X_{d}}{2}n_{b} = \frac{k_{p}^{3}}{3\pi^{2}} \nonumber \\
%n_{d}^{(v)} &=& \frac{X_{d}}{2} n_{b} \: .
%\end{eqnarray}
%which are expressed in terms of the corresponding momentum $k_q$ ($q = p, n$) in the nucleonic case, %assuming natural units such that $\hbar = c = 1$.
%}

Following the framework illustrated in Ref. \cite{paisNPCCP2017,Pais:2017}, the thermodynamic properties of nuclear matter are completely determined once the grand canonical potential density $\tilde{\omega} (T, \{\mu_i\})$ is specified. %\red{It} \gre{The latter} 
It depends, apart from the temperature, on the chemical potentials $\mu_{i}$ of all constituents $i$.
%\red{($i = n, p, d$)} \red{, characterized by their baryon numbers $A_i$ %and their charge numbers $Z_{i}$}. 
In the present work, three types of mesons are considered: an isoscalar scalar $\sigma$ meson to describe the attraction between nucleons, an isoscalar vector $\omega$ meson for their repulsion and an isovector vector $\rho$ meson for the isospin dependence of the strong force. 
%\blu{Should we say something regarding the $\delta$-meson?} 
%\blu{I think that we don't need to mention it here.}
The interaction between a baryon $i$ and a meson %\red{$m$} \gre{$j$} 
$j$ ($j = \sigma, \omega, \rho$) is realized by a minimal coupling with a strength that is given by the product of a scaling factor %\red{$\chi_{im}$} \gre{$\chi_{ij}$}
$\chi_{ij}$, the mass number $A_{i}$, and a coupling 
%\red{$\Gamma_{m}$} \gre{$\Gamma_{m}$}
$\Gamma_{j}$. The latter quantity depends on the baryon density $n_{b}$ to describe the medium dependence of the effective interaction. 
Different prescriptions might be adopted as recently discussed in Ref.\ \cite{typelterreroEPJA2020}.
%\blu{Should we mention the density dependent form we are using in our work or, at least, refer to previous works where these forms are given?}
In this work the functional form of the couplings as introduced in Ref.\ \cite{typelNPA1999}
is used.
In the application of the GRDF to homogeneous nuclear matter, only the ratio 
%\red{$\Gamma_{m}/m_{m}$} \gre{$\Gamma_{m}/m_{m}$}
$\Gamma_{j}/m_{j}$
of coupling and mass $m_j$ of the mesons $j$ is relevant. Hence it is convenient to introduce the coefficients
\begin{equation}
%\red{C_m = \frac{\Gamma_m^2}{m_m^2}} \gre{C_j = \frac{\Gamma_j^2}{m_j^2}}
C_j = \frac{\Gamma_j^2}{m_j^2}
\end{equation}
and %\red{the} \gre{their} 
their derivatives %\gre{$C_j^{\prime}$}
\begin{equation}
%\red{C_m^{\prime} = \frac{dC_{m}}{dn_{b}}}
C_j^{\prime} = \frac{dC_{j}}{dn_{b}}
 = 2 \frac{\Gamma_{j}}{m_{j}^{2}} \frac{d\Gamma_{j}}{dn_{b}}
\end{equation}
with respect to the baryon density.

%\textcolor{blue}{(later, to be removed)
%Moreover, we restrict to the zero temperature case, where a term that considers the contribution of bosonic particles in condensates should be added, with respect to the grand canonical potential density given in Ref.\ \cite{Pais:2017}.} % \cite{paisNPCCP2017}.}

%\textcolor{blue}{(later, to be removed) A useful check of the thermodynamic consistency of the used approach is to verify that the usual definitions of thermodynamic quantities and relations for derivatives hold after the modification of the original thermodynamic potential.}

The total grand canonical potential density of the system can be written as 
\begin{eqnarray}
\label{eq:omega_tot}
\lefteqn{\tilde{\omega} (T, \{\mu_i\}) =}
\\ \nonumber & & \sum_i \tilde{\omega}_i + \tilde{\omega}_d^{(\mathrm{cond})} + \tilde{\omega}_{\mathrm{meson}} - \tilde{\omega}_{\mathrm{meson}}^{(\mathrm{r})} - \tilde{\omega}_{\mathrm{mass}}^{(\mathrm{r})}
\end{eqnarray}
containing the standard expression for the single quasi-particle (non-mesonic) contribution
\begin{equation}
\label{eq:omega_sp}
\tilde{\omega}_i = -T\dfrac{g_i}{\sigma_i} \int \dfrac{d^3 k}{(2\pi)^3} \ln \left[ 1 + \sigma_i \exp \left( -\dfrac{E_i -  \mu_i^{\ast}}{T} \right) \right]
\end{equation}
with the well-known integral over the momentum $k$ that appears in the quasi-particle energy
\begin{equation}
E_i = \sqrt{k^2 + \left(m_i^{\ast}\right)^2} 
\end{equation}
assuming natural units such that $\hbar = c = 1$.
In Eq.\ \eqref{eq:omega_sp},
the sign factor $\sigma_i$ distinguishes
the particle statistics ($\sigma_i =1$ for fermions and $\sigma_i =-1$ for bosons, respectively)
%\red{.
%\sout{and} The degeneracy factor $g_i=(2J_{i}+1)$ is determined by the %particle spin $J_{i}$} \gre{and $g_i$ is the spin-degeneracy factor}.
and %\red{$g_i=2J_{i}+1$} 
$g_i$ is the spin-degeneracy factor. 
%\blu{This is a well known I guess. Moreover, the letter $J$ is now used for the symmetry energy.}
The boson condensate term
\begin{equation}
\label{eq:omega_cond}
\tilde{\omega}_d^{(\mathrm{cond})} = %n_b \dfrac{X_d}{2} 
\frac{1-\sigma_{i}}{2}
n_{d}^{(\mathrm{cond})}
(m^{\ast}_d - \mu_d^{\ast}) 
\end{equation}
in Eq.\ (\ref{eq:omega_tot}) 
with the density of the condensate, $n_{d}^{(\mathrm{cond})}$, is only relevant for deuterons. 
%\blu{The definition of $n_{d}^{(\mathrm{cond})}$ is missing!}

The effective chemical potential
\begin{equation}
\label{eq:mustar}
\mu_i^{\ast} = \mu_i - V_i.
\end{equation}
and the effective mass
\begin{equation}
\label{eq:mstar}
%\red{m_i^{\ast} = m_i - S_i}
m_i^{\ast} = m_i + \Delta m_i - S_i
\end{equation}
depend on scalar 
\begin{equation}
\label{eq:spot}
%\red{S_i = \chi_{i \sigma} A_i C_{\sigma} n_{\sigma} - \Delta m_i}
S_i = \chi_{i \sigma} A_i C_{\sigma} n_{\sigma}
\end{equation}
and vector
\begin{equation}
\label{eq:vpot}
V_i = \chi_{i \omega} A_i C_{\omega} n_{\omega} + \chi_{i \rho} A_i C_{\rho} n_{\rho} + A_i V^{(r)} + W_i^{(r)}
\end{equation}
potentials, respectively. They are defined in terms of the different source densities %$n_j$ ($j = \sigma, \omega, \rho$), given by
\begin{eqnarray}
%\label{eq:meson_dens}
\label{eq:nsigma}
n_{\sigma} &=& \sum_i \chi_{i \sigma} A_i n_{i}^{(s)}  \\
\label{eq:nomega}
n_{\omega} &=& \sum_i \chi_{i \omega} A_i n_{i}^{(v)}  \\
\label{eq:nrho}
n_{\rho} &=& \sum_i \chi_{i \rho} A_i n_{i}^{(v)}
\end{eqnarray}
where the single-particle scalar densities ($n_{i}^{(s)}$) and vector densities ($n_{i}^{(v)}$) appear.

The mass shift $\Delta m_{i}$
in the %\red{scalar potential (\ref{eq:spot})} 
effective mass (\ref{eq:mstar}) appears only for the deuteron and is assumed to depend 
on the baryon density $n_{b}$.
%like the %\gre{nucleon-nucleon-meson couplings.
This dependence leads to
rearrangement contribution
\begin{equation}
    W_i^{(r)} = n_{d}^{(s)} \frac{\partial \Delta m_{d}}{\partial n_{i}^{(v)}}
\end{equation}
in the vector potentials \eqref{eq:vpot} of nucleons and the deuteron.

The further rearrangement contribution
\begin{equation}
\label{eq:reapot}
V^{(r)} = \frac{1}{2} \left( C_{\omega}^{\prime} n_{\omega}^{2} + C_{\rho}^{\prime} n_{\rho}^{2} - C_{\sigma}^{\prime} n_{\sigma}^{2} \right)
\end{equation}
in the vector potential (\ref{eq:vpot}) is due to the density dependence of the couplings 
%\red{$\Gamma_{m}$} \gre{$\Gamma_{j}$} and can \red{eb} \gre{be} 
$\Gamma_{j}$ and can be expressed with the coefficients 
%\red{$C_{m}$} \gre{$C_{j}$} 
$C_{j}$ and source densities %\red{$n_{m}$} \gre{$n_{j}$} 
$n_{j}$ of the three mesons considered here. 

Corresponding to the two rearrangement contribution in (\ref{eq:vpot}), there are also two
such terms in the total grand potential density (\ref{eq:omega_tot}): the meson term
\begin{equation}
    \tilde{\omega}_{\mathrm{meson}}^{(r)} = V^{(r)} n_{b}
\end{equation}
and the mass shift term
\begin{equation}
    \tilde{\omega}_{\mathrm{mass}}^{(r)} = \sum_{i} n_{i}^{(v)} W_i^{(r)}
\end{equation}
Furthermore, the meson contribution in (\ref{eq:omega_tot}) is given by
\begin{equation}
\label{eq:omega_meson}
\tilde{\omega}_{\mathrm{meson}} = - \frac{1}{2} \left( C_{\omega} n_{\omega}^{2} + C_{\rho} n_{\rho}^{2} - C_{\sigma}n_{\sigma}^{2} \right)
\end{equation}
similar in structure to (\ref{eq:reapot}).
%\blu{What do you mean with similar?}.

The single-particle number densities can be derived from (\ref{eq:omega_tot}) using the thermodynamic definitions
\begin{eqnarray}
\label{eq:vec_dens}
n_{i}^{(v)} &=& - \left. \dfrac{\partial \tilde{\omega}}{\partial \mu_i}\right|_{T, \{\mu_j\}_{j \ne i}} %= g_i \left[ \int \dfrac{d^3 k_i}{(2\pi)^3} d_i + \dfrac{1 - \sigma_i}{2} \xi_i \right] 
\\
\label{eq:sca_dens}
n_{i}^{(s)} &=&  \left. \dfrac{\partial \tilde{\omega}}{\partial m_i}\right|_{T, \{\mu_j\}} %= g_i \left[ \int \dfrac{d^3 k_i}{(2\pi)^3} \dfrac{m_i^{\ast}}{E_i} d_i + \dfrac{1 - \sigma_i}{2} \xi_i \right] 
\end{eqnarray} 
that give
\begin{eqnarray}
n_{i}^{(v)} &=& g_i \int \dfrac{d^3 k}{(2\pi)^3} d_i 
+ \frac{1 - \sigma_i}{2} n_{i}^{(\mathrm{cond})}
\\
n_{i}^{(s)} &=& g_i \int \dfrac{d^3 k}{(2\pi)^3} \dfrac{m_i^{\ast}}{E_i} d_i 
+ \frac{1 - \sigma_i}{2} n_{i}^{(\mathrm{cond})}
\end{eqnarray} 
with a thermal and a condensate contribution, once the distribution function
\begin{equation}
\label{eq:distr}
d_i (T, k, m_i^{\ast}, \mu_i^{\ast}) = \left[ \exp\left( \dfrac{E_i - \mu_i^{\ast}}{T}\right) + \sigma_i \right]^{-1}
\end{equation}
is defined. These expressions are consistent with the usual definitions of the densities.
The vector densities can be expressed as
\begin{eqnarray}
n_{n}^{(v)} &=& \frac{1+\beta-X_{d}}{2}n_{b} % = \frac{k_{n}^{3}}{3\pi^{2}} \\
\\
n_{p}^{(v)} &=& \frac{1-\beta-X_{d}}{2}n_{b} %= \frac{k_{p}^{3}}{3\pi^{2}} \\
\\
n_{d}^{(v)} &=& \frac{X_{d}}{2} n_{b} \label{eq:vec_dens_d}
\end{eqnarray}
using the baryon density $n_{b}$, the asymmetry $\beta$, and the deuteron fraction $X_{d}$.

The deuteron fraction has to stay below a maximum value of
\begin{equation}
\label{eq:Xdmax}
    X_{d}^{(\mathrm{max})} = 
    %\mbox{min} \left\{ 1, 
    \frac{m_{\mathrm{nuc}}}{\chi C_{\sigma} n_{b}} 
    %\right\}
\end{equation}
in order to ensure positive effective masses of the nucleons. 
%The first limit for $X_{d}^{\mathrm{max}}$ can be reached only when there are no free nucleons in %the system. 
%the second 
This limit is reached when the effective masses of the nucleons are zero, i.e.,
$S_{n} = S_{p} = C_{\sigma} n_{\sigma} = m_{\mathrm{nuc}}$ 
with the source density $n_{\sigma} = 2 \chi n_{d}^{(s)} = \chi X_d^{\rm (max)} n_{b}$ of the $\sigma$ meson. It only has a contribution from the quasi-deuterons because the scalar densities of the nucleons
vanish for $m_{\mathrm{nuc}}^{\ast}=0$. A second limitation of the deuteron fraction arises from the fact that 
%the minimum of the neutron and proton densities determines the maximum density of the quasi-deuterons in the matter because 
for every neutron a proton is needed, or vice versa, to form the cluster. This translates to the condition $X_{d} \leq 1-|\beta|$ depending on the isospin asymmetry $\beta$. So, in total, one has
$0 \leq X_{d} \leq \mbox{min} \left\{ X_{d}^{(\mathrm{max})}, 1 - |\beta|\right\}$.

At high densities or temperatures, %\red{there will be} 
a mixture of deuterons, neutrons and protons might be expected. 
If all three particle species have non-zero densities, the condition of chemical equilibrium 
\begin{eqnarray}
\label{eq:chem_equilibrium}
\mu_{d} = \mu_{n} + \mu_{p}, 
\end{eqnarray}
applies between the chemical potentials of the degrees of freedom involved. 
Using the definitions \eqref{eq:mustar} and \eqref{eq:mstar} with the potentials
\eqref{eq:spot} and \eqref{eq:vpot}, this relation can be written
as 
\begin{equation}
\label{eq:deltamd}
 \Delta m_{d}  = \mu_{n}^{\ast} + \mu_{p}^{\ast} - m_{d} + S_{d} +  V_{n} + V_{p} - V_{d},
\end{equation}
%or 
%\begin{eqnarray}
%\label{eq:deltamd}
%\lefteqn{\Delta m_{d}^{(\mathrm{high})}}
%\\ \nonumber & = & \mu_{n}^{\ast} + \mu_{p}^{\ast} - m_{d} 
%+ 2\chi C_{\sigma} n_{\sigma} 
%+ 2(1-\chi) C_{\omega} n_{\omega} 
%\end{eqnarray}
and thus an expression for the deuteron mass shift is obtained.

All thermodynamic quantities of the system
can be easily obtained from the grand canonical thermodynamic
potential \eqref{eq:omega_tot}.
For instance, the pressure is given by
\begin{equation}
\label{eq:pressure}
    P = - \omega(T,\{\mu_{i}\})
\end{equation}
and the free energy density %\red{$f=F/V$} \cya{
$\mathcal{F}$ can be expressed as
\begin{equation}
\label{eq:fed}
%    \red{f} \cya{
\mathcal{F} = \sum_{i=n,p,d} \mu_{i} n_{i}^{(v)} + \omega
    = \left( \mu_{b} + \frac{1-\beta}{2} \mu_{c} \right) n_{b} - P
\end{equation}
with the baryon chemical potential $\mu_{b}=\mu_{n}$ and the charge chemical potential
$\mu_{c} = \mu_{p}-\mu_{n}$. The entropy density finally can be written as
\begin{eqnarray}
\label{eq:sdens}
% \red{s} \cya{
\mathcal{S} & = & %\red{\frac{S}{V} =} 
- \left. \frac{\partial \omega}{\partial T}\right|_{\{\mu_{i}\}}
 + \mathcal{S}_{\mathrm{cond}}
 \\ \nonumber & = &
 - \sum_{i} g_{i} \int \frac{d^{3}k}{(2\pi)^{3}} \:
 \left[ d_{i} \ln d_{i} + \frac{1-\sigma_{i} d_{i}}{\sigma_{i}}
 \ln \left( 1 - \sigma_{i} d_{i}\right)
 \right]
 \\ \nonumber & & 
 + n_{d}^{(\mathrm{cond})} \ln g_{d}
\end{eqnarray}
with the distribution functions \eqref{eq:distr}
and a contribution of the condensed deuterons. The latter contribution arises because deuterons have spin $1$ and thus the ground state of nuclear matter at $T=0$ contains a mixture of the different spin substates.

%%%%%%%%%%%%%%%%%%%%%%%%%%%%%%%%%%%%%%%%%%%%%%%%%%%%%%%%%%%%%%%%%%%%%
\subsection{Zero temperature and quasi-deuteron condensation}

If the temperature vanishes, %is restricted to $zero$, 
the vector and scalar densities of the the nucleons $q=n,p$ can be expressed in analytical form as
\begin{equation}
\label{eq:nqv}
    n_{q}^{(v)} = \frac{g_{q}}{6\pi^{2}} k_{q}^{3}
\end{equation}
with $g_{q}=2$ and
\begin{equation}
\label{eq:nqs}
n_{q}^{(s)} = \frac{g_{q}m_{q}^{\ast}}{4\pi^{2}} \left[ k_{q} \mu_{q}^{\ast} - (m_{q}^{\ast})^{2} \ln \frac{k_{q}+\mu_{q}^{\ast}}{m_{q}^{\ast}}\right]
\end{equation}
with the Fermi momentum $k_{q}$ and the effective chemical potential
\begin{equation}
\label{eq:mu_q_eff}
 \mu_{q}^{\ast}  = \sqrt{k_{q}^{2}+\left( m_{q}^{\ast}\right)^{2}} \: .
\end{equation}
Since quasi-deuterons are bosons, they can exist only as a condensate
in the zero temperature case we are focusing on.
For the deuteron there is no thermal contribution to the density but the condensate term
\begin{equation}
\label{eq:d_dens}
    n_{d}^{(v)} = n_{d}^{(s)} = n_{d}^{(\mathrm{cond})}
\end{equation}
with equal vector and scalar densities and the effective chemical potential becomes
\begin{equation}
\label{eq:mu_d_eff}
 \mu_{d}^{\ast} = m_{d}^{\ast} \: .
\end{equation}

%\subsection{Quasi-deuteron condensation}

In nuclear matter at very low densities, it is advantageous to form
quasi-deuterons, which still have a positive binding energy, 
to gain energy as
compared to a system composed of neutrons and protons only. In SNM, all protons and neutrons will be bound in quasi-deuterons in the low-density limit and no free nucleons remain. The binding energy per nucleon will approach half of the deuteron binding energy for $n_{b} \to 0$ and not zero as for homogeneous nucleonic matter without clusters. For ANM, only quasi-deuterons and neutrons (protons) will be the active constituents in case of positive (negative) isospin asymmetry $\beta$. 

With the help of the nucleon and deuteron densities, a simple expression for the pressure $P$ is found. Since the the momentum
integral \eqref{eq:omega_sp} can be calculated explicitly after partial integration, one obtains 
\begin{eqnarray}
    P & = & \sum_{q=n,p}
    \frac{1}{4} \left( \mu_{q}^{\ast} n_{q}^{(v)}- m_{q}^{\ast} n_{q}^{(s)}\right) 
    \\ \nonumber & & 
    + \frac{1}{2} \left( D_{\omega} n_{\omega}^{2} + D_{\rho} n_{\rho}^{2} - D_{\sigma}n_{\sigma}^{2} \right)
    %\\ \nonumber & &    
    %+ \frac{1}{2} \left( C_{\omega} n_{\omega}^{2} + C_{\rho} n_{\rho}^{2} - %C_{\sigma}n_{\sigma}^{2} \right)
    %\\ \nonumber & & 
    %+ \frac{1}{2} \left( C_{\omega}^{\prime} n_{\omega}^{2} + C_{\rho}^{\prime} n_{\rho}^{2} - %C_{\sigma}^{\prime} n_{\sigma}^{2} \right) n_{b}
    %\\ \nonumber & & 
    %+ n_{d}^{(s)} \sum_{i} n_{i}^{(v)} \frac{\partial \Delta m_{d}}{\partial n_{i}^{(v)}}
    + \sum_{i=n,p,d} n_{i}^{(v)} W_{i}^{(r)}
\end{eqnarray}
with
\begin{equation}
\label{eq:Ddef}
    D_{i} = C_{i} + C_{i}^{\prime} n_{b}
\end{equation}
for the meson couplings ($i=\sigma$, $\omega$, and $\rho$). Because the entropy density \eqref{eq:sdens} vanishes for $T=0$, the internal energy density  %\red{$\varepsilon=E/V$} \cya{
$\mathcal{E}$ %} 
is identical to the free energy density \eqref{eq:fed}.

\subsection{Coupling strength scaling factors}

In Eqs.\ %\red{\eqref{eq:spot}, \eqref{eq:vpot} and \eqref{eq:meson_dens}}
\eqref{eq:spot} - \eqref{eq:nrho}, the scaling factors %\red{$\chi_{im}$} \gre{$\chi_{ij}$} 
$\chi_{ij}$ appear. These factors are always unitary for nucleons and, in particular, 
\begin{eqnarray}
\chi_{n\sigma} &=& \chi_{n\omega} = \chi_{n\rho} = 1 \nonumber \\
\chi_{p\sigma} &=& \chi_{p\omega} = - \chi_{p\rho} = 1
\end{eqnarray}
for the three mesons.
For the nucleons bound in clusters, the choice of the scaling factors is instead widely debated \cite{ferreiraPRC2012}.

It is a natural choice to assume that the nucleons inside the deuteron couple to the mesons with the same strength as the unbound nucleons. %\sout{considerably similifies the theoretical framework depicted in Section \ref{sec:grdf}.}
Nevertheless, previous studies have already shown that, to take into account in-medium effects in calculation of the EoS for astrophysical applications, a universal scaling factor smaller than $1$ should be assumed for the cluster-meson coupling strength \cite{paisPRC2018, paisPRC2019}. A reduced value for the coupling of the $\sigma$-meson to different light clusters, including deuterons, allows %\gre{also} 
also a good description of the chemical equilibrium constants determined from the NIMROD data \cite{qinPRL2012}. However, recent Bayesian analysis \cite{paisPRL2020, paisJPG2020} have highlighted that a larger value should be taken for the cluster $\sigma$-meson coupling, to describe recent results from INDRA collaboration as well \cite{bougaultJPG2020}. Moreover, a possible model-dependence of this result was recently assessed \cite{custodioEPJA2020}, calling for further microscopic analysis to get more stringent constraints. On the other hand, the possible choice of different values for the $\sigma$ and $\omega$ coupling strength factors is known to produce a strong imbalance between the corresponding scalar and the vector potentials. %\red{, which} \gre{This would in turn}
This would, in turn, reflect itself in an unrealistic behavior of the nuclear EoS, 
%\gre{owing to the corresponding strong change of the central potential,} %\red{because the central potential}, 
owing to a corresponding change of the central potential given by the difference $V_{i}-S_{i}$ of very large potentials in lowest order non-relativistic approximation. %\red{, would change strongly}. 
A reasonable choice would be therefore to assume the same scaling factor for both $\sigma$ and $\omega$ meson ($\chi_{d \sigma} = \chi_{d \omega} = \chi$) and explore the sensitivity of our results to this ingredient ($\chi_{d\rho}=0$ because the deuteron has zero isospin.) This is thus the strategy that will be adopted in the following.

%%%%%%%%%%%%%%%%%%%%%%%%%%%%%%%%%%%%%%%%%%%%%%%%%%%%%%%%%%%%%%%%%%%%%
\section{Quasi-deuteron mass shift constraints}
\label{sec:mass_shift_constraints}

In the GRDF, a mass shift $\Delta m_d$ is introduced in the 
%\red{scalar potential defined in Eq. \eqref{eq:spot}} 
effective mass defined in Eq.\ \eqref{eq:mstar} in order to suppress the cluster formation at supra-saturation densities. This mass shift may generally receive several contributions. For example, in compact star matter, where electrons are included to fulfill the requirement of charge neutrality, a possible contribution comes from the screening of the Coulomb potential produced by the electronic background.
%\red{in compact star matter, where electrons are included to fulfill the requirement of charge neutrality}. 

In the nuclear matter case, the mass shift is usually provided just from Pauli blocking, which plays a crucial role in suppressing the cluster formation. Indeed, the Pauli exclusion principle implies that a single-particle state in momentum space would not be longer available for formation of a cluster when it is already occupied by nucleons of the medium. The Pauli blocking of states strongly reduces at high temperatures with increasing diffuseness of the Fermi sphere or when the c.m.\ momentum of the cluster is much larger than the typical radius of the Fermi sphere. This effect can be represented effectively as a repulsive, medium-dependent potential or a change of the cluster binding energy. A quantitative value can be calculated by solving the many-body Schr\"odinger equation when proper potentials for the nucleons in matter are introduced.
%\blu{The sentence in cyan is not clear to me!} 
The corresponding results, which are explicitly calculated for various conditions of temperature, density and isospin asymmetry of the medium, are then usually approximated by suitable parameterizations in a wide range of thermodynamic variables \cite{typelPRC2010, ropkePRC2009, ropkeNPA2011, ropkePRC2015}. 

The change of the cluster binding energy, calculated in this approach, is valid only at sub-saturation densities and has to be extrapolated to higher densities, in particular above the cluster dissociation (Mott) density, where the binding energy vanishes, and only a many-body correlation in the continuum remains. For this purpose, suitable heuristic parameterizations were introduced within the GRDF \cite{paisNPCCP2017,Typel:2018wmm}.
Focusing on deuteron-like correlations, the aim of our work is to provide a unified parameterization of the quasi-deuteron binding energy shifts, so that it can be used as effective means
%\red{means} \gre{mean} \blu{No, "means"="instrument" is correct (generic plural), "mean"="average"} \blu{Ok, sorry! So, maybe one should remove "an", isn't it?} 
to treat SRCs at supra-saturation densities. In this context, the pioneering analysis developed in Ref.\ \cite{typelEPJA2020} already suggests that a substantial modification of the cluster mass shift has to be expected in the regime beyond the deuteron dissociation density, in comparison to the traditionally adopted form. 

Such a parameterization should be appropriately chosen to interpolate between the low-density limit constrained by microscopic many-body calculations and the high-density behavior %derived 
postulated under the assumed boson condensation condition. 
%\red{In order to keep a proper description of well-constrained nuclear matter quantities around the %equilibrium saturation point of symmetric nuclear matter (SNM), some %prescriptions conditions have to be fulfilled by the proposed parameterization.
%\blu{In the last version, this part is not included anymore in the present secttion. I think it can be removed, since its introduction at this level is too premature.}
%at saturation. 
In the following subsections the different constraints to be imposed to the deuteron binding-energy shifts will be discussed. As will be seen, various scenarios will 
emerge depending also on the value of the deuteron-meson scaling factors $\chi = \chi_{d\sigma} = \chi_{d\omega}$ introduced in Eqs.\ \eqref{eq:spot} - \eqref{eq:nrho}.

%because
%\begin{equation}
%    W_{n}^{(r)} + W_{p}^{(r)} - W_{d}^{(r)}  = n_{d}^{(s)}
%    \frac{\partial \Delta m_{d}}{\partial n_{b}}
%  \left( 1 + 1 - 2 \right) = 0
%\end{equation}
%due to the definition \eqref{eq:ndeff} of the effective density.

%%%%%%%%%%%%%%%%%%%%%%%%%%%%%%%%%%%%%%%%%%%%%%%%%%%%%%%%%%%%%%%%%%%%%
\subsection{Low-density constraint}

Several parameterizations of the binding energy or mass shift of the deuteron 
at sub-saturation densities have been developed to account for the Pauli blocking effects. In particular, when neglecting the dependence of the mass shift on the momentum of the deuteron with respect to the medium and limiting to the zero temperature case, the simplified functional form 
\begin{equation}
\label{eq:deltamd_low}
\Delta m_d^{(\mathrm{low})} = \delta B_{d} (0) n_d^{(\mathrm{eff})}
\end{equation}
can be assumed, where the effective vector density $n_d^{(\mathrm{eff})}$, as defined in Ref.\ \cite{typelEPJA2020}
\begin{equation}
\label{eq:ndeff}
n_d^{(\mathrm{eff})} =
\frac{2}{A_{d}} \left[ N_{d}\left(n_{n}^{(v)}+n_{d}^{(v)}\right) + Z_{d}\left(n_{p}^{(v)}+n_{d}^{(v)}\right)\right] %= n_{n}^{(v)} + n_{p}^{(v)} + 2 n_{d}^{(v)} = n_{b}, 
\end{equation}
with $N_{d} = Z_{d}=1$ and $A_{d}=2$, turns out to be equal to the total baryon density $n_b$, independent of the global isospin asymmetry $\beta$ of the system. %, owing to the symmetry of the cluster here considered.
The quantity $\delta B_{d}$ generally regulates the temperature dependence; its zero temperature limit which appears in Eq.\ \eqref{eq:deltamd_low} is $\delta B_{d} (0) = 3634.16$ MeV fm$^3$, as given in Ref.\ \cite{typelPRC2010}, where explicit expressions for finite temperature calculations are also provided.

A linear increase of the mass shift %with 
proportional to $n_b$ at low baryon densities is consistent with the results obtained in \cite{typelPRC2010, ropkePRC2009, ropkeNPA2011, ropkePRC2015}, at least for density values lying below the dissociation or Mott density defined as
\begin{equation}
\label{eq:ndiss}
n_d^{(\mathrm{diss})} = \dfrac{B_{d}}{\delta B_d (0)},
\end{equation}
where $B_{d} = m_n+m_p-m_d = 2.225$ MeV \cite{wangCPC2012} is the deuteron binding energy in vacuum.

%%%%%%%%%%%%%%%%%%%%%%%%%%%%%%%%%%%%%%%%%%%%%%%%%%%%%%%%%%%%%%%%%%%%%
\subsection{High-density limit}
\label{sec:highdens}

In the traditional treatment of cluster dissolution using the concept of mass shifts, a heuristic density dependence stronger than linear %\cya{in} 
in $n_b$ is customarily assumed at baryon densities above the dissociation density to prevent the clusters to reappear. A divergence of the mass shift ensures in particular the deuteron removal from the system when approaching saturation density, resulting in pure nucleonic matter above saturation. 
The possibility of using quasi-deuterons to effectively embed nuclear SRCs at supra-saturation densities requires thus a proper change of the usual parameterization 
adopted in the GRDF, 
%. In particular, only a much weaker increase of the mass shift at high densities allows the clusters to survive, 
as first noticed in Ref.\ \cite{typelEPJA2020}.

%%%%%%%%%%%%%%%%%%%%%%%%%%%%%%%%%%%%%%%%%%%%%%%%%%%%%%%%%%%%%%%%%%%%%
\subsubsection{Deuteron mass shift}

Assuming a dependence %\red{of} \cya{on} 
on the effective density $n_{d}^{(\mathrm{eff})}=n_{b}$, as defined in Eq.\ \eqref{eq:ndeff}, the quasi-deuteron mass shift 
expression derived in Eq.\ \eqref{eq:deltamd} simplifies to
\begin{eqnarray}
\label{eq:deltamd_high}
\lefteqn{\Delta m_{d}^{(\mathrm{high})}}
\\ \nonumber & = & \mu_{n}^{\ast} + \mu_{p}^{\ast} - m_{d} 
+ 2\chi C_{\sigma} n_{\sigma} 
+ 2(1-\chi) C_{\omega} n_{\omega} 
\end{eqnarray}
because
\begin{equation}
    W_{n}^{(r)} + W_{p}^{(r)} - W_{d}^{(r)}  = n_{d}^{(s)}
    \frac{\partial \Delta m_{d}}{\partial n_{b}}
  \left( 1 + 1 - 2 \right) = 0 \: .
\end{equation}
Eq.\ \eqref{eq:deltamd_high} can be used to calculate the deuteron mass shift as a function of the baryon density $n_{b}$,
the deuteron fraction $X_{d}$ %\red{, and given} 
and the isospin asymmetry $\beta$.

\begin{figure*}[tbp!]
\includegraphics[width=.45\textwidth]{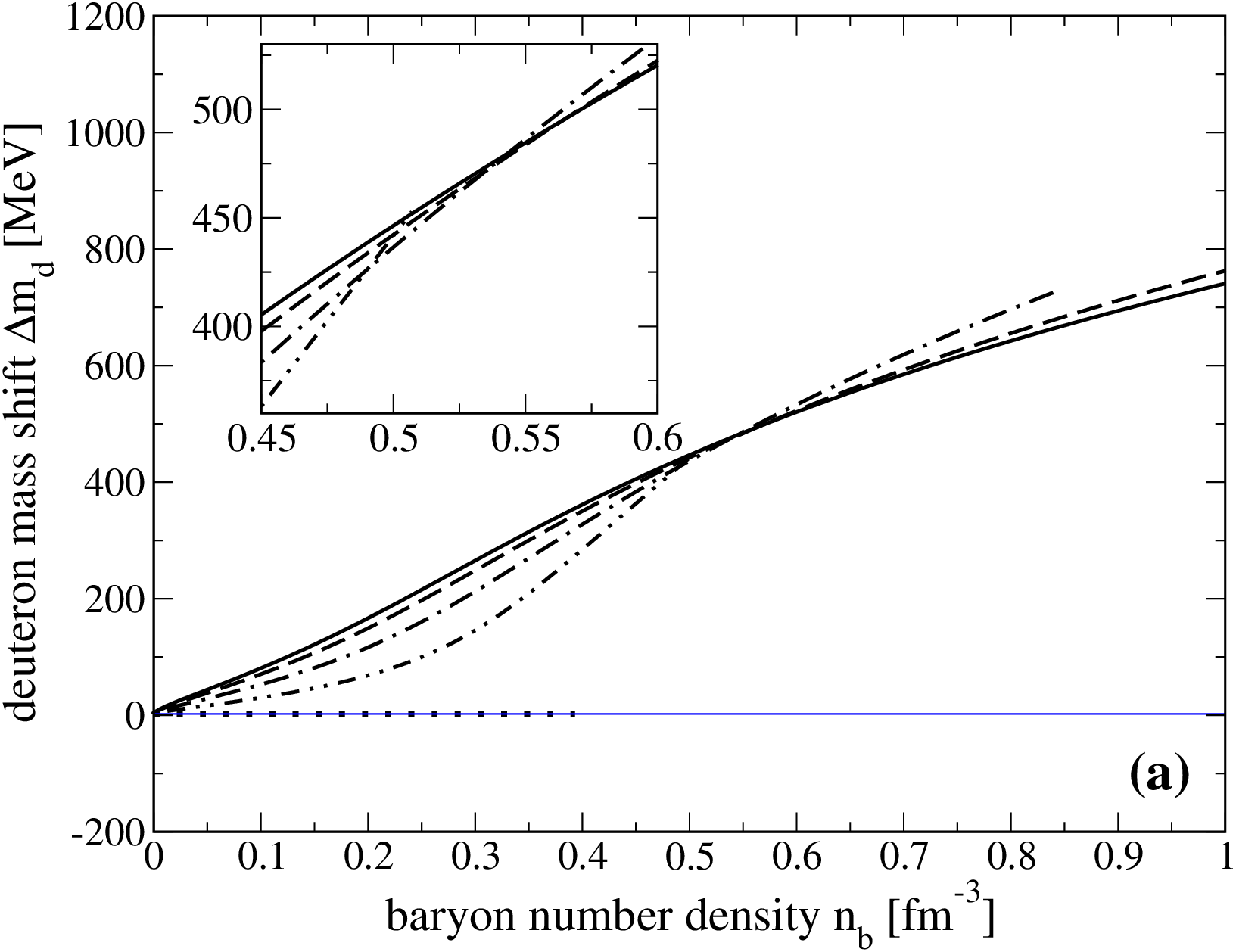} \qquad \includegraphics[width=.45\textwidth]{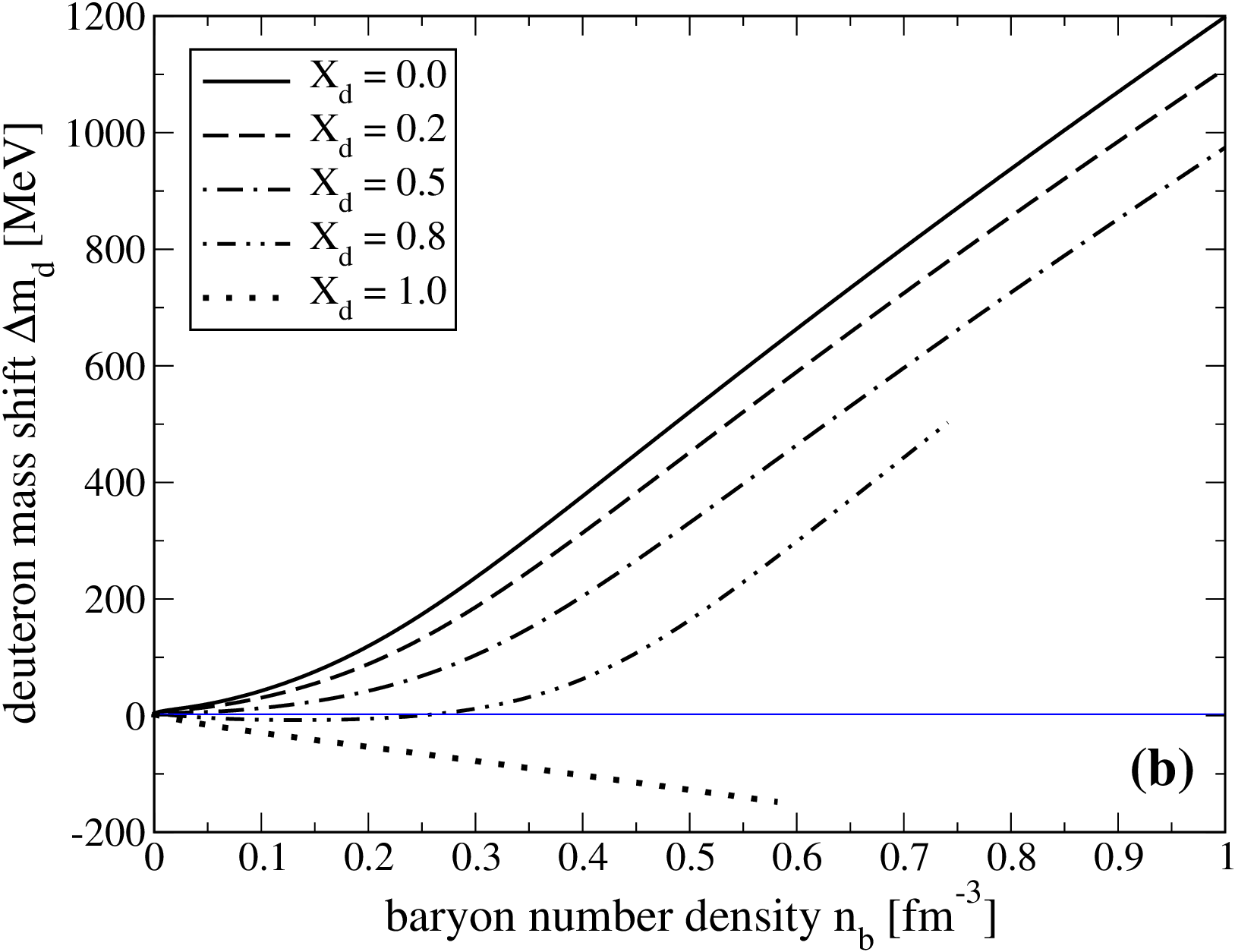}
\caption{\label{fig:deltamd_high} Panel (a): Deuteron mass shift as function of the baryon density, as determined according to Eq.\ \eqref{eq:deltamd_high} in the SNM case, by assuming a unitary scaling factor $\chi$ for the deuteron-meson coupling strengths. Panel (b): the same as in panel (a), but assuming a reduced scaling factor $\chi = 1/\sqrt{2}$. In both panels, the DD2 parameterization \cite{typelPRC2010} of the nucleon-meson effective interaction is adopted.
The inset in panel (a) shows a zoom around the density values where the curves cross each other. The thin blue line indicates in both panels the binding energy of the deuteron in vacuum $B_{d}$.}
\end{figure*}

An impression of the density dependence of the quasi-deuteron mass shift is depicted in Fig.\ \ref{fig:deltamd_high} for given, constant mass fractions $X_{d}$ and two values of the scaling factor $\chi$ of the deuteron. 
%}
%
%\red{In Fig. \ref{fig:deltamd_high}, the baryon density behavior of the deuteron mass shift, as determined by Eq. \eqref{eq:deltamd_high}, is shown for two different values of the scaling factor:} 
A unitary value of $\chi$ is considered in panel (a), whereas a reduced scaling factor of $\chi=1/\sqrt{2}$ is assumed in panel (b). The choice of the latter $\chi$ value  %\red{adopted in panel (b), {\it i.e.} $\chi = \frac{\sqrt{2}}{2}$,} 
will be %\red{justified} \gre{
explained below. In both panels, for sake of simplicity, the SNM case is considered. 
Moreover, the DD2 parameterization \cite{typelPRC2010} of the nucleon-meson effective interaction is adopted.

%\gre{
%\red{The behavior of the deuteron mass shift obviously depends on the choice of $\chi$.} \blu{This is also stated in the next paragraph, when discussing the high-density behavior. Moreover, I do not find that it is so obvious that the deuteron mass shift should be qualitatively different depending on the choice of $\chi$.}
%\red{As one intuitively expects, Fig.\ } \cya{Figure}
Fig.\ \ref{fig:deltamd_high} shows that the deuteron mass shift
$\Delta m_{d}$ depends quite strongly on the value assumed for the scaling factor $\chi$. This is deduced by comparing the results of panel (a) with the corresponding ones of panel (b), especially in the high-density regime, we are focusing on. 
Indeed, 
%\red{in the general case in which the deuteron mass fraction $X_{d}$ assumes any value smaller than 1,} \blu{I do not remember why I wrote this sentence.} 
for %\red{$n_{b}\to \infty$} %\red{larger} 
large densities, 
%\blu{For $n_{b}\to \infty$, only $X_{d} = 0$ can be considered.}, 
the effective chemical potentials of the nucleons in Eq.\ \eqref{eq:deltamd_high} are dominated by the Fermi momenta $k_{q}$ as defined in Eq.\ \eqref{eq:nqv} 
%\blu{Eq. \ \eqref{eq:vec_dens_d} does not include anymore the Fermi momenta}
and thus exhibit a
$n_{b}^{1/3}$ behavior. Since the meson couplings $C_{\sigma}$ and $C_{\omega}$ approach constants
at high densities, the mesonic contributions to the mass shift are determined by the behavior of the source densities $n_{\sigma}$ and $n_{\omega}$. For $\chi \neq 1$, the source density %\red{$n_{\omega}=n_{b}$ is} 
$n_{\omega}$ rules the dominating term. 
Since the deuteron fraction has to vanish for $n_b \to \infty$
due the constraint \eqref{eq:Xdmax}, a linear asymptotic %\red{and a linear} 
increase of $\Delta m_{d}$ with the baryon density is expected. 
Indeed, such a dependence is observed in panel (b) of Fig.\ \ref{fig:deltamd_high} for $\chi = 1/\sqrt{2}$. A softer increase of the mass shift with the baryon density is seen in panel (a) because the last term in Eq.\ \eqref{eq:deltamd_high}
does not contribute for $\chi=1$ and the high-density behavior is driven by the 
%\red{contribution of the $\sigma$ meson and an} 
$\sigma$ meson term. 
%\red{Since the deuteron fraction has to vanish for $n_b \to \infty$ due the constraint \eqref{eq:Xdmax},} 
However, the source density $n_{\sigma}$ may asymptotically receive contributions only from the scalar densities of the nucleons because the deuteron fraction asymptotically approaches zero. 
An asymptotic dependence proportional to $n_b^{2/3}$ is 
%\red{expected since the 
%contribution of the deuteron density to $n_{\sigma}$ is limited if %the deuteron fraction is constrained by $X_{d}^{(\mathrm{max})}$}
then expected for the $\sigma$ meson term and the deuteron mass shift of Eq.\ \eqref{eq:deltamd_high}.

In the (unrealistic) case when all nucleons are bound inside the deuteron ($X_{d} = 1$), independent of the density, both the scalar and vector densities of the nucleons vanish. The mass shift defined in Eq.\ \eqref{eq:deltamd_high} assumes the following form
\begin{equation}
\Delta m_{d}^{(\mathrm{high})} (X_{d} = 1) = B_d(0) + 2 \chi (1 - \chi) (C_{\omega} - C_{\sigma})n_{b} \: ,
\end{equation}
i.e., a linear dependence on $n_b$ is generally predicted, 
for any finite value of the scaling factor $\chi$, 
as shown by the dotted line of panel (b). %\red{The only} 
An exception is the case with $\chi = 1$, when the deuteron mass-shift coincides with the deuteron binding energy, see dotted line in panel (a).

Furthermore, Fig.\ \ref{fig:deltamd_high} shows that the largest deuteron fraction generally
corresponds to the lowest mass shift, %\gre{
with a clear ordering in panel (b). However, in the case when a unitary scaling factor is adopted, such a statement actually holds only 
%below \red{a crossing point %\gre{of the lines} \cya{the region where the lines cross each other},
below the region where the lines cross each other, which is observed 
%\red{at $n_c \simeq 0.55$ fm$^{-3}$} 
above $0.45$~fm$^{-3}$.
This region is also emphasized in the inset of panel (a) of Fig.\ \ref{fig:deltamd_high} to evidence the fact that there is no single crossing point between the different curves.

The emergence of this crossing can be easily understood when looking at the explicit form of Eq.\ \eqref{eq:deltamd_high}. %\blu{Should we also include a figure for the deuteron mass shift as function of the deuteron mass fraction, at least in the case with $\chi = 1$ for selected values of the baryon density and in the simplified case of SNM?} 
Indeed, for $\chi = 1$, the vector contribution vanishes, and there exists a delicate interplay between the remaining terms. At low densities, the leading role is played by the effective chemical potentials, which reduce when the deuteron mass fraction increases. However, at larger densities, the importance of the term involving the source density of the scalar meson is enhanced. Since the corresponding contribution increases with the deuteron fraction, a crossing among the curves 
is observed at a certain baryon density $n_b^{\mathrm{cross}}$ and an inversion of their previous ordering is expected at higher densities. 
On the other hand, for smaller values of the scaling factor, as the one considered in panel (b) of Fig.\ \ref{fig:deltamd_high}, the vector term plays also a role. The source density of the $\omega$ meson field does not change with the deuteron mass fraction at constant $n_{b}$. 
%\blu{I also do not remember what means this sentence we wrote.} 
%\gre{The source density of the $\omega$ meson field is independent of the deuteron mass fraction at constant $n_{b}$.} 
%\blu{Maybe this is more clear.}
Since the vector term dominates asymptotically over the scalar one, in light of the power of density involved there, for scaling factor values small enough, such as the one considered in panel (b) of Fig.\ \ref{fig:deltamd_high}, no crossing is observed and the
same ordering %\red{as observed} \cya{exhibited} in the low-density regime 
for all allowed densities is preserved.
%\blu{The curves do not cross all at the same point. Question: why do the curves cross? Let us try to look at the deuteron mass shift derivative with respect to the mass fraction, at constant total baryon density.}

Moreover, it is worthwhile to notice in panel (a) that the dotted-dashed line, which corresponds to the deuteron mass shift obtained for a fixed deuteron mass fraction $X_d = 0.5$, ends %\red{is interrupted} 
at a density around $0.84$~fm$^{-3}$. Beyond this value, %\red{in fact,} 
the deuteron mass fraction would %\gre{
exceed its maximum allowed value, see Eq.\ \eqref{eq:Xdmax}. 
The same occurs also for the two largest deuteron mass fraction values considered in
both panels of Fig.\ \ref{fig:deltamd_high}, although at  smaller densities.
%\begin{equation}
%\label{eq:xdmax}
%X_d^{\rm max} = \dfrac{m}{\chi C_{\sigma} n_b}    
%\end{equation} 
%\red{and a negative value for the Dirac effective is predicted.}
%\footnote{Equation \eqref{eq:xdmax} is %indeed derived by using Eqs. \eqref{eq:mstar}, 
%\eqref{eq:meson_dens}, \eqref{eq:vec_dens_d} and \eqref{eq:d_dens}, assuming a vanishing value for %the effective mass.}
It is also worth noticing that Eq.\ \eqref{eq:Xdmax} implies that any finite asymptotic value for the deuteron mass fraction is not allowed. As a result, the cluster is forced to dissolve asymptotically, at least as $n_b^{-1}$, for any finite value of $\chi$.

\begin{figure*}[tbp!]
\includegraphics[width=.45\textwidth]{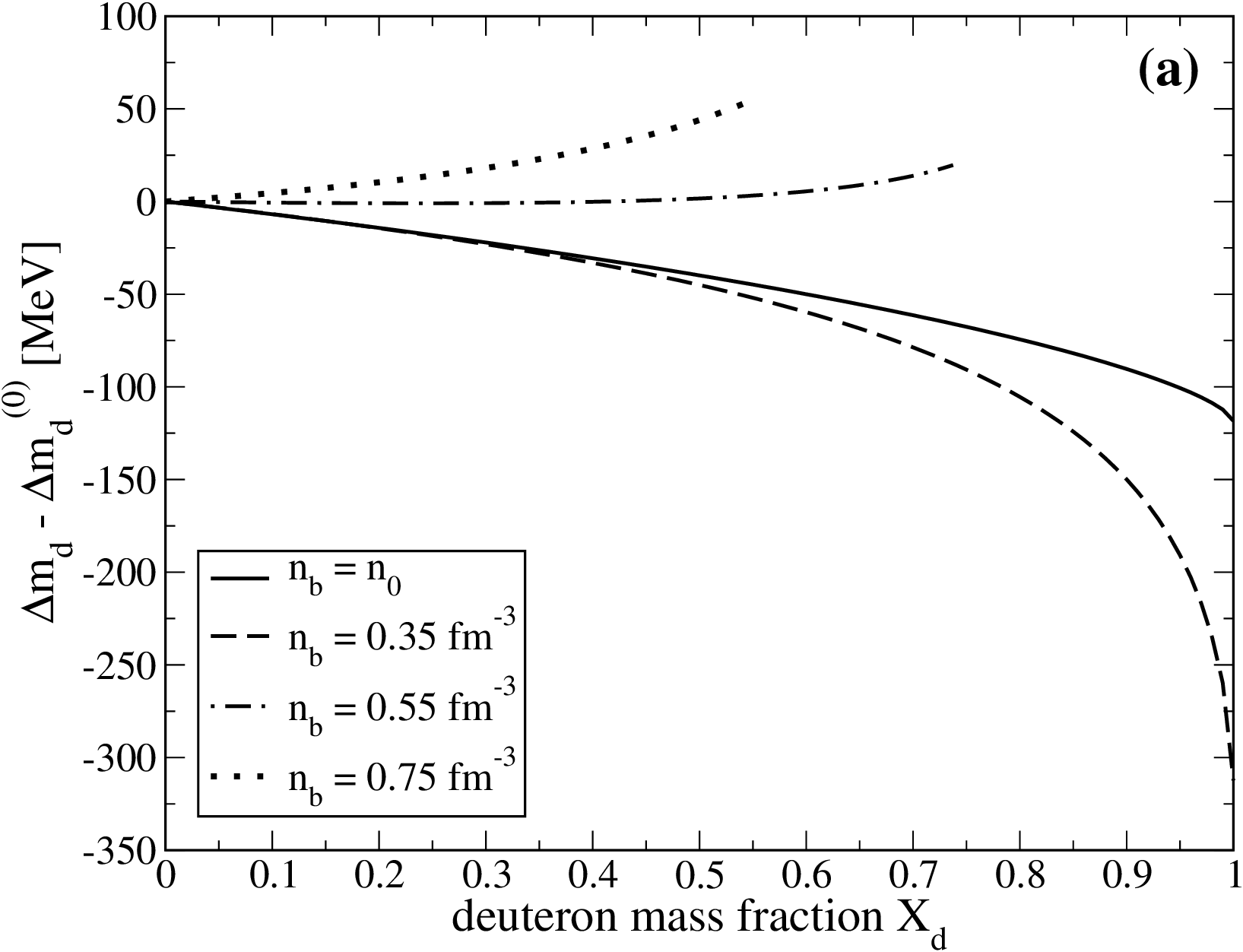} \qquad \includegraphics[width=.45\textwidth]{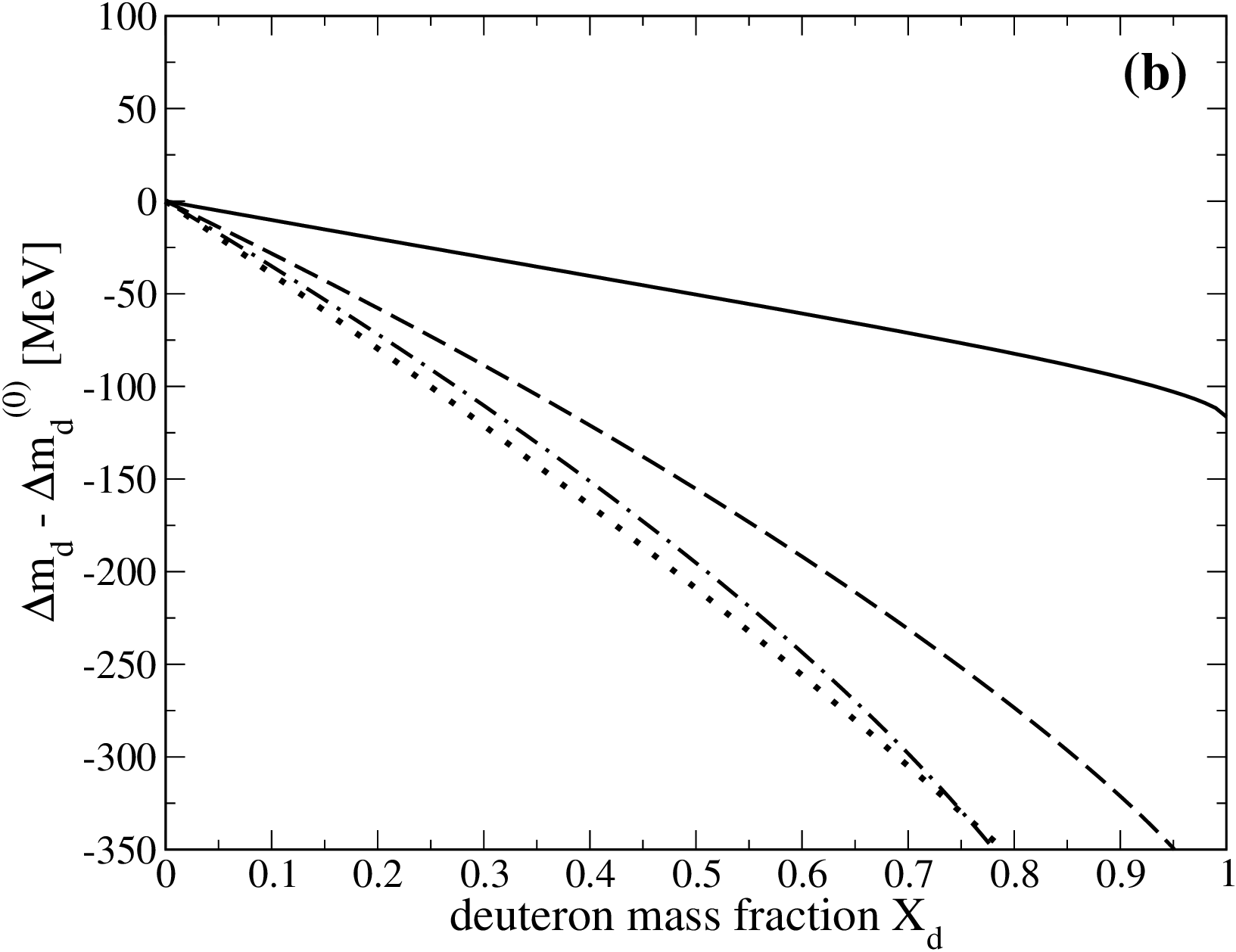}
\caption{\label{fig:deltamd_xd_high} Panel (a): Difference between the deuteron mass shift $\Delta m_{d}$, as determined according to Eq.\ \eqref{eq:deltamd_high} 
and its value in the case with zero deuteron mass fraction $\Delta m_{d}^{(0)}$, 
as function of the deuteron mass fraction $X_{d}$ 
%\red{as determined according to Eq.\ \eqref{eq:deltamd_high} in the SNM case, by assuming a}
for different, but constant, values of the baryon number density $n_{b}$. The SNM case is considered and a unitary scaling factor $\chi$ for the deuteron-meson coupling strengths 
are assumed.
%\red{and different, but constant values of the baryon number density $n_{b}$}. 
Panel (b): the same as in panel (a), but assuming a reduced scaling factor $\chi = 1/\sqrt{2}$. In both panels, the DD2 parameterization \cite{typelPRC2010} of the nucleon-meson effective interaction is adopted.
%The inset in panel (a) shows a zoom around the density values where the curves cross each other.
}
\end{figure*}

The observation of the line crossing in panel (a) of 
%\red{Fig.\ \ref{fig:ddeltamd_high}} 
Fig.\ \ref{fig:deltamd_high} motivates
to study the dependence of the deuteron mass shift on the deuteron mass fraction for constant
baryon density $n_{b}$ and constant asymmetry $\beta$. A deeper insight into this behavior can be achieved, by looking at Fig.\ \ref{fig:deltamd_xd_high}, where $\Delta m_{d}$ is plotted as a function of the deuteron mass fraction for different baryon density values. In order to facilitate the comparison among the curves, in Fig.\ \ref{fig:deltamd_xd_high} the differences with respect to the deuteron mass shift evaluated in the limiting case without deuteron (denoted as $\Delta m_d^{(0)}$) are actually considered. These differences turn out to be systematically lower than zero, at least for the reduced value of the scaling factor considered in panel (b). A change of sign is instead observed for $\chi = 1$ at larger densities, {\it i.e.}, beyond the crossing point observed in the panel (a) of Fig.\ \ref{fig:deltamd_high} and already discussed above.
Quite interestingly, one observes in both panels that the (negative) slope of the curves strongly increases for those lines that approach $X_{d}=1$. This represents the ideal, but unrealistic, case where no free nucleons exist in the system also at supra-saturation densities. The reason behind this behavior will be clarified in the following, when the mass shift derivatives are investigated in detail.

%%%%%%%%%%%%%%%%%%%%%%%%%%%%%%%%%%%%%%%%%%%%%%%%%%%%%%%%%%%%%%%%%%%%%
\subsubsection{Mass shift derivatives}

%\blu{This subsection might be eventually moved to as a paragraph of the previous subsection.}
%\red{The observation of the line crossing in panel (a) of 
%%\red{Fig.\ \ref{fig:ddeltamd_high}} 
%Fig.\ \ref{fig:deltamd_high} motivates
%to study the dependence of the deuteron mass shift on the deuteron mass fraction for constant
%baryon density $n_{b}$ and asymmetry $\beta$.} 
In Appendix \ref{sec:ddeltamd_dxd}, 
an explicit, general expression for the mass fraction derivative of the mass shift is derived.
For sake of simplicity, the case of SNM is considered here, where
the Fermi momenta, effective masses and effective chemical potentials
of the nucleons are identical, i.e., $k_{\mathrm{nuc}}=k_{n}=k_{p}$, 
$m_{\mathrm{nuc}}^{\ast}=m_{n}^{\ast}=m_{p}^{\ast}$, 
$\mu_{\mathrm{nuc}}^{\ast}=\mu_{n}^{\ast}=\mu_{p}^{\ast}$. Then 
the simple form
\begin{eqnarray}
\label{eq:ddmddxd_SNM}
  \lefteqn{\left. \frac{\partial \Delta m_{d}^{(\mathrm{high})}}{\partial X_{d}} \right|_{n_{b},\beta=0} =}
  \\ \nonumber & & 
  \left[ \frac{2C_{\sigma}}{1 + f_{\mathrm{nuc}}   C_{\sigma}}
 \left( \chi - 
 \frac{m_{\mathrm{nuc}}^{\ast}}{\mu_{\mathrm{nuc}}^{\ast}}\right)^{2}
     -\frac{\pi^{2}}{k_{\mathrm{nuc}}} \frac{1}{\mu_{\mathrm{nuc}}^{\ast}} \right] n_{b}
\end{eqnarray}
with the factor
\begin{equation}
\label{eq:fnuc}
    f_{\mathrm{nuc}} = 3 \left( \frac{n_{\mathrm{nuc}}^{(s)}}{m_{\mathrm{nuc}}^{\ast}}
    - \frac{n_{\mathrm{nuc}}^{(v)}}{\mu_{\mathrm{nuc}}^{\ast}}\right)
\end{equation}
and the total nucleon densities
$n_{\mathrm{nuc}}^{(v)}=n_{n}^{(v)}+n_{p}^{(v)}$ and 
$n_{\mathrm{nuc}}^{(s)}=n_{n}^{(s)}+n_{p}^{(s)}$ is obtained.
%\blu{I would keep the following sentence.}
%\gre{
The derivative \eqref{eq:ddmddxd_SNM} is the difference of two
positive contributions and
depending on the choice of $\chi$ there can be a zero at a certain %\red{positive} 
baryon density. %} % $n_{b}^{\mathrm{cross}}$.
%\red{Below this value an ordering of the lines as observed in the low-density region of panel (a) %and in panel (b) of Fig.\ \ref{fig:deltamd_high} follows, whereas an inversion will appear above %$n_{b}^{\mathrm{cross}}$
%when a unitary scaling factor of the deuteron-meson coupling strengths, see Fig.\ %\ref{fig:deltamd_high}, panel (a).}
\begin{figure*}[tp!]
\includegraphics[width=.45\textwidth]{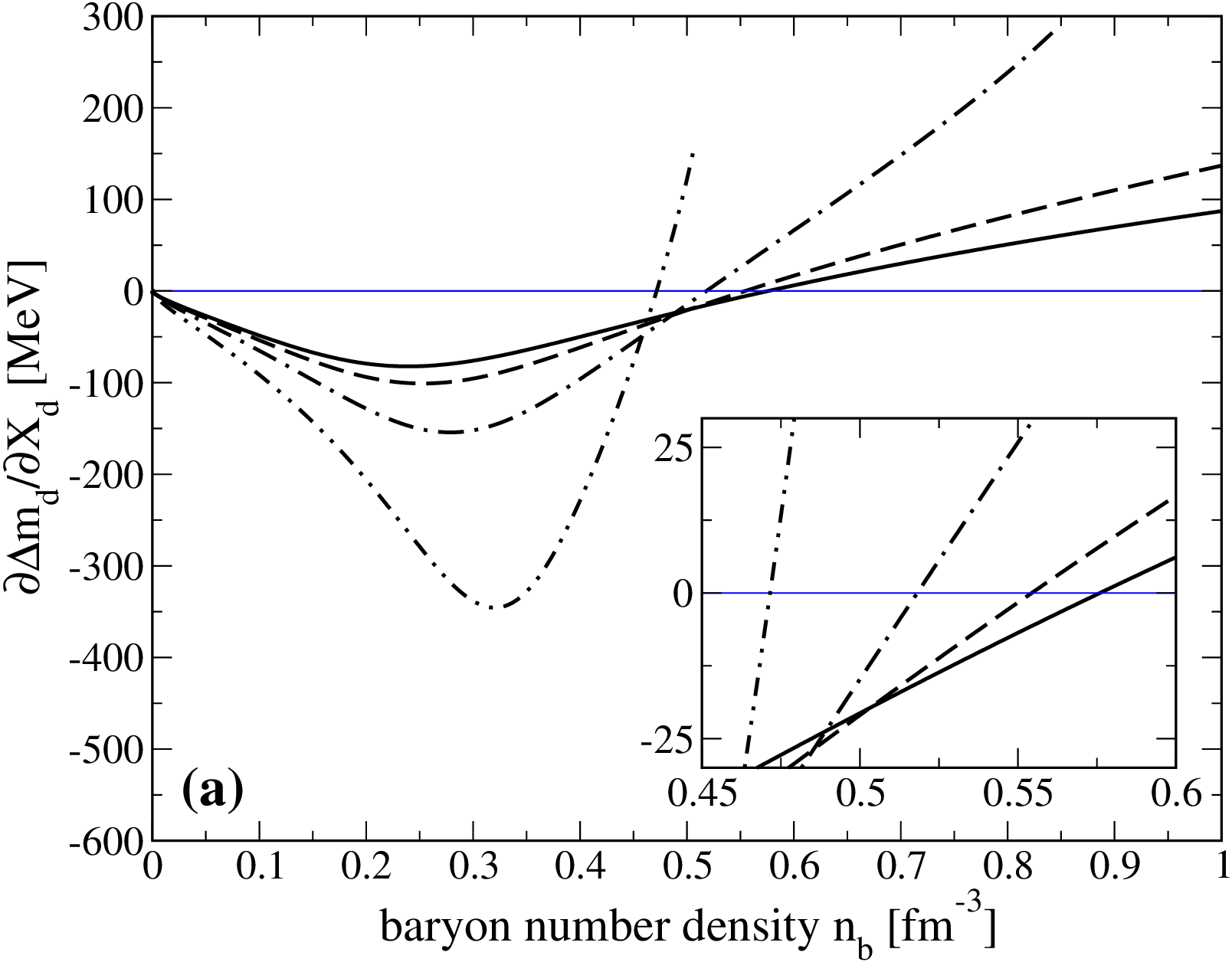} \qquad \includegraphics[width=.45\textwidth]{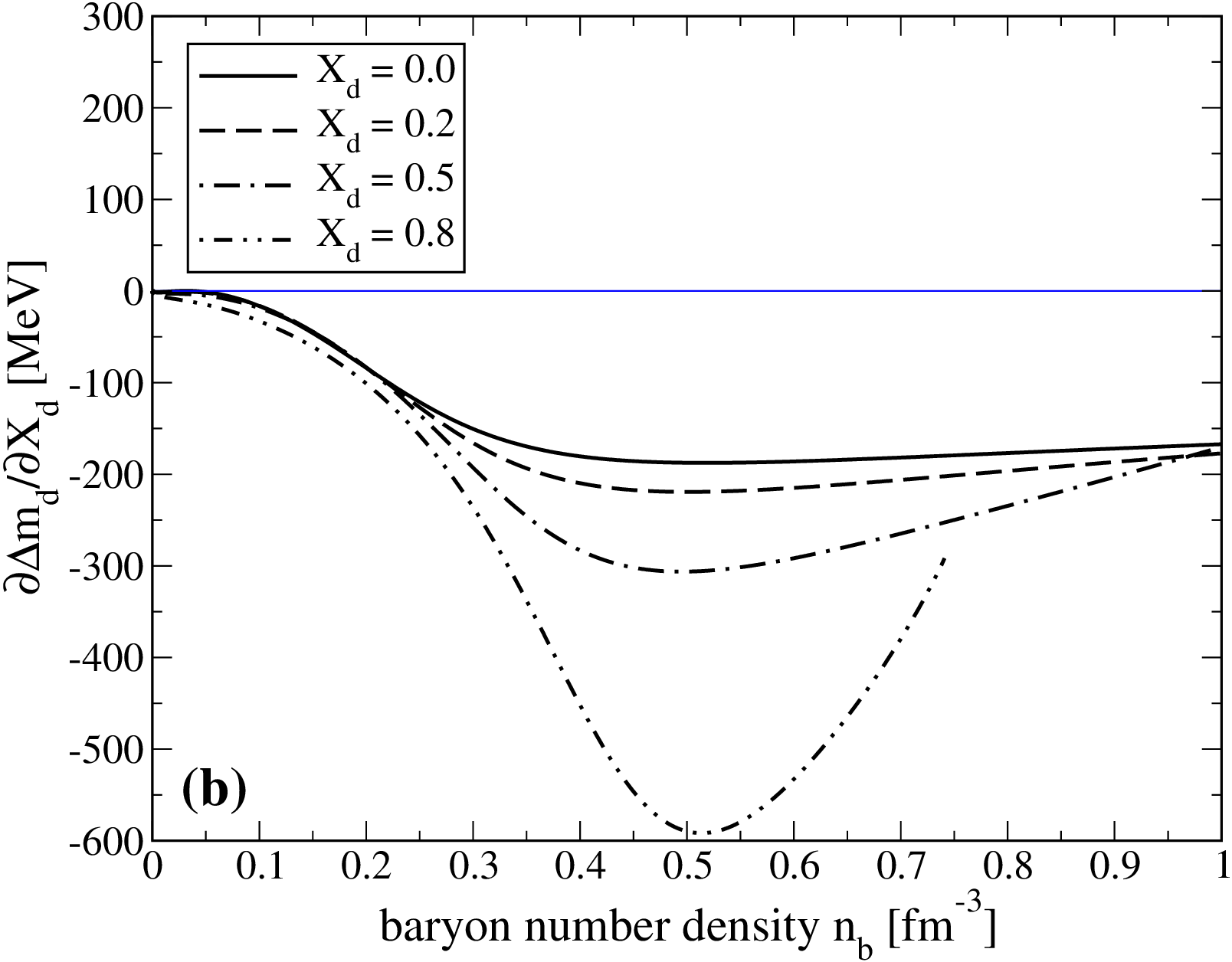}
\caption{\label{fig:ddmdx} Panel (a): Mass fraction derivative of the deuteron mass shift as function of the baryon density, as determined according to Eq.\ \eqref{eq:ddmddxd_SNM}
%\red{ in the SNM case, by assuming a unitary scaling factor $\chi$ for the deuteron-meson coupling strengths and}
for different, but constant values of the deuteron mass fraction $X_d$. 
The SNM case is considered and a unitary scaling factor $\chi$ for the deuteron-meson coupling strengths is assumed. Panel (b): The same as in panel (a), but assuming a reduced scaling factor $\chi = 1 / \sqrt{2}$. In both panels, the DD2 parameterization \cite{typelPRC2010} of the nucleon-meson effective interaction is adopted. The thin blue line indicates the zero of the deuteron mass fraction derivative of the mass shift. The inset in panel (a) shows a zoom around the density where the derivative vanishes.}
\end{figure*}
The dependence of the derivative \eqref{eq:ddmddxd_SNM} on the baryon density $n_{b}$ is depicted in Fig.\ \ref{fig:ddmdx} panel (a)
for $\chi=1$ and in panel (b) for $\chi=1/\sqrt{2}$. %\red{and for the} \gre{
The same constant values of $X_{d}$ %\red{are considered} 
as in Fig.\ \ref{fig:deltamd_high} are considered.  
Only the limit case with $X_{d} = 1$ is not shown %\red{. Indeed, for a unitary} \gre{
because for this %} 
deuteron mass fraction, $k_{\rm nuc}$ vanishes and a (negative) divergent mass shift derivative is obtained. %\red{through Eq.\ \eqref{eq:ddmddxd_SNM}, for the densities for which the corresponding $X_{d}^{\rm (max)}$ exceed 1.}
Moreover, as already observed in Fig.\ \ref{fig:deltamd_xd_high}, this result is independent of the adopted value for the scaling factor. %\red{three constant values of $X_{d}$. The curves cross the zero line at similar values above $0.45$~fm${}^{-3}$ with a weak dependence on the deuteron fraction, highlighted in the inset.}

Even though we are interested here in the investigation of the high-density limit, it is instructive to study both the low- and high-density behaviors of
the derivative
\eqref{eq:ddmddxd_SNM}. One observes that, for $k_{\mathrm{nuc}}\to 0$, 
%\red{For $k_{\mathrm{nuc}}\to 0$, one has
%$n_{\mathrm{nuc}}^{(s)} \to n_{\mathrm{nuc}}^{(v)}$ 
%and
%$\mu_{\mathrm{nuc}}^{\ast} \to m_{\mathrm{nuc}}^{\ast}$. Then} 
%\blu{In my opinion, these formulas are obvious and tend to overload the discussion.}
the derivative is dominated by the negative, diverging contribution inside the brackets and
the derivative approaches thus zero from negative values, as shown
in both panels of Fig.\ \ref{fig:ddmdx}.
For $k_{\mathrm{nuc}}\to \infty$, the total nucleonic vector and scalar %\red{densites} 
densities scale as
%\red{$n_{\mathrm{nuc}}^{(v)} \to [2/(3\pi)^{2}]k_{\mathrm{nuc}}^{3}$} 
$n_{\mathrm{nuc}}^{(v)} \sim [2/(3\pi^{2})]k_{\mathrm{nuc}}^{3}$ and
$n_{\mathrm{nuc}}^{(s)} \sim (1/\pi^{2})m_{\mathrm{nuc}}^{\ast}k_{\mathrm{nuc}}^{2}$,
respectively. 
Furthermore, %\red{$m_{\mathrm{nuc}}^{\ast} \to 0$ and} 
%\blu{$m_{\mathrm{nuc}}^{\ast}$ should tend to a constant value, right?} 
$n_{b} \sim n_{\rm nuc}^{(v)}$, $m_{\mathrm{nuc}}^{\ast} \sim \pi^{2}/(C_{\sigma} k_{\mathrm{nuc}}^{2})$, and
$\mu_{\mathrm{nuc}}^{\ast} \sim  k_{\mathrm{nuc}}$, so that
$f_{\mathrm{nuc}} \sim k_{\mathrm{nuc}}^{2}/\pi^{2}$ and the simple
asymptotic form
\begin{equation}
\label{eq:ddmddxd_SNM_asym}
  \left. \frac{\partial \Delta m_{d}^{(\mathrm{high})}}{\partial X_{d}} \right|_{n_{b},\beta=0} 
  \sim
  \left( 2 \chi^{2} - 1 \right)
    %\red{\frac{\pi^{2}}{k_{\mathrm{nuc}}^{2}} n_{b}} 
    \frac{2}{3} k_{\mathrm{nuc}}
\end{equation}
remains. For $\chi < 1/\sqrt{2}$ the derivative in the high-density
limit is negative as for $k_{\mathrm{nuc}}\to 0$ and no zero at finite baryon densities is expected. In contrast, for $\chi=1$, the 
derivative approaches asymptotically a positive value and a zero at a certain baryon value, $n_{b}^{\rm cross}$, appears, as depicted in Fig.\ \ref{fig:ddmdx}, panel (a). 
The curves cross the zero line at similar values above $0.45$~fm${}^{-3}$ with a weak dependence on the deuteron fraction, highlighted in the inset. 
Below the crossing an ordering of the lines as observed in panel (b) of Fig.\ \ref{fig:deltamd_high} follows, whereas an inversion will appear above $n_{b}^{\mathrm{cross}}$. 
%\red{This case corresponds to the behavior as shown in 
%panel (a) of Fig.\ \ref{fig:deltamd_high} with the inversion of the ordering of the lines with %respect to the chosen value of $X_{d}$.} 
%\blu{We have already stated that several times.}
For $\chi=1/\sqrt{2}$, however, the derivative of the mass shift with respect to the deuteron mass fraction of Eq.\ \eqref{eq:ddmddxd_SNM} is always %\red{negative and} 
non-positive, as evidenced in the panel (b) of Fig.\ \ref{fig:ddmdx}. Then, the zero line is only asymptotically approached and $\chi = 1/\sqrt{2}$ constitutes the largest value for which the ordering of the lines with respect to $X_{d}$, depicted in Fig.\ \ref{fig:deltamd_high}, panel (b)
persists, for all baryon densities. %\gre{either below the crossing points (panel (a))

\begin{figure*}[tbp!]
\includegraphics[width=.45\textwidth]{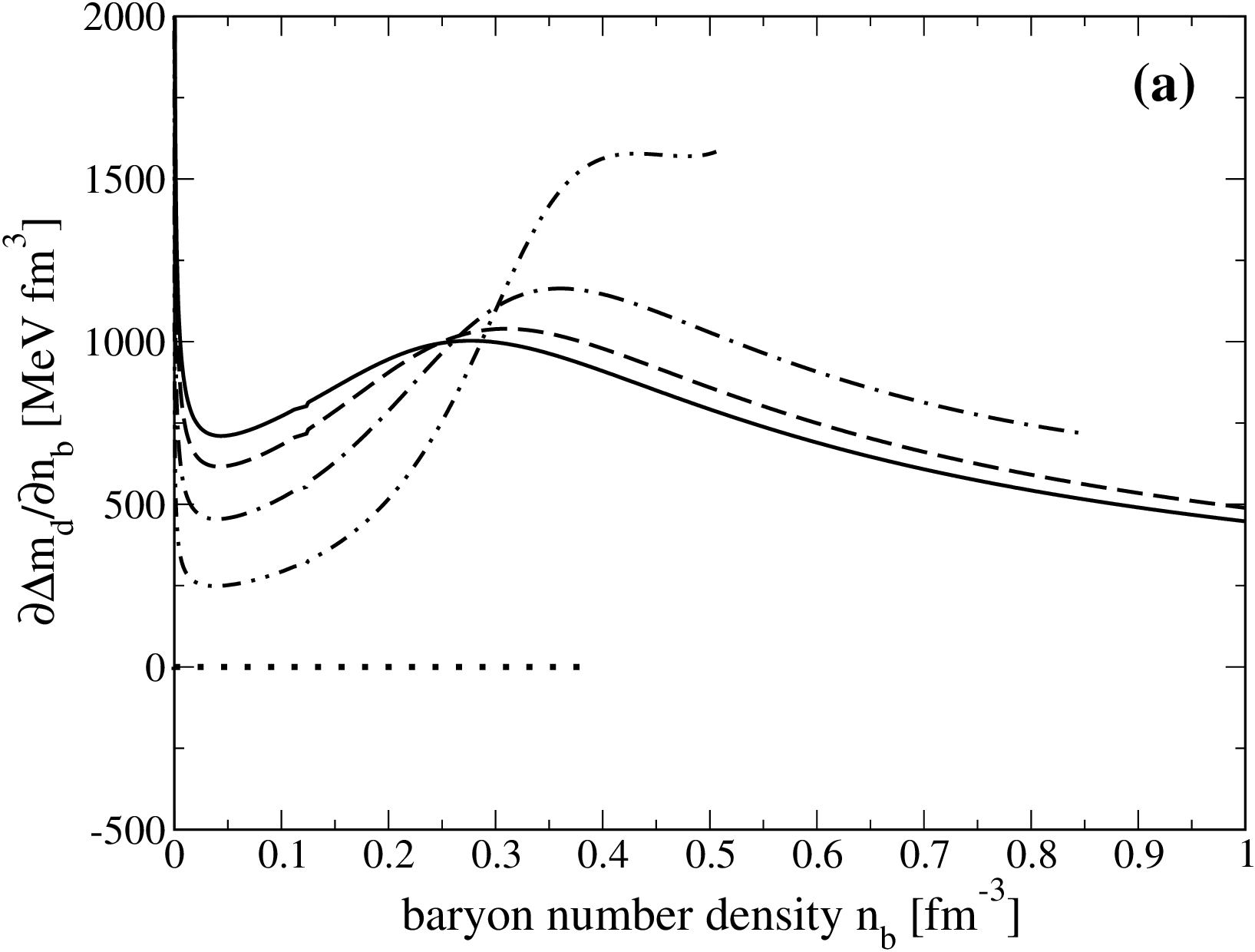} \qquad \includegraphics[width=.45\textwidth]{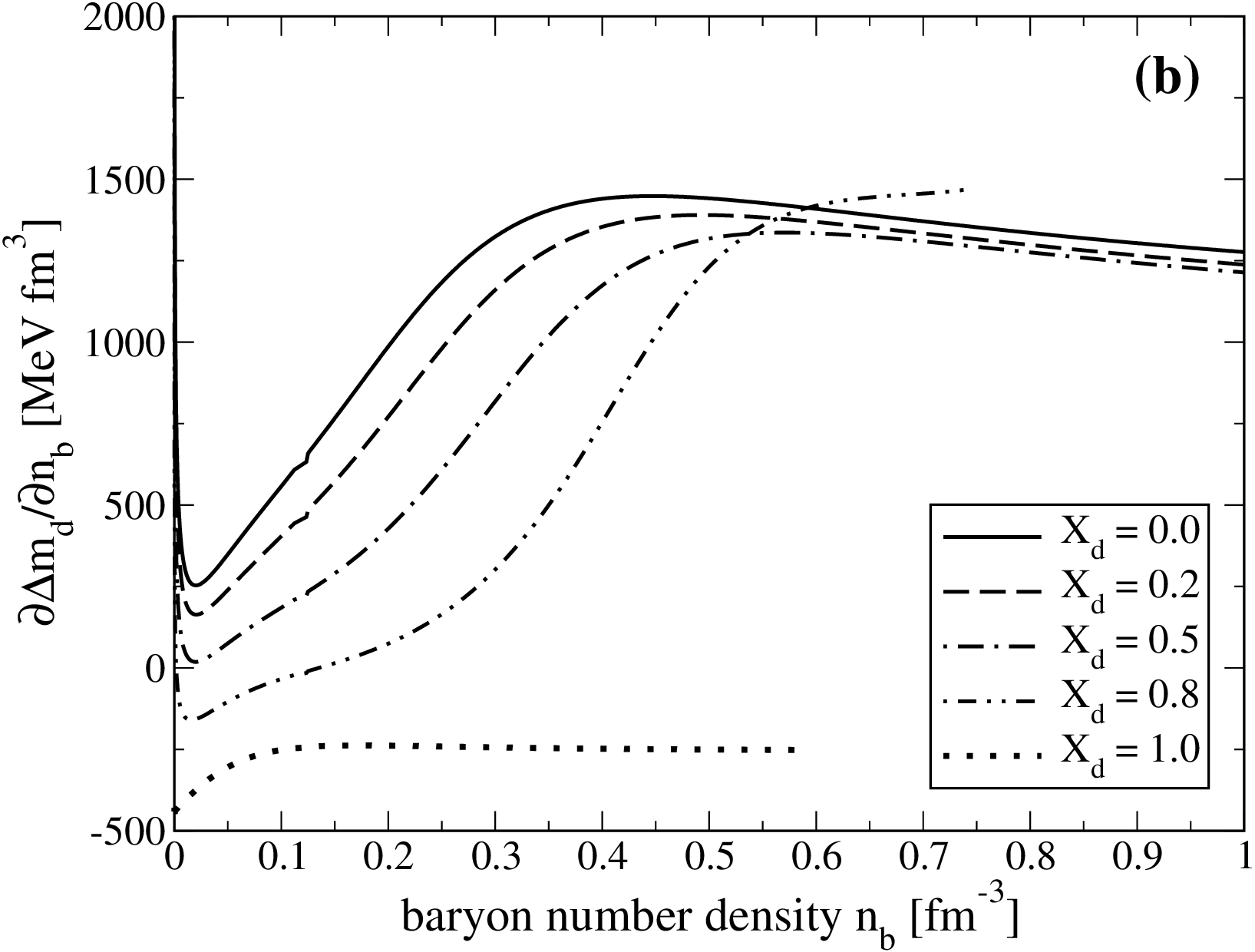}
\caption{\label{fig:ddeltamd_high} Panel (a): %\red{Slope} 
Baryon density derivative of the deuteron mass shift as function of the baryon density, as determined %\red{in the SNM case, by assuming a unitary scaling factor $\chi$ for the deuteron-meson coupling strengths and different, but constant  values of the deuteron mass fraction $X_d$.} 
according to Eq.\ \eqref{eq:ddmdnb_SNM}
for different, but constant values of the deuteron mass fraction $X_d$. 
The SNM case is considered and a unitary scaling factor $\chi$ for the deuteron-meson coupling strengths is assumed. Panel (b): The same as in panel (a), but assuming a reduced scaling factor $\chi = 1 / \sqrt{2}$. In both panels, the DD2 parameterization \cite{typelPRC2010} of the nucleon-meson effective interaction is adopted. %\blu{How much should we extend the x-scales of these figures? I think we should use the same as in Fig. \ref{fig:deltamd_param_chismall}. What do you think?}
}
\end{figure*}
In the following, also the
derivative of the mass shift with respect to the baryon density will be studied. It can be calculated explicitly from Eq.\ \eqref{eq:deltamd_high}. The general case for arbitrary values of $\beta$ is treated in Appendix \ref{sec:ddeltamd_dnb}.
Again, the result for the simplified case of SNM is given here.
It can be written for $\chi=\chi_{d\omega}=\chi_{d\sigma}$ as
\begin{equation}
\label{eq:ddmdnb_SNM}
    \left. 
    \frac{\partial \Delta m_{d}^{(\mathrm{high})}}{\partial n_{b}} 
    \right|_{\beta=0}
    = \mathcal{W}_{d}^{\mathrm{SNM}} - \mathcal{Z}_{d}^{\mathrm{SNM}} %Y_{d}
    %\left( X_{d} + n_{b} \frac{\partial X_{d}}{\partial n_{b}} \right) 
    \left. \frac{\partial (n_{b}X_{d})}{\partial n_{b}}
    \right|_{\beta=0}
\end{equation}
with the quantities
\begin{eqnarray}
\label{eq:ZdSNM}
    \mathcal{Z}_{d}^{\mathrm{SNM}} & = &
    \frac{\pi^{2}}{\mu_{\mathrm{nuc}}^{\ast}k_{\mathrm{nuc}}}
   + 2  \left(1-\chi %_{d\omega}
   \right)^{2} C_{\omega} 
   \\ \nonumber & & 
  - \frac{2C_{\sigma}}{1 + f_{\mathrm{nuc}} C_{\sigma}}
  \left( \chi %_{d\sigma} 
  - \frac{m_{\mathrm{nuc}}^{\ast}}{\mu_{\mathrm{nuc}}^{\ast}} \right)^{2}
\end{eqnarray}
and
\begin{eqnarray} 
\label{eq:WdSNM}
  \lefteqn{\mathcal{W}_{d}^{\mathrm{SNM}}   = 
  \frac{\pi^{2}}{\mu_{\mathrm{nuc}}^{\ast}k_{\mathrm{nuc}}} 
  + 2(1-\chi %_{d\omega}
  ) \left( C_{\omega} 
  +  C_{\omega}^{\prime} n_{\omega} \right)}
  \\ \nonumber & &
  +   \frac{2}{1 +f_{\mathrm{nuc}} C_{\sigma}}
%  \\ \nonumber & &
  \left( C_{\sigma}^{\prime} n_{\sigma}
  + C_{\sigma} \frac{m_{\mathrm{nuc}}^{\ast}}{\mu_{\mathrm{nuc}}^{\ast}} \right)
  \left( \chi %_{d\sigma}
  - \frac{m_{\mathrm{nuc}}^{\ast}}{\mu_{\mathrm{nuc}}^{\ast}} \right)
\end{eqnarray}
that contain again the factor \eqref{eq:fnuc}.
The dependence of the derivative \eqref{eq:ddmdnb_SNM} for the SNM case is depicted in the two panels of Fig.\ \ref{fig:ddeltamd_high} for the selected scaling factors of $\chi=1$
(left) and $\chi=1/\sqrt{2}$ (right).
In the zero-density limit, a divergent behavior is observed, owing to the contribution originating from the density derivatives of the effective chemical potentials that lead to the terms proportional to $k_{\mathrm{nuc}}^{-1}$ in $\mathcal{Z}_{d}^{\mathrm{SNM}}$
and $\mathcal{W}_{d}^{\mathrm{SNM}}$. 
%\red{For constant $X_{d}=1$, the} 
The divergence 
of \eqref{eq:ddmdnb_SNM} for $n_{b} \to 0$ will disappear only for constant $X_{d}=1$. The relative importance of the terms proportional to $k_{\mathrm{nuc}}^{-1}$ strongly reduces with increasing density, so that a rise %\gre{of \eqref{eq:ddmdnb_SNM}},
driven by the $\sigma$ meson contribution is observed at larger $n_{b}$ values, until a maximum is reached. Then,
%\red{After an initial rise with increasing $n_{b}$, a maximum is %reached, followed by} 
a continuous decrease 
%\red{of $\Delta m_{d}^{\mathrm{high}}$} \blu{What do you mean? $\Delta m_{d}^{\mathrm{high}}$ should never decrease, right?} 
of the density derivative of the deuteron mass shift emerges, which asymptotically approaches zero or a constant value for the unitary or the reduced value of $\chi$, respectively. An interplay among the different involved terms takes place, analogously to the one illustrated to describe the results of Fig.\ \ref{fig:deltamd_high}. As a consequence, 
the ordering of the %\red{mass shifts} 
curves in Fig.\ \ref{fig:deltamd_high} is reflected
in Fig.\ \ref{fig:ddeltamd_high}. As discussed before, 
%\red{for $\chi = 1$, the line for a constant deuteron fraction of $X_{d}=0.5$} 
for any finite deuteron fraction value $X_{d}$, the corresponding curve ends again at the baryon density where $X_{d}^{(\mathrm{max})}$, as a function of $n_{b}$, attains this value.

%%%%%%%%%%%%%%%%%%%%%%%%%%%%%%%%%%%%%%%%%%%%%%%%%%%%%%%%%%%%%%%%%%%%%
\section{Mass shift parameterization}
\label{sec:mass_shift_param}

%\blu{I would prefer to keep only parameterization in the singular.}
The condensation condition allows to calculate, for a given mass-fraction function, the quasi-deuteron mass shift and its derivatives at high-densities through Eqs.\ \eqref{eq:deltamd_high}, \eqref{eq:ddmddxd_SNM} and \eqref{eq:ddmdnb_SNM}, respectively. 
However, the density dependence of the mass fraction $X_{d}$ is not known a priori, in particular at supra-saturation densities. It should originate from microscopic calculations in a similar %\red{fashion} 
manner as the fractions of light clusters are determined by their mass shifts in the low-density domain, see Ref.\ \cite{typelPRC2010}.
Since calculations of the deuteron mass shift using proper interactions and many-body methods are only available at very low densities, it is %\red{reasonable to choose} 
necessary to resort to exemplary forms of the mass shift in the full range of baryon densities 
to study the properties of the system.
Instead of calculating the mass shift and its derivatives for a given mass fraction function, the main aim is then to choose a density dependent parameterization of the mass shift
and to determine the deuteron fraction, not only for symmetric but also asymmetric matter.
Although such a function is not yet available from microscopic models, the proposed form should comply with the available constraints.

%%%%%%%%%%%%%%%%%%%%%%%%%%%%%%%%%%%%%%%%%%%%%%%%%%%%%%%%%%%%%%%%%%%%%
\subsection{Piecewise mass-shift parameterization}

%\blu{Why did you change the title and postpone the "determination of the deuteron mass fraction from the mass-shift" later? Actually, I already solved the equation for $X_{d}$ starting from the piecewise parameterization. Indeed, between the dissociation density and the contact point of the two considered regimes, the deuteron fraction is not fixed, being neither 1, nor equal to the constant value assumed at high-density. This is why that part was included before in this subsection.}
\begin{figure*}[tbp!]
\includegraphics[width=.45\textwidth]{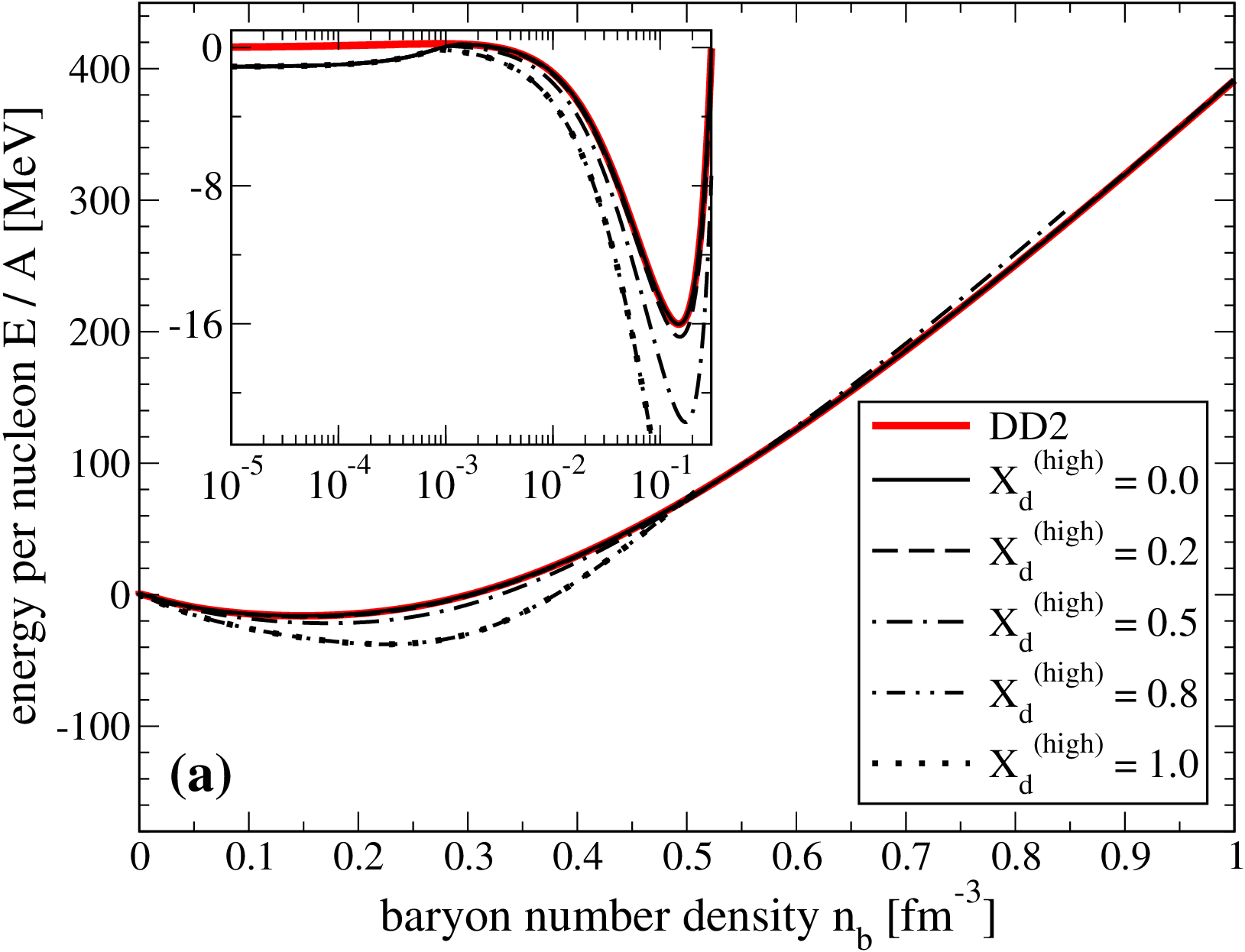} \qquad \includegraphics[width=.45\textwidth]{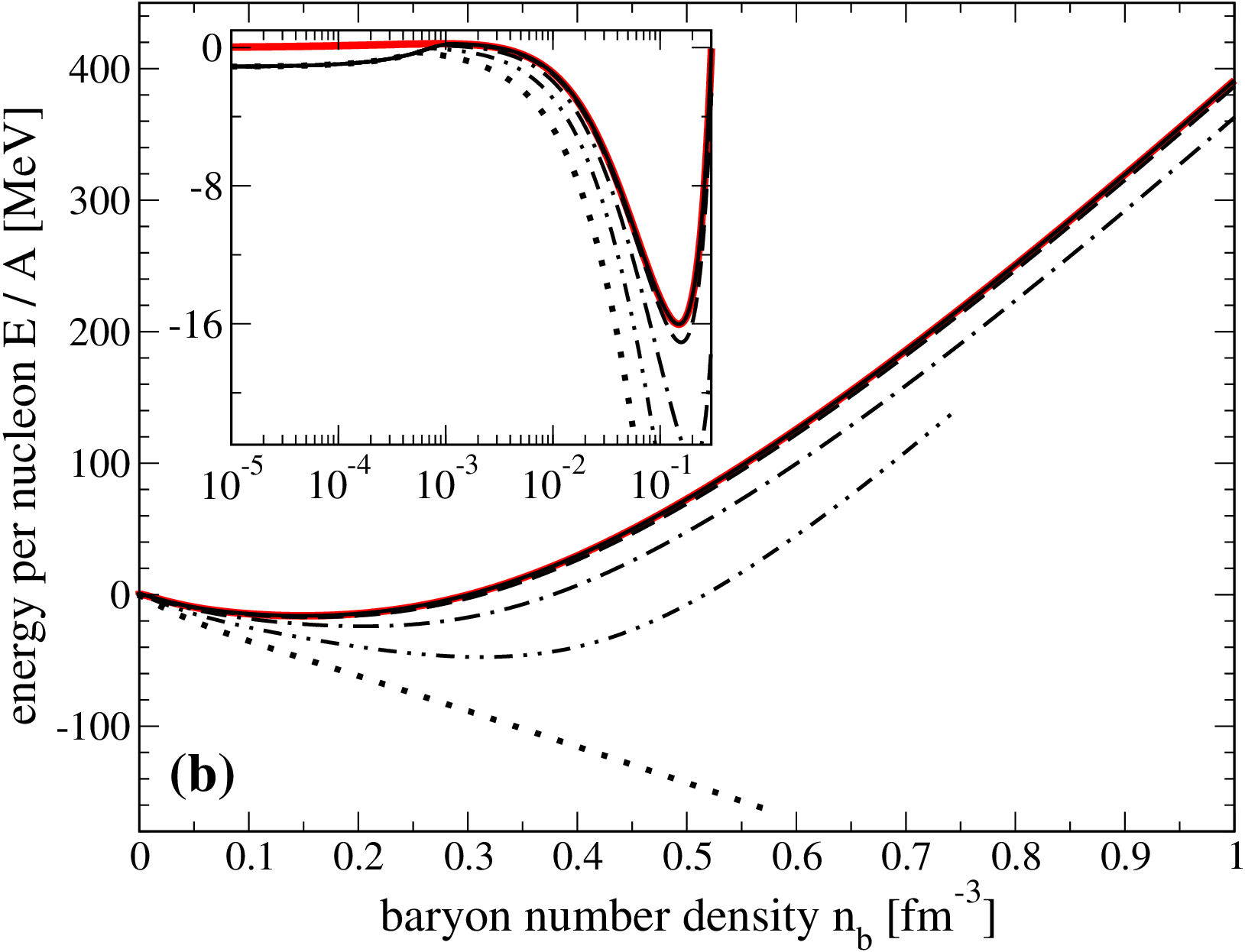}
\caption{\label{fig:energy_rho_xdc} Panel (a): Energy per nucleon as function of the baryon density as determined in the SNM case, by assuming a unitary scaling factor $\chi$ for the deuteron-meson coupling strengths. A simple piecewise parameterization, as given by Eq.\ \eqref{eq:interpolation} and interpolating between the low- and high-density constraints for the mass shift
%\red{, as given by Eq.\ \eqref{eq:interpolation},} 
is adopted. %\red{, which  provides different} 
Different, but constant, values of the deuteron mass fraction $X_{d}^{\rm (high)}$ at high density are considered. Panel (b): The same as in panel (a), but assuming a reduced scaling factor $\chi = 1 / \sqrt{2}$. In both panels, the DD2 parameterization \cite{typelPRC2010} of the nucleon-meson effective interaction is adopted.}
\end{figure*}

\begin{comment}
\red{The condensation condition allows to calculate, for a given mass-fraction function, the quasi-deuteron mass shift and its derivatives at high-densities through Eqs.\ \eqref{eq:deltamd_high}, \eqref{eq:ddmddxd_SNM} and \eqref{eq:ddmdnb_SNM}, respectively. 
However, the density dependence of the mass fraction $X_{d}$ is not known a priori, in particular at supra-saturation densities. It should originate from microscopic calculations in a similar %\red{fashion} 
manner as the fractions of light clusters are determined by their mass shifts in the low-density domain, see Ref.\ \cite{typelPRC2010}.
Since calculations of the deuteron mass shift using proper interactions and many-body methods are only available at very low densities, it is %\red{reasonable to choose} 
necessary to resort to exemplary forms of the mass shift in the full range of baryon densities 
to study the properties of the system.}
\end{comment}

The most simple choice for the deuteron mass shift function is the piecewise parameterization 
\begin{eqnarray}
 \label{eq:interpolation}
 \lefteqn{\Delta m_{d} (n_b) =}
 \\ \nonumber & &
 \mbox{min}\left\{\Delta m_d^{({\rm low})} (n_b), \Delta m_d^{({\rm high})} (n_b, X_d) \right\} 
\end{eqnarray}
that combines the low-density form \eqref{eq:deltamd_low}
with the high-density function \eqref{eq:deltamd_high} assuming, e.g., a constant deuteron fraction in the high-density region
as discussed in the previous subsection.
Although Eq.\ \eqref{eq:interpolation} provides a continuous function, the same does not apply for its derivatives. 
In addition, a constant deuteron fraction at high densities is not compatible with the asymptotic constraint which imposes a vanishing mass fraction $X_{d}$ for $n_{b} \to \infty$.

Despite these shortcomings, it is informative to investigate the effect of %\red{considering} 
finite deuteron fractions on the energy per nucleon $E/A$. Fig.\ \ref{fig:energy_rho_xdc} depicts the density behavior of $E/A$ in SNM as obtained with the piecewise mass shift parameterization \eqref{eq:interpolation} for different, but constant mass fractions in the high-density regime.
For comparison, the energy per nucleon obtained from the deuteron free case of the standard DD2 parameterization is also shown.  One observes that an extra binding is predicted in the energy per nucleon around saturation density, with respect to the case without deuterons. This result is independent of the adopted choice of the scaling factor $\chi$ of the deuteron-meson couplings.
Moreover, Fig.\ \ref{fig:energy_rho_xdc} shows that the overbinding persists 
%\red{up to very high densities} 
much beyond the saturation density, especially in panel (b), where a reduced value of $\chi$ is considered.
However, the stiffness of these curves in the high-density regime increases with the deuteron mass fraction. As a result, a crossing among the curves is generally observed at larger densities in case when $\chi \geq 1/\sqrt{2}$, except for  deuteron fraction values for which the corresponding curves end at lower densities.

Both insets of Fig.\ \ref{fig:energy_rho_xdc} highlight that the correct low-density limit is %\red{instead} 
reproduced because the low-density constraint is properly taken into account.
As a result, differently from the standard DD2 model without clusters, the curves tend to half the deuteron binding energy
in the limit $n_{b} \to 0$.

The overbinding observed in Fig.\ \ref{fig:energy_rho_xdc} in a broad range of densities around saturation seems to be surprising 
in light of the large positive mass shift of the deuteron 
%\red{is considered as} 
depicted in Fig.\ \ref{fig:deltamd_high}. The increase of the binding energy per nucleon is a result of two main effects.   
When nucleons are replaced by quasi-deuterons at a given baryon density, the value of $n_{\omega}$ does not change when $\chi = 1$.
%\blu{$n_{\omega} = n_{b}$ is only true for $\chi = 1$, isn't it? If so, does this explanation still hold?} 
At the same time, the source density $n_{\sigma}$ increases because the scalar and vector densities of the condensed deuterons are identical whereas $n_{q}^{(s)}/n_{q}^{(v)}<1$ for nucleons. Hence, on the one hand, a stronger attraction from the $\sigma$ meson is induced. 
This effect will be smaller for $\chi < 1$ when nucleons are replaced by deuterons. Here, the source density of the $\omega$ meson will reduce but the corresponding decrease of the $\sigma$ meson source density is also less strong.
On the other hand, nucleons at the Fermi surface with energies close to the nucleon chemical potential
\begin{equation} 
\mu_{q}^{(0)}= \sqrt{\left(k_{q}^{(0)}\right)^{2}+\left(m_{q}-S_{q}^{(0)}\right)^{2}}+V_{q}^{(0)}
\end{equation}
of the deuteron-free system, indicated by a superscript $(0)$,
are replaced by deuterons with energy
\begin{equation} 
E_{d}=\mu_{d} = m_{d}-S_{d}+V_{d}+\Delta m_{d}=\mu_{n}+\mu_{p} \: .
\end{equation}
Since $k_{q} < k_{q}^{(0)}$, $S_{q} > S_{q}^{(0)}$, and $V_{q} \approx V_{q}^{(0)}$,
the chemical potentials of the nucleons $\mu_{q}$ are lowered,
i.e., $\mu_{q} < \mu_{q}^{(0)}$,
even for a positive deuteron mass shift $\Delta m_{d}$.
This corresponds to a stronger binding of the system.

%%%%%%%%%%%%%%%%%%%%%%%%%%%%%%%%%%%%%%%%%%%%%%%%%%%%%%%%%%%%%%%%%%%%%
\subsection{Specific features in the determination of deuteron mass fractions from mass shifts}
\label{sec:pole}
\begin{comment}
\red{The condensation condition allows one to calculate, for a given mass fraction function, the mass shift and its derivatives at high-density, through Eqs.\ \eqref{eq:deltamd_high}, \eqref{eq:ddmddxd_SNM} and \eqref{eq:ddmdnb_SNM}, respectively. However, the density dependence of the deuteron mass fraction is not well known.}
\red{
An alternative strategy is instead to choose a given mass shift parameterization and to correspondingly determine the deuteron fraction, for example by using at high-density the condensation condition \eqref{eq:deltamd_high} to find $X_{d}$ for a given $\Delta m_{d}^{(\mathrm{high})}$. A proper interpolation of the deuteron mass shift between the low-density limit from the microscopic Pauli blocking calculation provided by 
Eq.\ \eqref{eq:deltamd_low} and the high-density limit of the
condensate model, as given in Eq.\ \eqref{eq:deltamd_high}, should be then introduced to get an extended GRDF
with quasi-clusters at supra-saturation densities.
In this context, further insight can be obtained from the explicit functional form of the mass shift derivative with respect to the density.}
\end{comment}

In principle, it would be sufficient to use the condensation condition \eqref{eq:deltamd_high} to find $X_{d}$ for a given $\Delta m_{d}^{(\mathrm{high})}$. In practice, however, it is found that, in the high-density regime, Eq.\ \eqref{eq:deltamd_high} can have multiple solutions when a
meson coupling scaling factor $\chi \geq 1/\sqrt{2}$ is considered. 
As a consequence, Eq.\ \eqref{eq:deltamd_high} can not be inverted uniquely in all cases. The correct solution has to be selected such that a continuous function of the 
%\red{mass fraction of the} 
density is obtained for the deuteron mass fraction.

\begin{comment}
\red{Nevertheless, the easiest choice} 
 \begin{equation}
% \label{eq:interpolation}
\red{\Delta m_{d} (n_b) = \mbox{min}\left\{\Delta m_d^{({\rm low})} (n_b), \Delta m_d^{({\rm high})} (n_b, X_d) \right\}}
 \end{equation}
to build piecewise the mass shift parameterization could not be adopted. Indeed, such a parameterization would imply a constant deuteron fraction at high-density, which is not compatible with the asymptotic constraint, which imposes a vanishing mass fraction $X_{d}$ for $n_b \to \infty$. Moreover, although Eq.\ \eqref{eq:interpolation} provides a continuous function everywhere, the same does not apply for its derivatives, at least in the contact point between the two density regimes.
\red{Then a suitable parameterization has to be adopted, in order to have a smooth behavior, asymptotically decreasing at least as $n_{b}^{-1}$. Moreover, to complicate matters, it is found that, in the high-density regime, Eq.\ \eqref{eq:deltamd_high} can have multiple solutions, when a scaling factor $\chi \geq 1/\sqrt{2}$ is considered. The correct one has to be selected to obtain a continuous function.
As a consequence, Eq.\ \eqref{eq:deltamd_high} can not be always inverted.
In this context, further insight can be obtained from the explicit functional form of the mass shift derivative with respect to the density.}
\end{comment}

In this context, further insight can be obtained from the explicit functional form of the mass shift derivative with respect to the density,
%\red{.The general expression of this quantity is derived in Appendix \ref{sec:ddeltamd_dnb} and given by Eq.\ \eqref{eq:ddmddnb_general} for nuclear matter of arbitrary asymmetry $\beta$.} 
given by Eq.\ \eqref{eq:ddmdnb_SNM}. It is, in fact, a first-order differential equation for $X_{d}$.
%, but for $Z_{d}=0$, the prefactor of the derivative $\partial %(n_{b}X_{d})/\partial n_{b}$ vanishes leading to a problem to find a %solution.
The expression can be analyzed most easily for SNM. 
%\red{For the further analysis,} 
It is convenient to write Eq.\
\eqref{eq:ddmdnb_SNM} in the form
\begin{eqnarray}
\label{eq:deq}
 \lefteqn{\left. \frac{\partial (n_{b}X_{d})}{\partial n_{b}} \right|_{\beta=0} =}
 \\ \nonumber & & 
 \frac{1}{\mathcal{Z}_{d}^{\mathrm{SNM}}}
 \left( \mathcal{W}_{d}^{\mathrm{SNM}} - 
 \left. \frac{\partial \Delta m_{d}}{\partial n_{b}} \right|_{\beta= 0} \right)
\end{eqnarray}
%\blu{I suggest to remove $(high)$ from the mass shift hereafter. Moreover, I find that using $Z_{d}$, which was already used for the deuteron proton number, is misleading. Maybe it would be better something like $W_{1,d}$ and $W_{2,d}$.}
with the functions $\mathcal{Z}_{d}^{\mathrm{SNM}}$ 
and $\mathcal{W}_{d}^{\mathrm{SNM}}$ as defined in Eqs.\ \eqref{eq:ZdSNM} and \eqref{eq:WdSNM}, respectively.
A special role is played by $\mathcal{Z}_{d}^{\mathrm{SNM}}$ that can be written as
\begin{equation}
\label{eq:ZdSNM2}
  \mathcal{Z}_{d}^{\mathrm{SNM}} = 
  2  \left(1-\chi  \right)^{2} C_{\omega} 
  - \frac{1}{n_{b}} 
  \left. \frac{\partial \Delta m_{d}}{\partial X_{d}} \right|_{n_{b},\beta=0}
\end{equation}
with the help of the mass fraction derivative defined in Eq.\ 
\eqref{eq:ddmddxd_SNM}. 
The superscript (high) is no longer added to the mass shift because a mass shift parameterization is considered now in the whole density range.
For $\chi=1$, the term from the $\omega$ meson does not contribute. The remaining term in Eq.\ 
\eqref{eq:ZdSNM2} develops a zero at a certain density $n_{b}^{\mathrm{cross}}$
as discussed
in section \ref{sec:highdens} close to Eq.\ \eqref{eq:ddmddxd_SNM_asym}. Thus the 
%\red{differential equation} 
density derivative
\eqref{eq:deq} develops a pole at $x_{b}^{\mathrm{cross}}$ and a continuous solution
of the differential equation can only be obtained if the term in 
parentheses in Eq.\ \eqref{eq:deq} vanishes at the same density.
For $\chi \leq 1/\sqrt{2}$, however, 
the mass-fraction derivative 
\eqref{eq:ddmddxd_SNM}
is negative, implying a positive $\mathcal{Z}_{d}^{\mathrm{SNM}}$ because also the contribution of the $\omega$ meson is positive. Then, a continuous solution of the differential equation \eqref{eq:deq} can be found for all densities. 
The discussion developed above justifies the choice of the reduced scaling factor. Thus, two different values of the scaling factor, namely $\chi= 1$ and $\chi = 1 / \sqrt{2}$, will be considered %\red{in the analysis} 
in the following analysis. It is worthwhile to notice that the value of $\chi = 1 / \sqrt{2}$ is significantly smaller than the universal scaling factor for the cluster-meson coupling strength proposed in some recent works \cite{paisPRC2018, paisPRC2019}.
Our choice complies, however, with the aim to consider two extreme values of $\chi$ with two distinct paths, bearing in mind that any intermediate behavior may also occur.

\subsection{Saturation constraints}

The overbinding observed around saturation density in both panels of Fig.\ \ref{fig:energy_rho_xdc} implies that a proper refit of the nucleon-meson couplings to the saturation properties of SNM is mandatory if one wants to keep nuclear matter quantities around the saturation point $n_0$ well constrained. 
%\blu{Should we show also other thermodynamical quantities around %saturation, such as pressure and/or incompressibility?}
%The requirement to satisfy Eq.moreover  \eqref{eq:chem_equilibrium} under the condensation condition, while fixing some general nuclear matter properties in correspondence of the equilibrium point $n_0$ of SNM produces useful constraints at saturation for the mass shift.
The actual deuteron fraction at saturation, $X_{d,0}$, 
which has to 
be specified to fix the couplings, can be imposed from recent
experimental investigations of SRCs by extrapolating results of nuclei to infinite nuclear matter.
%\red{Moreover, further constraints on the deuteron mass fraction around saturation might be imposed by extrapolating the results from some recent experimental investigations to infinite nuclear matter.}
They assess that SRCs pairs amount to approximately $20\%$
of the nucleon density \cite{egiyanPRL2006, subediSCI2008, henSCI2014}.
Here, as in the following, the index $0$ on the quantities indicates the values at saturation.

\begin{table*}[t]
\centering
\caption{\label{tab:conv}Values at saturation density of 
%\red{scalar $\Gamma_{\sigma, 0}$ and vector $\Gamma_{\omega, 0}$} 
nucleon-meson coupling strengths $\Gamma_{j,0}$ ($j=\sigma, \omega, \rho)$), deuteron mass shift $\Delta m_{d, 0}$ and its slope, for the two different scaling factors $\chi$ considered in this work. The mass shift and its slope are expressed in MeV and MeV fm$^3$, respectively, whereas all other quantities are dimensionless.
%\blu{Add the values of $\Gamma_{\rho,0}$ from refitting $J_{0}$.}
} 
\begin{tabular}{*{6}{c}}
\toprule
$\chi$ & $\Gamma_{\sigma, 0}$ & $\Gamma_{\omega, 0}$ & $\Gamma_{\rho,0}$ &
$\Delta m_{d, 0}$ [MeV] & $\left. \frac{d \Delta m_d}{d n} \right|_{n_0,\beta=0}$ [MeV fm${}^{3}$] \\
% & & & \gre{[MeV]} & \gre{[MeV fm${}^{3}$]} \\
\hline
\hline
%1 & $10.580609$ & $13.218144$ & $104.92831$ & $813.728623$ \\
1 & $10.580042$ & $13.217226$ & $3.556424$ & $104.92$ & $813.98$ \\
\hline
%$1/\sqrt{2}$ & $10.920551$ & $13.720259$ & $58.242010$ & $570.563968$ \\
$1/\sqrt{2}$ & $10.919963$ & $13.719324$ & $3.400187$ & $58.23$ & $570.80$ \\
\bottomrule
\end{tabular}
\end{table*}

In order to reproduce the properties
%\red{, e.g., of the DD2 model,} 
at saturation in SNM, e.g., of the DD2 model, the binding energy per nucleon %\red{$B_{0}=16.0224$~MeV} 
$B_0$ 
and the effective nucleon mass $m_{\mathrm{nuc},0}^{\ast}$ %\red{$m_{\mathrm{nuc},0}^{\ast}=0.562544~ m_{\mathrm{nuc}}$} $m_{\mathrm{nuc},0}^{\ast}$ 
at the saturation density 
%\red{$n_{0}=0.149065$~fm${}^{-3}$} $n_{0}$ 
should be obtained also in the model with a finite deuteron fraction $X_{d,0}$. 
%\blu{How many of these digits are really significant?} 
%\blu{I put always 6 digits to be consistent.} 
This will be realized by rescaling the meson-nucleon couplings assuming no change in their density dependence as given in 
a reference parameterization. 
%\red{Here, as in the following, the index $0$ on the quantities indicates the values at saturation.} 
The values of $n_0$, $B_{0}$, $m_{\mathrm{nuc},0}$ and $X_{d,0}$ together with the pressure $P_{0}=0$~MeV~fm$^{-3}$ give four conditions that allow to determine the rescaled couplings $\Gamma_{\sigma,0}$, $\Gamma_{\omega,0}$, the deuteron mass shift $\Delta m_{d,0}$ and its derivative $\left. d\Delta m_{d}/dn_{b} \right|_{n_{0}}$ at saturation.
Owing to the rescaling of the $\sigma$ and $\omega$ coupling strengths, also the energy per nucleon of PNM would be 
%\red{consequently} 
modified. A proper rescaling also of the $\rho$-nucleon coupling strength is hence in order if one wants to keep the symmetry energy at saturation $J_{0}$ unaltered, with respect to the selected reference parameterization.  %\red{considered}. 
Here, the symmetry energy is calculated in the parabolic approximation as the difference between the energies per nucleon in PNM and SNM. This recipe gives finite values at sub-saturation densities in models with clusters or liquid-gas phase transition %\red{as opposed to} 
differently than the original definition using second derivatives of the energy per nucleon in SNM with respect to the isospin asymmetry, see, e.g., Ref.\  \cite{Typel:2013zna}.
%\red{unaltered, the symmetry energy at saturation $J_{0}$ as, e.g., determined in the parabolic approximation, with the reference parametererization considered.}
The full conversion procedure is illustrated in detail in Appendix \ref{app:conv}. 
%\blu{I would prefer to keep the text more general up to this point, rather than constantly refer to the standard DD2 parameterization.}
The actual values of these quantities are given in 
Table \ref{tab:conv} for the two considered values of the deuteron coupling scaling factor $\chi$ and a deuteron fraction $X_{d,0}=0.2$. The standard DD2 model is chosen as reference parameterization with
$n_{0}=0.149065$~fm${}^{-3}$, $B_{0}=16.0224$~MeV, 
$m_{\mathrm{nuc},0}^{\ast}=0.562544~m_{\mathrm{nuc}}$ and $J_{0}=32.73$~MeV. 
The $\sigma$, $\omega$, and $\rho$ nucleon-meson couplings are given by $\Gamma_{\sigma,0}=10.686681$, 
$\Gamma_{\omega,0}=13.342362$ and $\Gamma_{\rho,0}=3.626940$ %\blu{Where did you take this value? In Ref.\ \cite{typelPRC2010}, I found $\Gamma_{\rho,0}=3.626940$. Could you please check that?} 
in this case. 
%\blu{Sorry, the value that I gave is just twice the value of the reference because I took it from a code where the source density for the rho meson is defined differently (with isospin quantum numbers) as compared to here. So the original value $\Gamma_{\rho,0}=3.626940$ is correct.}
%\red{The $\sigma$
%and $\omega$ couplings can be compared to
%the original DD2 values of $\Gamma_{\sigma,0}=10.686681$ 
%and $\Gamma_{\omega,0}=13.342362$.}

Quite interestingly, one observes that, in the case with a unitary scaling factor $\chi$, a small reduction of %\red{both $\sigma$- and $\omega$-nucleon} 
the three nucleon-meson coupling strengths is obtained at saturation, with respect to the original DD2 parameterization. On the other hand, these quantities turn out to be larger 
for both $\sigma$- and $\omega$-meson, when the smaller $\chi$ value is considered. 
%\blu{Do you have any idea for that?}
%\gre{
This change reflects the balance between the couplings and scaling factors to achieve the same strength of the effective interaction at saturation density. 
%}
%\red{For what it concerns the} 
%\red{values of the} 
Concerning the mass shifts, 
%\red{in contrast to what was observed in Fig.\ \ref{fig:deltamd_high} at high-density}
%\blu{?} \cya{
%\red{as it was already observed in Fig.\ \ref{fig:deltamd_high} around saturation density}, a larger value and a stronger density 
%\red{dependence is predicted for $\chi = 1/\sqrt{2}$,} 
larger values and stronger density slopes are predicted at saturation 
%\red{slope is predicted} 
when assuming that the nucleons bound inside the deuterons couple to the mesons with the same strength as the unbound nucleons, see also Fig.\ \ref{fig:deltamd_high}.

A possible mass shift parameterization will be proposed in the following section. %\red{under this assumption.} 
It will be constrained at saturation density, in the low-density limit by microscopic many-body calculations and at high-density by an assumed mass-fraction behavior that respects the condensation condition and the constraint on the maximum deuteron fraction. 
%\blu{this is not clear to me.}

%%%%%%%%%%%%%%%%%%%%%%%%%%%%%%%%%%%%%%%%%%%%%%%%%%%%%%%%%%%%%%%%%%%%%
\subsection{Unified mass shift parameterization}
\label{sec:unified_mass_shift}

%\red{Let us focus, as a first step, to the SNM case. Let us introduce a unified mass shift parameterization for the SNM case. The adopted choice might then be employed also for ANM, to achieve predictions for matter of any asymmetry. The simplest option is to assume that such a  parameterization depends only on the baryon density $n_b$.}
Different forms of the mass shift density parameterization might be employed to interpolate among the low- and the high-density constraints discussed in the previous section, while keeping the required saturation properties. 
A unified form, adopted for the SNM case, might be then employed also for ANM to obtain predictions at arbitrary isospin asymmetries.
%\red{Our approach was to combine a parameterization, mainly active in the dilute regime where it reproduces by construction the linear increase foreseen in the zero-density limit, with a suitable functional form, providing the expected asymptotic density dependence.}
The mass shift parameterization introduced in this work
combines two limiting dependencies on the density that reproduce
by construction the linear increase in the zero-density limit of the dilute region and the high-density asymptotic behavior.
Taking into account also the two constraints imposed at saturation, such a parameterization 
%\red{should then have, at least, 5} \red{has then four fixed} 
has to depend on at least four parameters and satisfy $\Delta m_{d}(0) = 0$. 
However, some additional parameters should enter in the proposed parameterization to guarantee a smooth transition between the different density regimes and leave some freedom in the high-density behavior of the deuteron fraction. %}

A possible choice, among others, is the function
\begin{equation}
\label{eq:deltamd_param}
\Delta m_d (x) = \Delta m_{d,1} (x) + \Delta m_{d,2} (x) + \Delta m_{d,3} (x)
\end{equation}
depending on $x = n_{b} / n_0$ with three contributions
\begin{eqnarray}
\Delta m_{d, 1} (x) &=& \dfrac{a x}{1 + b x}  \\
\Delta m_{d, 2} (x) &=& c x^{\eta + 1}\left[1 - \tanh{(ex)} \right]\\
\Delta m_{d, 3} (x) &=& fx^{\gamma}\tanh{(gx)}
\label{eq:deltamd123}
\end{eqnarray}
with $\gamma = 1$ %\red{and} 
or $2/3$ for $\chi = 1/\sqrt{2}$ %\red{and} 
or $1$, respectively, and seven %\red{positive} 
coefficients
$a$, $b$, $c$, $\eta$, $e$, $f$, $g$, which allow to comply with the constraints by a proper choice. 
%\blu{Maybe also the choice of the letter $d$, used for deuterons, for this exponent is questionable. Shall we use $\eta$, for example?}

\begin{comment}
\blu{Another possibility would be to consider 
\begin{equation}
\Delta m_{d, 3} (x) = fx^{\alpha}\tanh{(gx)}
\end{equation}
with $\alpha = 1, 2/3$, for $\chi = 1/\sqrt{2}, 1$, respectively, if we want to show also one case where the pole/multiple solutions exist.}
\end{comment}

The different terms in %\red{Eq.\ \eqref{eq:deltamd_param}}
Eqs.\ \eqref{eq:deltamd_param} - \eqref{eq:deltamd123}
are chosen so that the density derivative of the mass shift
is given by
\begin{equation}
  \left. \frac{d \Delta m_d}{d n_{b}} \right|_{n_{b}=0} = \frac{a}{n_{0}}
\end{equation}
at vanishing $n_{b}$ and the high-density behavior is
dominated by the third term $\Delta m_{d,3}$.
The second contribution acts %\red{only} 
mainly in an intermediate density range.
%\red{due to the suppression at high densities from the $\tanh$ function}.  
The coefficient $a$ is determined as $\delta B_{d}(0) n_{0}$ by
the limiting form of the deuteron mass shift parameterization from microscopic calculations,
c.f., Eq.\ \eqref{eq:deltamd_low}. There is no a priori constraint for the parameter $b$. Here, it is set to 
$b=a/B_{d}=n_{0}/n_{d}^{(\mathrm{diss})}$ so that $\lim_{n_{b}\to \infty} \Delta m_{d,1} = B_{d}$ with
the deuteron binding energy $B_{d}$ and the dissociation density $n_{d}^{(\mathrm{diss})}$ defined in Eq.\ \eqref{eq:ndiss}.
%\red{For $n_{b}\to \infty$} 
In the asymptotic limit, for $\chi = 1/\sqrt{2}$, the mass shift 
approaches a linear function in the baryon density 
%\red{with}
%\begin{equation}
%\red{\Delta m_d (n_{b}) \sim f \frac{n_{b}}{n_{0}}}
%\end{equation}
and a slope determined by the ratio of $f$ and the saturation
density $n_{0}$.
%}
On the other hand, the asymptotic form
\begin{equation}
\Delta m_d (n_{b}) \sim f \left(\frac{n_{b}}{n_{0}}\right)^{2/3}
\end{equation}
is expected for the deuteron mass shift, in the case when $\chi = 1$.

The coefficients $c$ and $\eta$, 
whose analytical expressions are given in Appendix\ \ref{sec:cd},
are finally determined by the constraints introduced on the mass shift and its derivative at saturation. 
%\blu{Remind to modify the formulas for $c$ and $d$ for the general case with $\gamma \neq 1$.}
%\blu{Should we give the analytical expressions derived for $c$ and $d$ parameters?}
%\blu{Maybe in an appendix if they are not too complicated.}
The remaining parameters $e$ and $g$ are free and allow to tune the relative role of the different contributions $\Delta m_{d, 2}$ and $\Delta m_{d, 3}$ in Eq.\ \eqref{eq:deltamd_param}.
%\red{In particular, the coefficient $b$ in the first term of \eqref{eq:deltamd_param} is chosen as $a/B_{d}= n_{0}/n_{d}^{(\mathrm{diss})}$ with the dissociation density $n_{d}^{(\mathrm{diss})}$ defined in Eq.\ \eqref{eq:ndiss}. In such a way, asymptotically, $\Delta m_{d, 1}$ tends to the deuteron binding energy $B_{d}$ which means the condition in which the deuterons are unbound.}
Only a tiny sensitivity of the results was assessed by varying the parameter $e$. Thus, this parameter was kept fixed to $1$ and only different values for the parameter $g$ were considered. Different choices of $g$ permit indeed to produce alternative supra-saturation scenarios while keeping the same asymptotic behavior. 

%%%%%%%%%%%%%%%%%%%%%%%%%%%%%%%%%%%%%%%%%%%%%%%%%%%%%%%%%%%%%%%%%%%%%
\subsubsection{Deuteron mass shift and mass fraction for $\chi = 1$}

\begin{table*}[t]
\centering
\caption{Values of the parameters in the deuteron mass shift parameterization defined by %\red{Eq.\ \eqref{eq:deltamd_param},  as obtained when} 
Eqs.\ \eqref{eq:deltamd_param} - \eqref{eq:deltamd123} for six different sets. They are obtained by employing the DD2 nucleon-meson effective interaction, with properly rescaled %\red{$\sigma$ and $\omega$}
meson coupling strengths at saturation. 
The parameters $a$, $c$ and $f$ are expressed in MeV, $b$, $\eta$, $e$, $g$ and $\gamma$ are dimensionless. The first three sets refer to the case with deuteron-meson coupling scaling factor $\chi=1$, the others to $\chi=1/\sqrt{2}$. %\blu{I have to double check the values for param1, param2, param3. Moreover, although it is not probably a problem, with these parameterizations $c$ turns out to be negative.}
} 
\begin{tabular}{*{9}{c}}
\toprule
& $a$ & $b$ & $c$ & $\eta$ & $e$ & $f$ &$g$ & $\gamma$ \\
\midrule 
\midrule
\hline
DD2 - d1 & $541.726060$ & $243.472387$ & $ -83.230901$
& $3.491787$ & $1.0$ & $214.368137$ & $0.65$ & $2/3$ \\
\hline
DD2 - d2 & $541.726060$ & $243.472387$ & $-98.923123$ & $3.200967$ & $1.0$ & $214.368137$ & $0.67632$ & $2/3$ \\
\hline
DD2 - d3 & $541.726060$ & $243.472387$ & $ -140.309501$ & $2.715545$ & $1.0$ & $214.368137$ & $0.75$ & $2/3$ \\           
\hline
\hline
DD2 - $\chi$d1 & $541.726060$ & $243.472387$ & $99.677247$ & $1.656159$ & $1.0$ & $181.113975$ & $0.18$ & $1$ \\
\hline
DD2 - $\chi$d2 &$541.726060$ & $243.472387$ & $70.476986$ & $1.230947$ & $1.0$ & $181.113975$ & $0.22$ & $1$ \\
\hline
DD2 - $\chi$d3 &$541.726060$ & $243.472387$ & $41.777908$ & $0.257252$ & $1.0$ & $181.113975$ & $0.26$ & $1$ \\
\bottomrule
\end{tabular}
\label{tab:deltamd_param}
\end{table*}

\begin{figure*}[tp!]
\includegraphics[width=.45\textwidth]{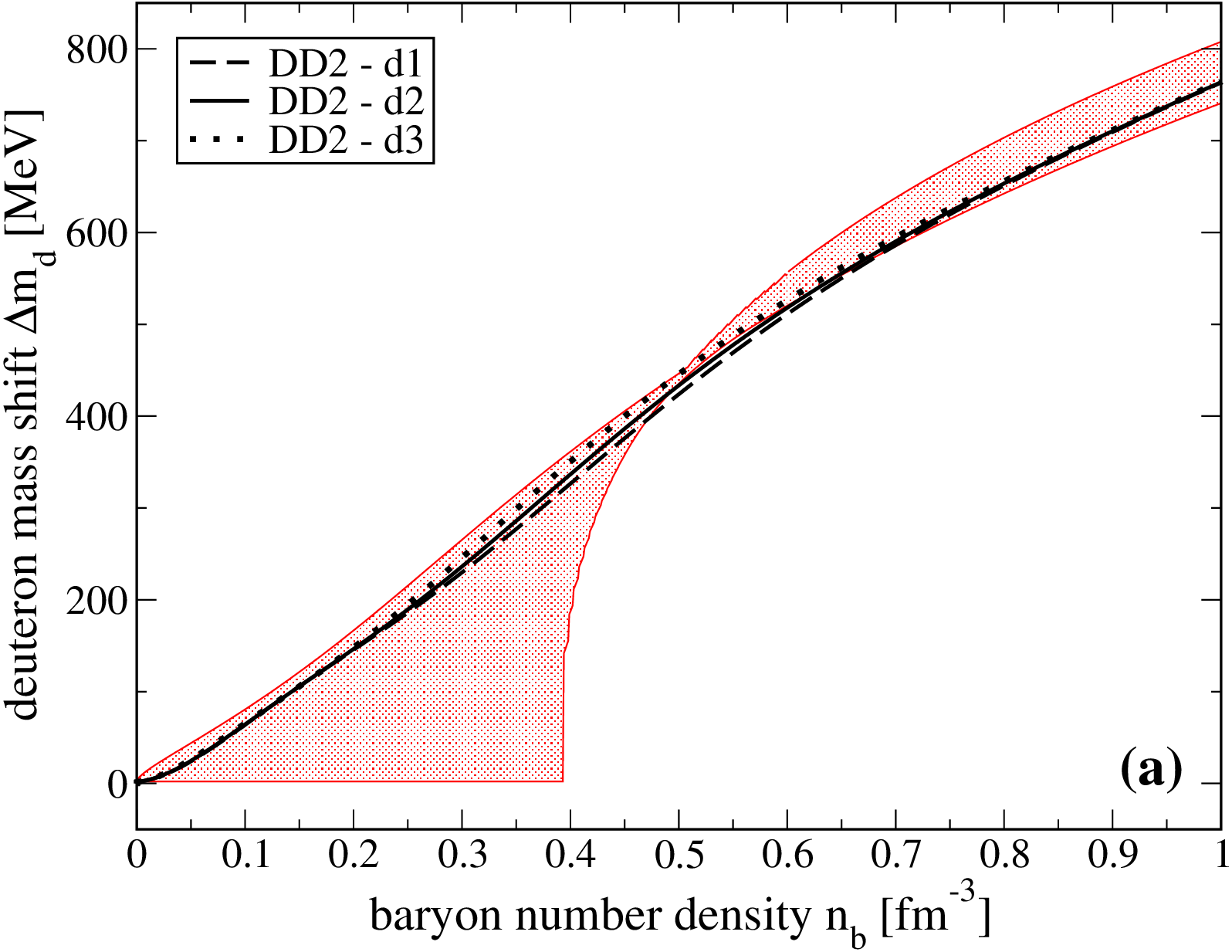} \qquad \includegraphics[width=.45\textwidth]{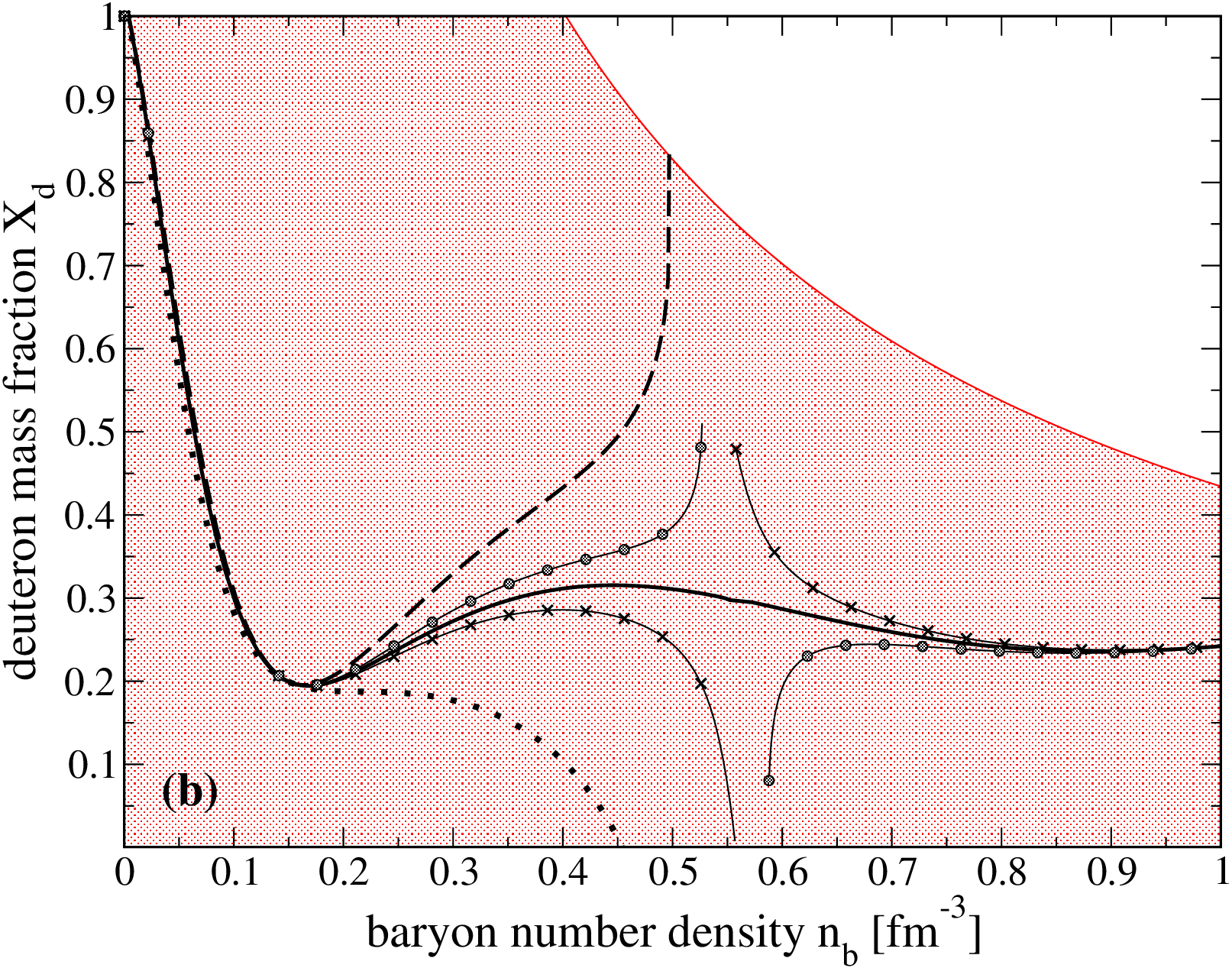}
\caption{\label{fig:deltamd_param_chi1} Panel (a): Deuteron mass shift as function of the baryon density, as determined according to the parameterization proposed in 
%\red{Eq.\ \eqref{eq:deltamd_param}, with $\gamma = 2/3$} 
Eqs.\ \eqref{eq:deltamd_param} - \eqref{eq:deltamd123}. The SNM case is considered and a scaling factor $\chi = 1$ is assumed for the deuteron-meson coupling strengths. Three different set of parameters were employed. 
%\blu{We should probably give the parameter values.}
The red shaded areas evidence the region of allowed mass shift or deuteron mass fraction values. Panel (b): Deuteron mass fraction $X_{d}$ as function of the baryon density, as determined by employing the same %\red{mass shift parameterizations} 
sets of parameters considered in panel (a). In both panels, the DD2 nucleon-meson effective interaction, with properly rescaled %\red{$\sigma$ and $\omega$} 
meson coupling strengths at saturation, is adopted. Two curves with symbols (not shown in panel (a)) are considered in panel (b), as a result of slightly varying the $g$ parameter of the set labeled as DD2-d2 (see text for more details).}
\end{figure*}

Let us consider first the case with $\chi = 1$.
For such a scaling factor, three different sets of parameters will be considered. They are determined according to the parameterization proposed in Eqs.\ \eqref{eq:deltamd_param} - \eqref{eq:deltamd123} with $\gamma=2/3$ and are labeled in the following as DD2-d1, DD2-d2 and DD2-d3.
The values of the parameters for these mass shift parameterizations are listed in the first three lines of Table \ref{tab:deltamd_param}. They were obtained %\red{when} 
by employing the DD2 nucleon-meson effective interaction with properly rescaled meson coupling strengths at saturation as given in Table \ref{tab:conv}.
Panel (a) of Fig.\ \ref{fig:deltamd_param_chi1} shows the deuteron mass shift as function of the baryon density 
%, \red{as determined according to the parameterization proposed in Eq.\ \eqref{eq:deltamd_param}, with $\gamma=2/3$ and by considering three different set of parameters} 
for these three different sets. The red shaded area corresponds to the range of possible mass shift values that are explored by assuming, for each density, deuteron fractions within the range %\red{$[0, X_{d}^{\rm(max)}]$} 
$[0, \mbox{min}\{ 1, X_{d}^{\rm(max)}\}]$. % \red{for each density}. 
First of all, Fig.\ \ref{fig:deltamd_param_chi1} highlights the validity of the proposed parameterization %\red{proposed} 
to comply with the constraints imposed on the deuteron mass-shift. 
The three %\red{parameterization proposed} \gre{
lines lie indeed within the red area for the whole range of displayed baryon densities. 

However, as shown in Table\ \ref{tab:deltamd_param}, 
a rather small range %\red{of values} 
of the parameter $g$ may be actually explored, owing to the shrinkage of the red area in the density region around the crossing points, which were observed in Fig.\ \ref{fig:deltamd_high}, panel (a). 
Despite their proximity, %\red{extremely different outcomes are obtained} %\red{, at least when approaching the region of the crossings,} 
when these parameterizations are employed in Eq.\ \eqref{eq:deq} to %\red{correspondingly} 
determine the density behavior of the deuteron fraction, 
extremely different outcomes are obtained,
at least when approaching the region of the crossings. 
As clearly %\red{evidenced} \gre{
depicted in Fig.\ \ref{fig:deltamd_param_chi1}, panel (b), the three adopted parameterizations %\red{adopted} 
correctly reproduce the low-density limit, corresponding to the situation in which the matter is entirely clusterized ($X_{d}=1$), and account also for the constraints at saturation. 

Nonetheless, highly diverse scenarios are %\red{foreseen} \gre{
predicted at high-densities. The dashed curve, which is the lowest one in the region around the shrinkage observed in panel (a), corresponds to a deuteron mass fraction which tends to %\red{overcome} \gre{
exceed, at a certain density, the maximum allowed value for $X_{d}^{\rm (max)}$. We remind that the density region beyond this point would be characterized by a negative value of the Dirac effective mass of the nucleons. To exclude this opportunity, the %\red{calculation} 
curve in Fig.\ \ref{fig:deltamd_param_chi1}, panel (b) is thus stopped beyond that density. 
Owing to the unfeasible high-density scenario, this set of parameter will not be further employed in the following.
Conversely, the DD2-d3 parameterization, plotted as the dotted line, lying above the other curves around $0.5$~fm$^{-3}$ in panel (a), predicts a sudden disappearance of the deuterons, when approaching the pole of Eq.\ \eqref{eq:deq}. 
Actually, even in this case, the curve is not plotted beyond the region of the crossing, to exclude the unrealistic situation in which the clusters reappear at very high densities. As already anticipated in Section \ref{sec:pole}, a quasi-continuous solution of Eq.\ \eqref{eq:deq} might be %\red{found} 
arranged for all densities with a fine tuning of the parameters. The corresponding curve is plotted in both panels of Fig.\ \ref{fig:deltamd_param_chi1} as the full line. 
In this case, a %\red{nice} 
smooth behavior is apparently recovered for the density behavior of the deuteron mass fraction. However, a strong sensitivity persists in correspondence of the pole, as manifested by the two thin lines with symbols plotted in panel (b). These curves are obtained with mass shifts functions that are found by varying the $g$ parameter only by 1\% of the %\red{parameterization} 
set labeled as %\red{param1} 
DD2-d2. Owing to the presence of the pole, the DD2-d2 set of parameters will be set aside hereafter too. A more refined method to find a continuous function would be to consider a more general form of the mass shift parameterization with respect to the one proposed in Eqs.\ \eqref{eq:deltamd_param}-\eqref{eq:deltamd123}. By enlarging the number of the involved parameters, it would be possible to constrain the mass-shift and its slope values such that the pole will be definitely washed out. Nonetheless, none of such possible parameterizations would 
%\red{Owing to the presence of the pole, such a parameterization does not}  
allow to accomplish our aim to extend the predictions to ANM.
Since the position of the pole evolves with the asymmetry, a divergence of the mass fraction would emerge once again as soon as the asymmetry of the matter is changed, despite its removal in the SNM case. %\red{by a fine tuning of the parameters}. 
The same holds, obviously, 
%\red{whatever form of the parameterization is proposed,} 
for any deuteron-meson coupling scaling factor $\chi > 1/\sqrt{2}$. 
However, one should bear in mind that, for values of the scaling factor
smaller than $1$, the pole is expected to appear at higher densities. Since for $\chi = 1$ the pole emerges already at rather large densities (around $3n_{0}$), one expects that for more realistic values of $\chi$, its position will be located much beyond the range that is relevant in the applications of the model. Then its emergence could be neglected in practice.
For $\chi=1$, only the DD2-d3 parameterization will be employed in the following, when the general properties of both SNM and ANM matter will be investigated.

%%%%%%%%%%%%%%%%%%%%%%%%%%%%%%%%%%%%%%%%%%%%%%%%%%%%%%%%%%%%%%%%%%%%%
\subsubsection{Deuteron mass shift and mass fraction for $\chi = 1/\sqrt{2}$}

As discussed in the previous sections, a possible way out to overcome the issue of the pole might be to assume a smaller scaling factor which is, at maximum, equal to $1/\sqrt{2}$.
For such a scaling factor, three different %\red{parameterizations} 
sets of parameters, labeled as %\red{DD2-d1, DD2-d2 and DD2-d3} 
DD2-$\chi$d1, DD2-$\chi$d2, and DD2-$\chi$d3, are proposed here.
The values of the parameters in %\red{Eq.\ \eqref{eq:deltamd_param}} 
Eqs.\ \eqref{eq:deltamd_param} - \eqref{eq:deltamd123} for these %\red{mass shift parameterizations} 
sets are listed again in Table \ref{tab:deltamd_param}. 
%\red{They were obtained %\red{when} by employing the DD2 nucleon-meson effective interaction with properly rescaled meson coupling strengths at saturation \cya{as} given in table \ref{tab:conv}.}

One observes a strong sensitivity to the $g$ parameter of the deduced values for $c$ and $\eta$. %\red{Indeed,} 
Moreover, when %\red{When} 
increasing $g$, both parameters to account for the saturation constraints decrease. A further increase of $g$ beyond a maximum value $g_{\rm max}\approx 0.26$ is
%\red{not permitted} 
excluded, since it would imply a negative value for $d$ and thus a dominant role of $\Delta m_{d, 2}$ in the zero-density limit, 
where a pole could even emerge. 
%\red{, where} 
However $\Delta m_{d, 1}$ returns already, by construction, the correct low-density trend. 

\begin{figure*}[tp!]
\includegraphics[width=.45\textwidth]{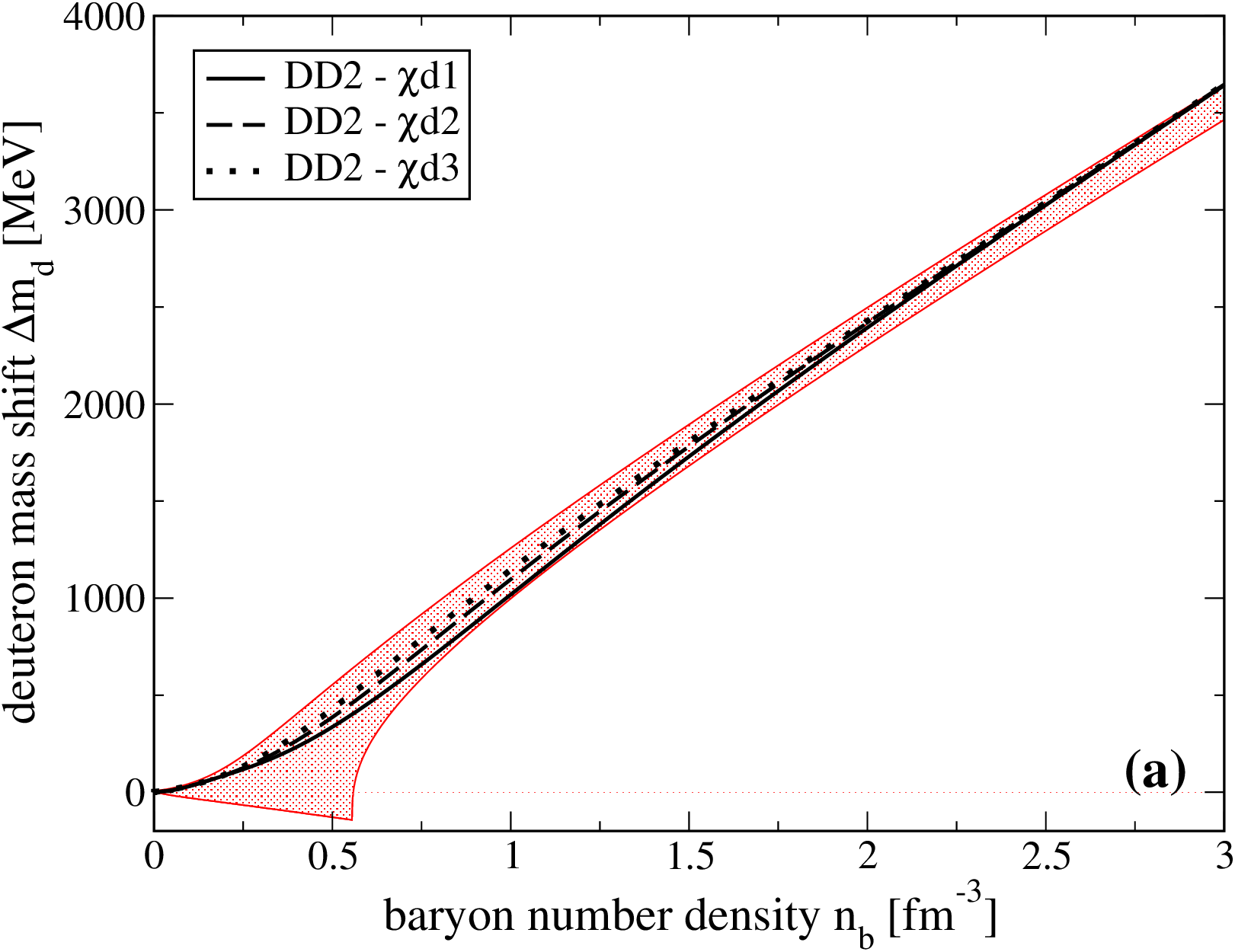} \qquad \includegraphics[width=.45\textwidth]{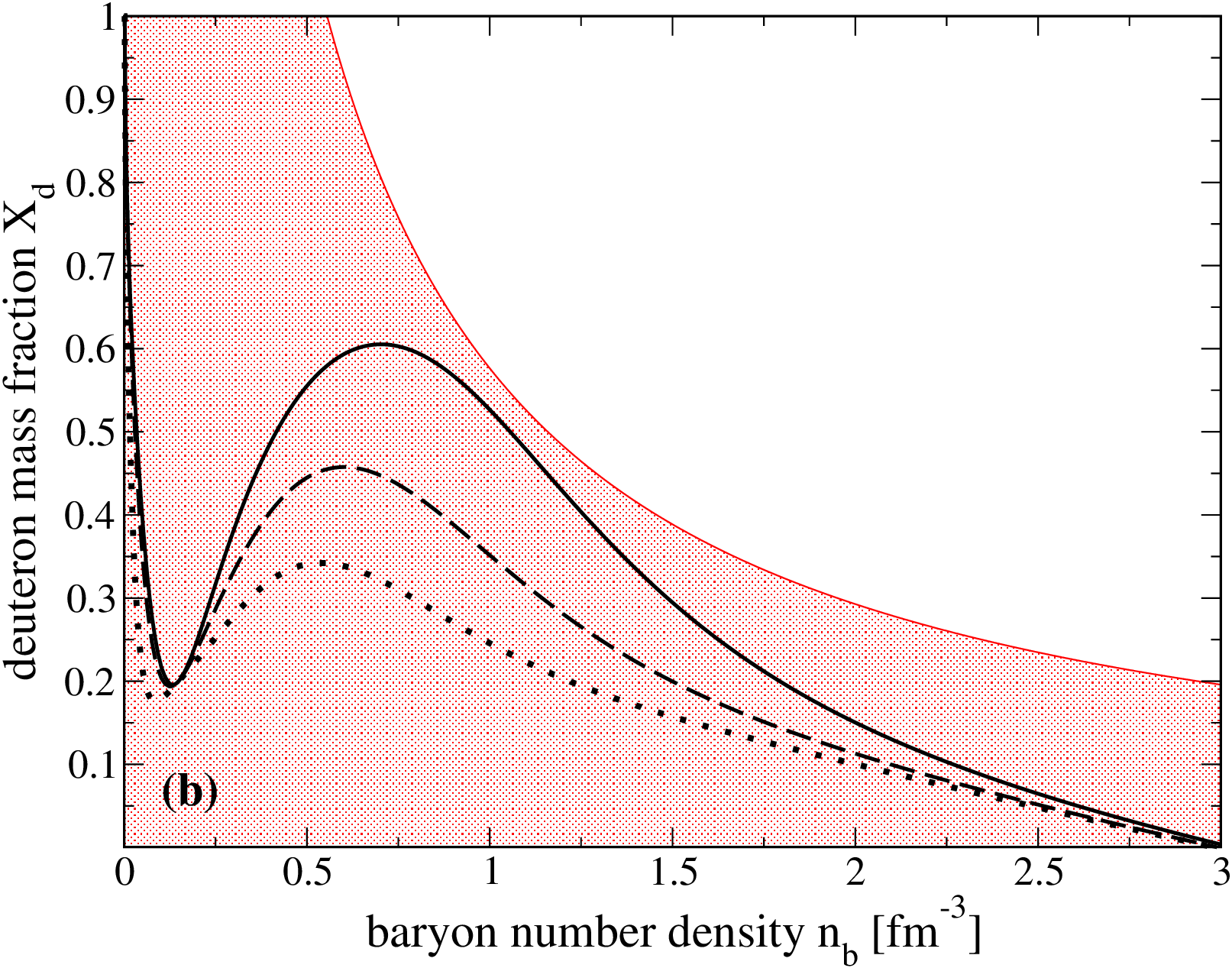}
\caption{\label{fig:deltamd_param_chismall} Panel (a): 
Deuteron mass shift as function of the baryon density, as determined according to the parameterization proposed in %\red{Eq.\ \eqref{eq:deltamd_param}} 
Eqs.\ \eqref{eq:deltamd_param} - \eqref{eq:deltamd123}. %\eqref{eq:deltamd_param}. 
The SNM case is considered and a scaling factor $\chi = 1/\sqrt{2}$ is assumed for the deuteron-meson coupling strengths. Three different set of parameters were employed. The red shaded area evidences the region between the (upper) curve, related to the deuteron-free case, {\it i.e.} $X_{d} = 0$, and the (lower) one, obtained by assuming $X_{d} =  \mbox{min} \left\{ 1,  X_{d}^{(\mathrm{max})}\right\}$. 
Panel (b): Deuteron mass fraction $X_{d}$ as function of the baryon density as determined by employing the same %\red{mass shift parameterizations} 
sets of parameters of panel (a). In both panels, the DD2 nucleon-meson effective interaction, with properly rescaled %\red{$\sigma$ and $\omega$} 
meson coupling strengths at saturation, is adopted.} 
\end{figure*}

Further insights may be achieved by looking at both panels of Fig.\ \ref{fig:deltamd_param_chismall}.
The density behaviors of these mass shifts are shown in Fig.\ \ref{fig:deltamd_param_chismall} panel (a). The same parameterizations are then employed in panel (b) of Fig.\ \ref{fig:deltamd_param_chismall} to determine the corresponding density behavior of the deuteron mass fraction $X_{d}$. In the two panels, the red shaded area evidences the allowed region between the (upper) curve, related to the deuteron-free case, {\it i.e.} $X_{d} = 0$, and the (lower) one, obtained by assuming $X_{d} =  \mbox{min} \left\{ 1, X_{d}^{(\mathrm{max})} \right\}$.

First of all, looking at panel (a), one notices that the black curves lie always within the red shaded area up to very large baryon densities, so validating the choice of the adopted parameter sets. In light of the constraints imposed in the extremely dilute regime and at saturation, the full, dashed and dottes black curves remain rather close up to $n_0$. All the curves also converge to the line characterized by $X_{d} = 0$, in the asymptotic limit. %\red{Significant} 
Some differences emerge instead in the high-density behavior, around and beyond $3n_{0}$. 

The observed differences in the mass shifts are then reflected in the density behavior of the deuteron mass fraction, which is plotted in panel (b). Independent on the parameterization, in the zero-density limit, the matter is completely clusterized and the deuteron mass fraction $X_{d}$ is %\red{$X_{d}=1$} 
equal to 1. With increasing density, a considerable reduction of $X_{d}$ is observed and a local minimum emerges in a density region around the saturation density. There the value $X_{d} = 0.2$ is reached, as required in agreement with the experimental evidences concerning the emergence of SRCs pairs. At supra-saturation densities, several scenarios take place. The curves never overshoot the line indicating the maximum allowed value for $X_{d}$, which ensures a non-negative value for the Dirac effective mass of the nucleons. Then, at higher densities, a decreasing trend is observed for all curves, which converge each other, approaching zero asymptotically.

It is worthwhile to notice that alternative scenarios, similar to the ones displayed 
by the dotted and the dashed lines in panel (b) of Fig.\ \ref{fig:deltamd_param_chi1}, would be possible also in the case with $\chi = 1/\sqrt{2}$. Solutions with deuteron mass fraction values which tend to exceed the maximum allowed at a certain density or abruptly vanishing might accidentally occur, when considering mass shift parameterizations which cross the lowest or the highest border, respectively, of the red shaded area shown in panel (a) of Fig.\ \ref{fig:deltamd_param_chismall}. Differently than the case with $\chi = 1$, these solutions are not connected to the emergence of any pole. 
%\red{As a result, they may be avoided with proper choices of the sets of parameters which, as seen in panel (b) of Fig.\ \ref{fig:deltamd_param_chismall},} 
The three sets of parameters proposed in panel (b) of Fig.\ \ref{fig:deltamd_param_chismall} avoid these scenarios. 
%\red{ and, conversely, return a smooth behavior for the whole range of baryon densities and for any asymmetry}. 
Solutions with the disappearance of the clusters at a certain baryon density would be likewise acceptable, even though none of the three chosen sets of parameters for $\chi = 1/\sqrt{2}$ provides a similar result. Indeed, for this class of solutions, the density behavior of the deuteron mass fraction would closely resemble the one obtained with the DD2-d3 parameterization. The main difference will be only the possible wider density range with a non-vanishing deuteron mass fraction values before the cluster is suppressed. For $\chi = 1/\sqrt{2}$, we will consider  three set of parameters, such that the mass shifts are characterized by similar smooth trends as functions of the baryon density, but different sizes of the deuteron mass fractions in the supra-saturation density regime. In such a way, we could assess and isolate the role of this ingredient.

The %\red{three} 
sets of parameters DD2-d3, DD2-$\chi$d1, DD2-$\chi$d2, and DD2-$\chi$d3 listed in Table \ref{tab:deltamd_param} return a smooth behavior of the mass fraction for the whole range of baryon densities and for any asymmetry. Thus, they will be employed in the next section to 
%\red{achieve the final results} 
study various properties, both for SNM and ANM.

%\blu{I think that also the case $\chi=1$ may be instructive to explore the different high-density scenarios. Alternatively, if it is possible to find it, for sake of illustration, one could show, still for $\chi = 1/\sqrt{2}$, another parameterization, with a critical density behavior of the deuteron mass fraction and thus a sudden disappearance of the deuterons, rather than only this nice smooth behavior.}

%%%%%%%%%%%%%%%%%%%%%%%%%%%%%%%%%%%%%%%%%%%%%%%%hese curves are obtained with mass shifts functionsthat are found by varying only by 1% thegparameterof the parameterization labeled asparam1. Owing to thepresence of the pole, such a parameterization does notallow to accomplishtoour aim tobeextendedtogetextend thepre%%%%%%%%%%%%%%%%%%%%%
\section{Thermodynamic quantities}
\label{sec:results}

Once the density behavior of the mass shift and of the corresponding deuteron mass fraction is determined, it is interesting to see how the %\red{appearance} 
embedding of quasi-clusters at supra-saturation densities affects some general thermodynamic quantities.

\subsection{SNM: EoS and incompressibility}
\begin{figure}[tbp!]
\includegraphics[width=.45\textwidth]{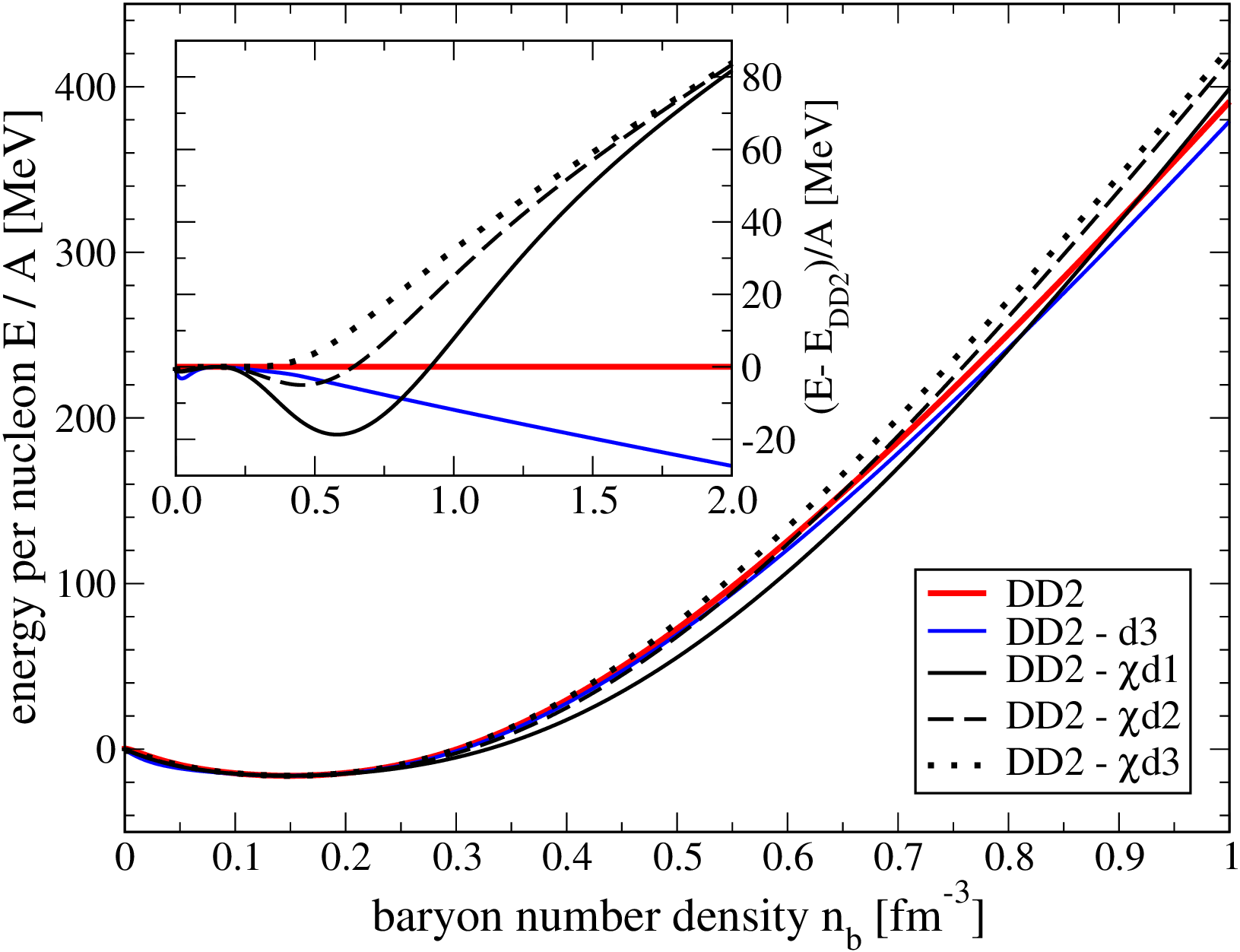} %\qquad \includegraphics[width=.44\textwidth]{compressibility_rho_param_SNM_chismall.eps}
\caption{\label{fig:energy_rho_SNM_param} Energy per nucleon as a function of the baryon density, as determined by employing four selected set of parameters listed in Table \ref{tab:deltamd_param}, for the mass shift parameterization given in Eqs.\ \eqref{eq:deltamd_param}-\eqref{eq:deltamd123}. For comparison, the curve obtained for the DD2 parameterization in the deuteron-free case is also shown, as the red full line. The inset shows the differences with respect to the predictions of the standard DD2 parameterization in a larger baryon density range.
}
\end{figure}

Let us focus in this section on the results for SNM.
The density dependence of the energy per nucleon $E/A$ is plotted in Fig.\ \ref{fig:energy_rho_SNM_param}. The sets of parameters DD2-d3, DD2-$\chi$d1, DD2-$\chi$d2, DD2-$\chi$d3 listed in Table\ \ref{tab:deltamd_param} are employed.
The standard DD2 parameterization, describing the deuteron free case, is also shown as reference for comparison. The inset of Fig.\ \ref{fig:energy_rho_SNM_param} depicts in particular the differences between the energy per nucleon derived with each %\red{parameterization} 
set and the chosen reference. %\blu{Which sets of parameters should we consider for SNM with $\chi = 1$?}\blu{Can we show all three?}\ora{In principle yes, of course. However, we already mentioned that we do not really trust some of them, expecially the one with a deuteron mass fraction exceeding $X_{d}^{\rm (max)}$. Moreover, if we show more lines, than two more figures are needed and we can not simply put it in Fig.\ \ref{fig:energy_compressibility_rho_SNM_smallchi}.} \blu{So maybe the dotted parametrization is a good choice, also for the incompressibility.}
First of all, one notices that, in light of the fit performed at saturation, all the curves remain rather close in the low-density regime. However, the parameterizations with deuterons differ from the standard DD2 result in the zero-density limit, 
approaching one half of the deuteron binding energy in vacuum. 
Moreover, remarkable differences also emerge in the high-density behavior of the energy per nucleon. Despite the fit performed at saturation, as %\red{shown} 
in Fig.\ \ref{fig:energy_rho_xdc}, a stronger binding is generally observed %\red{above} 
in the neighbourhood of $n_{0}$ in the parameterizations accounting for the presence of deuterons. 
This stronger attraction persists also 
%\red{at most} 
to all densities in the case of the DD2-d3 set, despite the disappearance of the deuterons which is expected to occur at $n_{b}$ around 0.45 fm$^{-3}$ with this parameterization (see Fig.\ \ref{fig:deltamd_param_chi1}). 
The reduction in the energy per nucleon observed at higher densities with the DD2-d3 is thus only driven by the changed balance between the scalar and vector components, which is a result of the rescaling of the meson coupling strengths at saturation. 

A different scenario manifests itself with the parameterizations derived with a reduced scaling factor $\chi$. There, 
%\red{despite the introduction of the quasi-deuterons and different from Fig.\ \ref{fig:energy_rho_xdc},} 
no systematic increased binding, as compared to the DD2 case, is observed. 
A delicate interplay takes place between the stronger attraction, which is produced by the presence of the deuterons, and the repulsion %\red{indicated} 
determined by the increased stiffness of the EoS.  This is a result of the 
%\red{changed balance between the scalar and vector components} 
modification introduced in the strengths of the $\sigma$ and $\omega$ meson couplings. A global repulsive contribution is %\red{asymptotically} predicted 
seen at large baryon densities, while %\red{, although} 
a significant reduction of the energy per nucleon might be observed up to approximately %\red{around} 
$3n_{0}$, depending on the value reached by the deuteron mass fraction in correspondence of the local maximum observed in Fig.\ \ref{fig:deltamd_param_chismall}. 
However, as highlighted in the inset of Fig.\ \ref{fig:energy_rho_SNM_param}, the three black curves converge in the asymptotic limit, where a smooth disappearance of the clusters %\red{is} 
was depicted in panel (b) of Fig.\ \ref{fig:deltamd_param_chi1}.

\begin{figure}[tbp!]
\includegraphics[width=.45\textwidth]{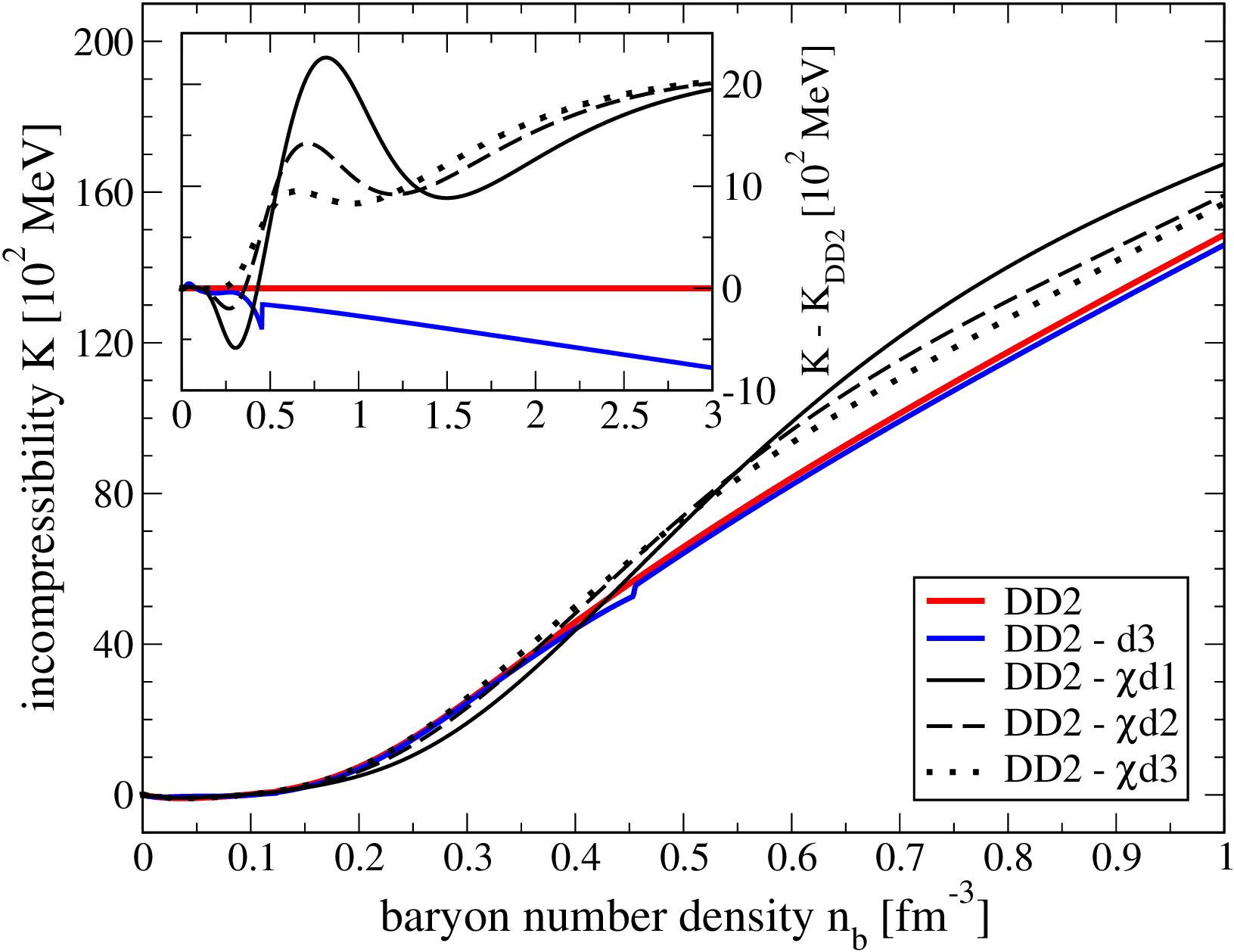} %\qquad \includegraphics[width=.44\textwidth]{compressibility_rho_param_SNM_chismall.eps}
\caption{\label{fig:incompressibility_rho_SNM_param} Incompressibility $K$ as a function of the baryon density as determined by employing four selected set of parameters listed in Table \ref{tab:deltamd_param} and for the mass shift parameterization given in Eqs.\ \eqref{eq:deltamd_param}-\eqref{eq:deltamd123}. For comparison, the curve obtained for the DD2 parameterization in the deuteron-free case is also shown as the red full line. The inset shows the differences with respect to the predictions of the standard DD2 parameterization in a larger baryon density range.
}
\end{figure}

The incompressibility characterizes the curvature of the energy per nucleon. It is defined here as
\begin{equation}
\label{eq:incompressibility}
K (n_{b}) = 9 n_b^2 \dfrac{\partial^2 (E/A)}{\partial n_{b}^2}
\end{equation}
through a second derivative with respect to the baryon density.\footnote{The original definition of $K$ uses a second derivative with respect to the Fermi momentum and gives different results as compared to the definition used here but can not be used at finite temperatures. However, both definitions coincide at saturation.} 
%\red{It is plotted in Fig.\ \ref{fig:incompressibility_rho_SNM_param} for four selected parameterizations of Table \ref{tab:deltamd_param} considered before. 
%The related inset shows the differences with respect to the standard DD2 reference, which is also shown in the main plot for comparison. } 
We recall that the incompressibility was not constrained within the approach adopted in this work. 
A constraint on the incompressibility would translate to a constraint on the second density derivative of the mass shift at saturation. In addition, it would require, as a further input at saturation, 
%\red{a change in the density dependence of the meson couplings or} \blu{?} 
the knowledge of the density derivative of the deuteron mass fraction. Although some numerical analyses have suggested that SRC pairs have a minimum in the neighbourhood of the saturation, owing to the interplay between the tensor component and the repulsive core of the nuclear force  \cite{riosPRC2014, yangPRC2019, liPRC2016}, we preferred to prescind from applying such a constraint. Instead, the predictions for the density behavior of $K$ are numerically extracted from the energy per nucleon. They are plotted in Fig.\ \ref{fig:incompressibility_rho_SNM_param} 
for the four selected parameterizations of Table \ref{tab:deltamd_param} considered before. 
The related inset shows the differences with respect to the standard DD2 reference, which is also shown in the main plot for comparison. 
As a general feature, one observes that
%\red{As a result, panel (b) highlights that,}
a softening of the EoS is recovered 
%\red{in the neighbourhood of saturation 
%independent of the parameterization incorporating the quasi-deuterons}
in the region beyond saturation, up to a density around $3n_{0}$. Furthermore, the size of this %\red{apart from the quite large increase of the incompressibility at higher densities. This} 
effect depends on the magnitude of the deuteron mass fraction. 

\begin{table}[t]
\centering
\caption{\label{tab:K0} Values of the incompressibility $K_{0}$, in MeV, at saturation density as derived according to Eq.\ \eqref{eq:incompressibility}, for four selected parameterizations employed in this work. The result for the DD2 is also given for comparison.
} 
\begin{tabular}{*{6}{c}}
\toprule
& DD2 & DD2-d3 & DD2-$\chi$d1 & DD2-$\chi$d2 & DD2-$\chi$d3 \\
\midrule
$K_{0}$ [MeV]& $242.7$ & $199.6$ & $185.3$ & $207.3$ & $240.3$ \\
\bottomrule
\end{tabular}
\end{table}

%\red{Nonetheless, as} 
As clearly evidenced in Table \ref{tab:K0},
the predictions for the incompressibility %\red{around} 
at saturation lie within or below the range of values generally assumed for this quantity 
\cite{Blaizot:1980tw,Youngblood:1999zza,Shlomo:2006,Stone:2014wza} 
for all the parameterizations here employed.
%\blu{Again, which sets of parameters should we consider for SNM with $\chi = 1$?}
Strong differences are observed in the high-density region, depending on the value for the scaling factor $\chi$. A much stiffer EoS is envisaged in particular at very large baryon densities, when a scaling factor $\chi = 1/\sqrt{2}$ is assumed, as a consequence of the significant change introduced in the balance between the scalar and the vector components in this case. 

Fig.\ \ref{fig:incompressibility_rho_SNM_param} also exhibits another aspect to be discussed. The blue curve, which corresponds to the DD2-d3 parameterization, reveals the emergence of a discontinuity at a baryon density around 0.45 fm$^{-3}$. 
This striking feature signals the abrupt disappearance of the cluster, which was observed in Fig.\ \ref{fig:deltamd_param_chi1}, panel (b) and already discussed before. It is worthwhile to mention that a discontinuity in the matter incompressibility or in any other quantity related to the second derivative of a thermodynamic potential is the signature for the possible emergence of a second-order phase transition. One notices, by the way, that this feature is in complete analogy to the disappearance of pairing correlations that was observed in previous works at low density
\cite{burrelloPRC2014, burrelloPRC2016}. Another discountinuity would emerge moreover at larger densities, if the calculation with the DD2-d3 parameterization is not stopped when the cluster dissolves, so that a further reappearance of the deuterons at higher density is allowed. %\blu{Maybe we can also show the result with this further reappearance of the cluster. What do you think?} 
The inset of Fig.\ \ref{fig:incompressibility_rho_SNM_param} shows that no discontinuity is instead observed when the deuteron mass fraction smoothly decreases with the density. The latter situation may however occur only for the parameterizations characterized by a reduced value of the scaling factor $\chi$.

\subsection{Predictions for ANM}
\begin{figure*}[tp!]
\includegraphics[width=.45\textwidth]{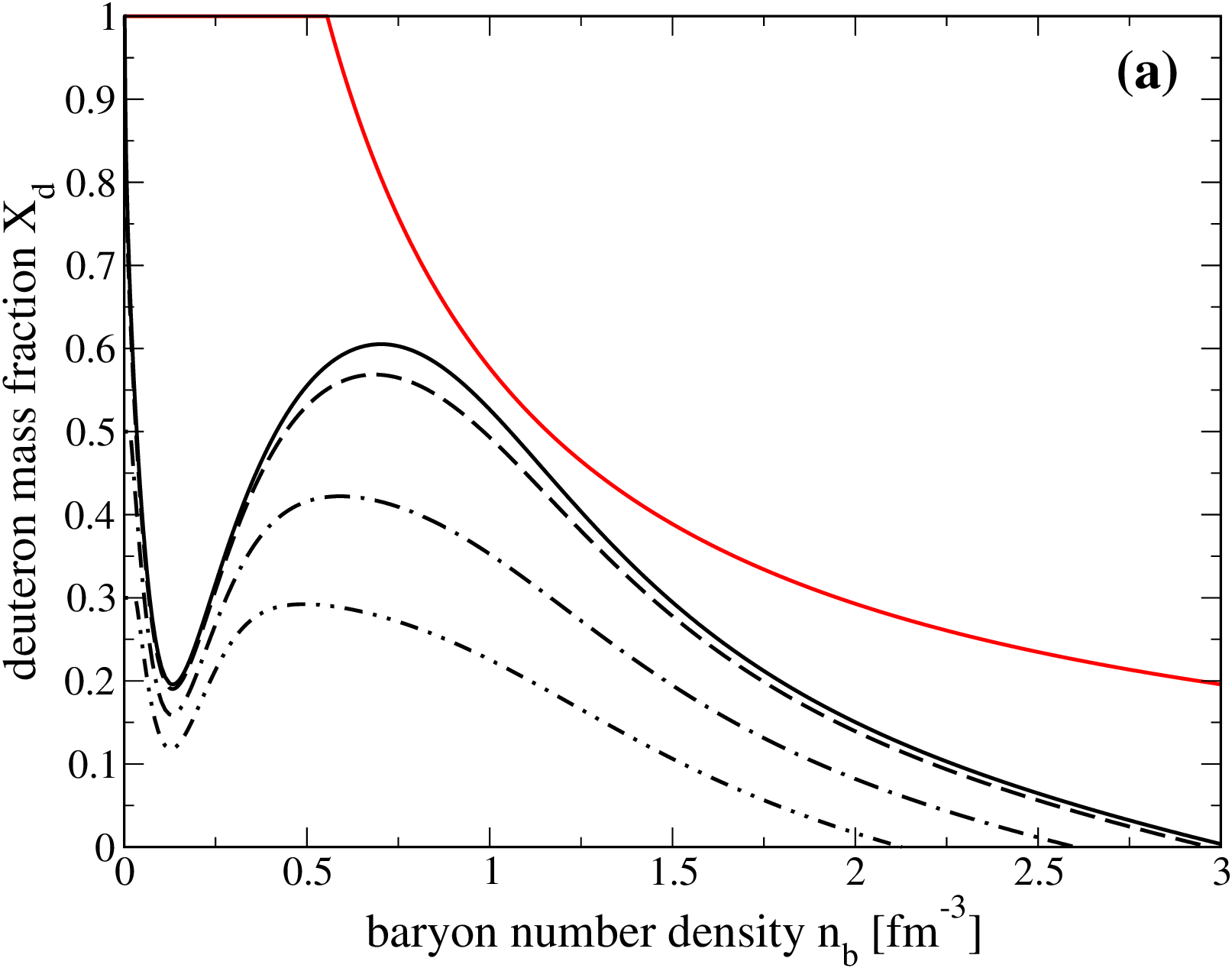} \qquad \includegraphics[width=.45\textwidth]{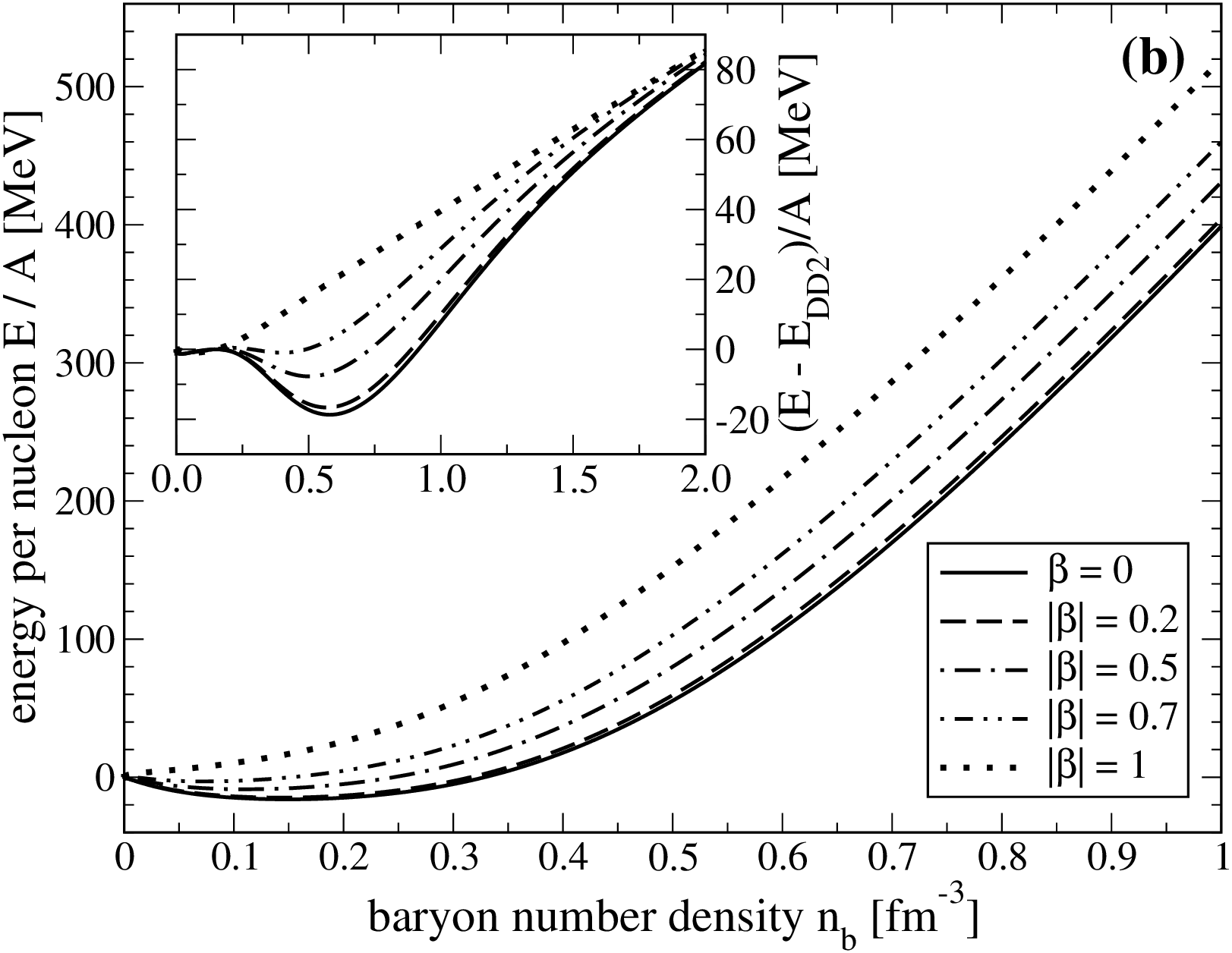}
\caption{\label{fig:xd_energy_rho_ANM_smallchi_param1} Panel(a): Deuteron mass fraction $X_{d}$ as function of the baryon density, as determined by employing the DD2-$\chi$d1 parameterization. The results for SNM ($\beta = 0$) are compared with the corresponding ones deduced with different values of the asymmetry $|\beta|$. Panel (b): Energy per nucleon as a function of the baryon density, as determined by employing the same parameterization and the same asymmetry values as in panel (a). The inset shows the difference of the energy per nucleon between the DD2-$\chi$d1 and DD2 parameterizations for the same asymmetry values, in 
%\red{the high-density regime} 
a wider range of baryon densities.}
\end{figure*}

\subsubsection{Deuteron mass fraction and EoS}
Let us finally concentrate on the predictions for ANM. In Fig.\ \ref{fig:xd_energy_rho_ANM_smallchi_param1}, panel (a), the density dependence of the deuteron mass fraction is plotted for different values of the isospin asymmetry $|\beta|$. The DD2-$\chi$d1 parameterization is employed for sake of illustration. The adopted parameterization allows one to get the largest value for the deuteron mass fraction around the local maximum, which was observed %\red{around} 
beyond 0.5 fm$^{-3}$ in panel (b) of Fig.\ \ref{fig:deltamd_param_chismall}. 
In such a way, the effect of embedding the quasi-deuterons at supra-saturation densities is better emphasized. However, similar results, at least from a qualitative point of view, would be obtained with the other sets of parameters accounting for the presence of the deuterons. 
The red line indicates the maximum allowed deuteron mass fraction values, which are compatible with a non negative value of the Dirac effective mass of the nucleons. Let us recall that this curve corresponds to $X_{d}= \mbox{min} \{ 1 - |\beta|,  X_{d}^{\rm (max)}\}$, so that it evolves with $|\beta|$. Then only the border for SNM ($\beta = 0$) is plotted in panel (a) of Fig.\ \ref{fig:xd_energy_rho_ANM_smallchi_param1} to avoid to overload the figure. 

As a quite interesting result, Fig.\ \ref{fig:xd_energy_rho_ANM_smallchi_param1} highlights that, although the mass shift function defined in Eqs.\ \eqref{eq:deltamd_param}-\eqref{eq:deltamd123} has no explicit dependence on the isospin asymmetry, the corresponding deuteron mass fraction evolves with $|\beta|$, giving rise to a continuous overall reduction when increasing the neutron-proton asymmetry of the matter. Moreover, the smooth transition to the cluster-free matter realized in SNM with the DD2-$\chi$d1 is preserved also in the ANM case. 

The density behavior of the energy per nucleon is depicted in panel (b) 
of Fig.\ \ref{fig:xd_energy_rho_ANM_smallchi_param1}, for the same parameterization and the same asymmetry values considered in panel (a). The inset of panel (b) displays moreover the difference in the energy per nucleon, as determined with the DD2-$\chi$d1 and the DD2 parameterization.
Although not clearly visible, differently than in the deuteron-free DD2 case, the zero-density limit does not approach zero, except for the PNM case ($\beta=1$) in which deuterons are obviously not formed. This feature will be more visible below, when studying the symmetry energy.

The results shown in the inset of panel (b) help to disentangle the effect induced on the stiffness of the EoS, owing to the rescaling of the meson couplings at saturation and the changes ascribable to the presence of the deuterons. The curve related to $|\beta|=1$, being characterized by $X_{d}=0$, demonstrates that, apart from a tiny enhancement of the attraction below saturation (not clearly visible in the figure), a much more repulsive PNM EoS 
%\red{for DD2-$\chi$d1} 
is produced for DD2-$\chi$d1 beyond saturation
as compared to DD2. This is the result of the modification induced on the effective interaction by changing the meson coupling strengths. However, for the curves characterized by smaller asymmetry values, such a repulsion is counterbalanced by the attraction produced by the deuterons. An interplay analogous to the one discussed in the SNM case takes place. In such a way, a reduction of the energy per nucleon of ANM might be observed with the DD2-$\chi$d1 parameterization, in the intermediate density region beyond saturation. This region may actually extend up to very large densities, %\red{ and}, 
close to $n_{b} = 0.9$ fm$^{-3}$ in SNM. On the other hand, in the asymptotic limit the quasi-deuterons tend to dissolve, so that their extra-binding vanishes and the black curves depicted in the inset converge to the $|\beta|=1$ one. 

\begin{figure*}[tbp!]
\includegraphics[width=.45\textwidth]{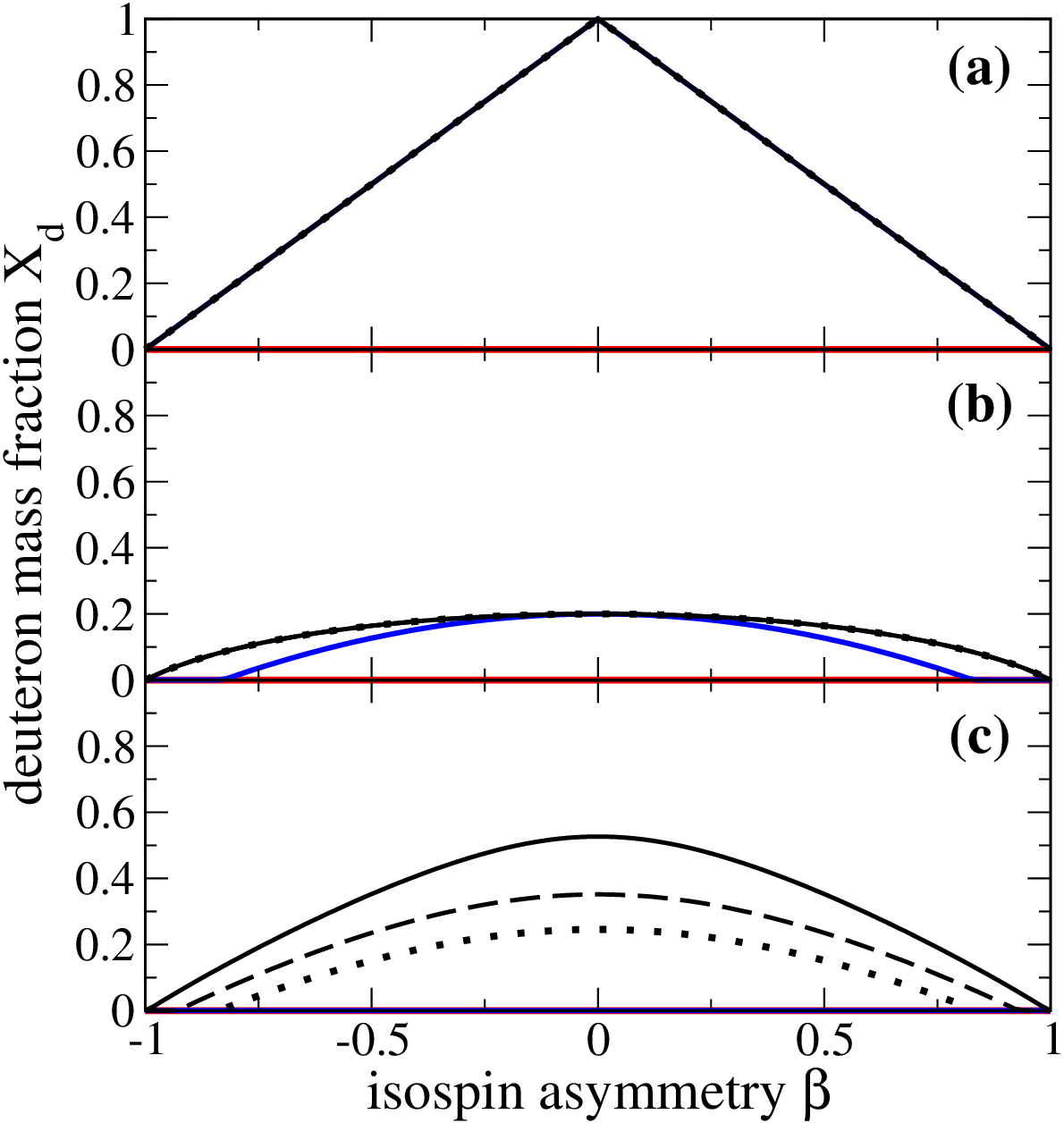} \qquad
\includegraphics[width=.45\textwidth]{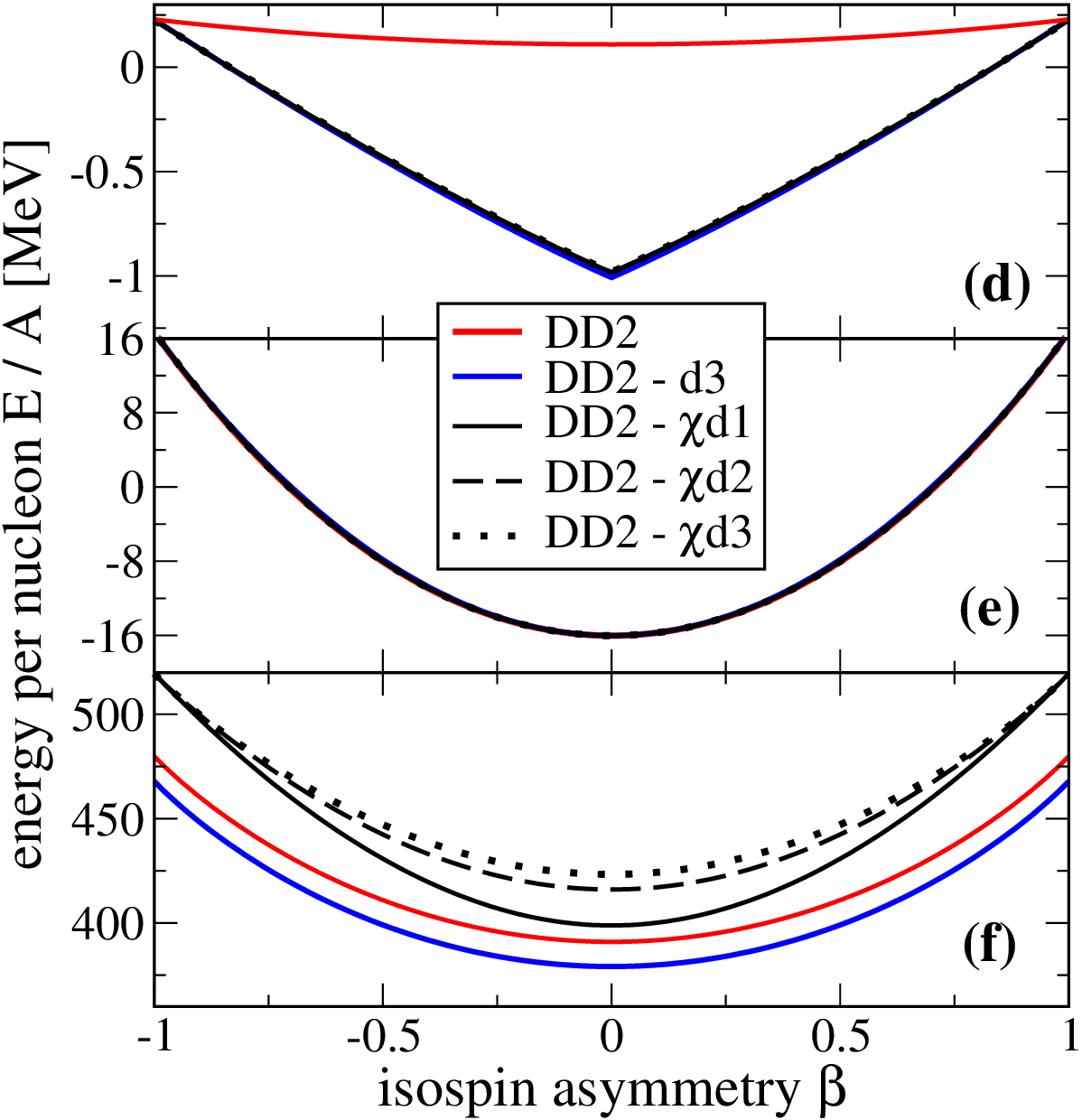} \caption{\label{fig:xd_energy_beta_ANM} Left panels: Deuteron mass fraction as function of the isospin asymmetry $\beta$, for four selected parameterizations accounting for the presence of the deuterons considered in this work. 
Right panels: Energy per nucleon as a function of $\beta$ as determined by employing the same parameterizations as in the left panels. For comparison, the curve obtained for the DD2 parameterization in the deuteron-free case is also shown. Three different values of the total baryon density are considered: $n_b^{\rm(low)} = 10^{-4}$ fm$^{-3}$ (panels (a) and (d)), $n_{0}$ (panels (b) and (e)) and $n_b^{\rm(high)} = 10^{0}$ fm$^{-3}$ (panels (c) and (f)).
}
\end{figure*}

A further insight into the dependence of the deuteron mass fraction and the energy per nucleon on the isospin asymmetry might be achieved by looking at Fig.\ \ref{fig:xd_energy_beta_ANM}. 
Three different values of the total baryon density are considered: $n_b^{\rm(low)} = 10^{-4}$~fm$^{-3}$ (panels (a) and (d)), $n_{0}$ (panels (b) and (e)) and $n_b^{\rm(high)} = 10^{0}$ fm$^{-3}$ (panels (c) and (f)). 
The standard DD2 parameterization is also plotted in the right panels.

First of all, one observes that all the quantities are symmetric with respect to the SNM ($\beta=0$) case. At the lowest density value considered in Fig.\ \ref{fig:xd_energy_beta_ANM}, $n_b^{\rm(low)}$, the deuteron fraction $X_{d}$ equals the maximum allowed fraction and behaves thus like $X_{d} = 1 - |\beta|$, for the parameterizations accounting for the presence of the deuterons. 
As a result, a characteristic triangular shape of $X_{d}$ and $E/A$ is observed in panels (a) and (d). The energy per nucleon does not follow the standard parabolic law which is predicted in the deuteron-free DD2 case and reaches %\red{considerably} 
smaller values %\red{for} 
in SNM. It approaches half of the deuteron binding energy in vacuum in the zero-density limit. The energy per nucleon determined with the parameterizations accounting for the deuterons coincides with the DD2 result only for matter composed exclusively of neutrons or protons. 

Secondly, a different picture is observed at the saturation density $n_{0}$. There, by varying the isospin asymmetry, the deuteron mass fraction %\red{, see panel (b),} 
departs from the value $X_{d} = 0.2$ imposed for $\beta = 0$
(see panel (b)). A mild dependence of the deuteron mass fraction on $\beta$ is assessed for the parameterization characterized by a reduced scaling factor $\chi$. A  larger sensitivity exists in the case of the DD2-d3 parameterization. Let us recall that, for this parameterization, the clusters disappear for SNM at density around 0.45 fm$^{-3}$. Panel (b) shows that the clusters may dissolve already %\red{at finite $\beta$ values} 
at saturation density for finite $\beta$ values, at least in the case of the DD2-d3 parameterization. On the other hand, the presence of the deuteron persists at $n_{0}$ 
%\red{and they dissolve only for} 
for all asymmetry values except $|\beta|=1$, 
for parameterizations with $\chi=1/\sqrt{2}$. 
However, the parabolic dependence of the energy per nucleon on the isospin asymmetry of the DD2 parameterization is perfectly reproduced with all the considered parameterizations. This result is clearly shown in panel (e). It originates from the requirement
to keep the energy per nucleon at saturation constrained, both for SNM and %\red{ANM} 
for matter composed exclusively of neutrons or protons.

Thirdly, it is interesting to discuss what happens at the highest density value, $n_{b}^{\rm (high)}$, considered in Fig.\ \ref{fig:xd_energy_beta_ANM}, panels (c) and (f). Here, different results are obtained among the %\red{parameterization} parameterizations. The same ordering as observed in the deuteron mass fraction in Fig.\ \ref{fig:deltamd_param_chismall} is preserved
%\red{among the} 
parameterizations characterized by a scaling factor $\chi = 1/\sqrt{2}$. However, for this density value, the deuterons survive at all asymmetries only with the DD2-$\chi$d1 parameterization, for which the largest value was already predicted in the SNM case. For the other two parameterizations with $\chi=1/\sqrt{2}$, the clusters dissolve already for $|\beta|$ values smaller than 1. On the other hand, since $n_{b}^{\rm (high)}$ lies beyond the density at which the cluster dissolution is predicted in SNM, the deuteron mass fraction identically vanishes in case of the DD2-d3 parameterization.  
The corresponding asymmetry dependence of the energy per nucleon, which is depicted in panel (f), is then driven only by the modification in the coupling strengths which was needed to keep the saturation properties well constrained. As in Fig.\ \ref{fig:energy_rho_SNM_param}, a slightly larger attraction is foreseen with the DD2-d3 parameterization, with respect to the DD2 reference case. The opposite happens instead when the parameterizations with a reduced value of the scaling factor are considered. In this case, a stronger repulsion is envisaged, partially mitigated, at least for small asymmetry values, by the stronger binding provided by the presence of deuterons. Quite interestingly, one observes that the parameterizations plotted by black curves always converge when approaching $|\beta|=1$, where $X_{d}=0$. %\red{Then, a change of parameterization is equivalent to vary the deuteron content at supra-saturation density. As a result, the main effect is a change in the curvature of the isospin asymmetry dependence of the energy per nucleon, that is the so-called symmetry energy. Further investigation on this quantity are then carried out in the following.} 
A change of the parameterization, which %\red{comes with} 
implies %\red{simultaneous} \red{a consequent}
an according change of the deuteron fraction at supra-saturation densities, 
affects the curvature of the dependence of the energy per nucleon on $\beta$ and thus the symmetry energy. The latter quantity 
%\red{and modifies the symmetry energy, which} 
will be studied in detail below. 

\subsubsection{Symmetry energy and its slope}
\begin{figure*}[tbp!]
\includegraphics[width=.45\textwidth]{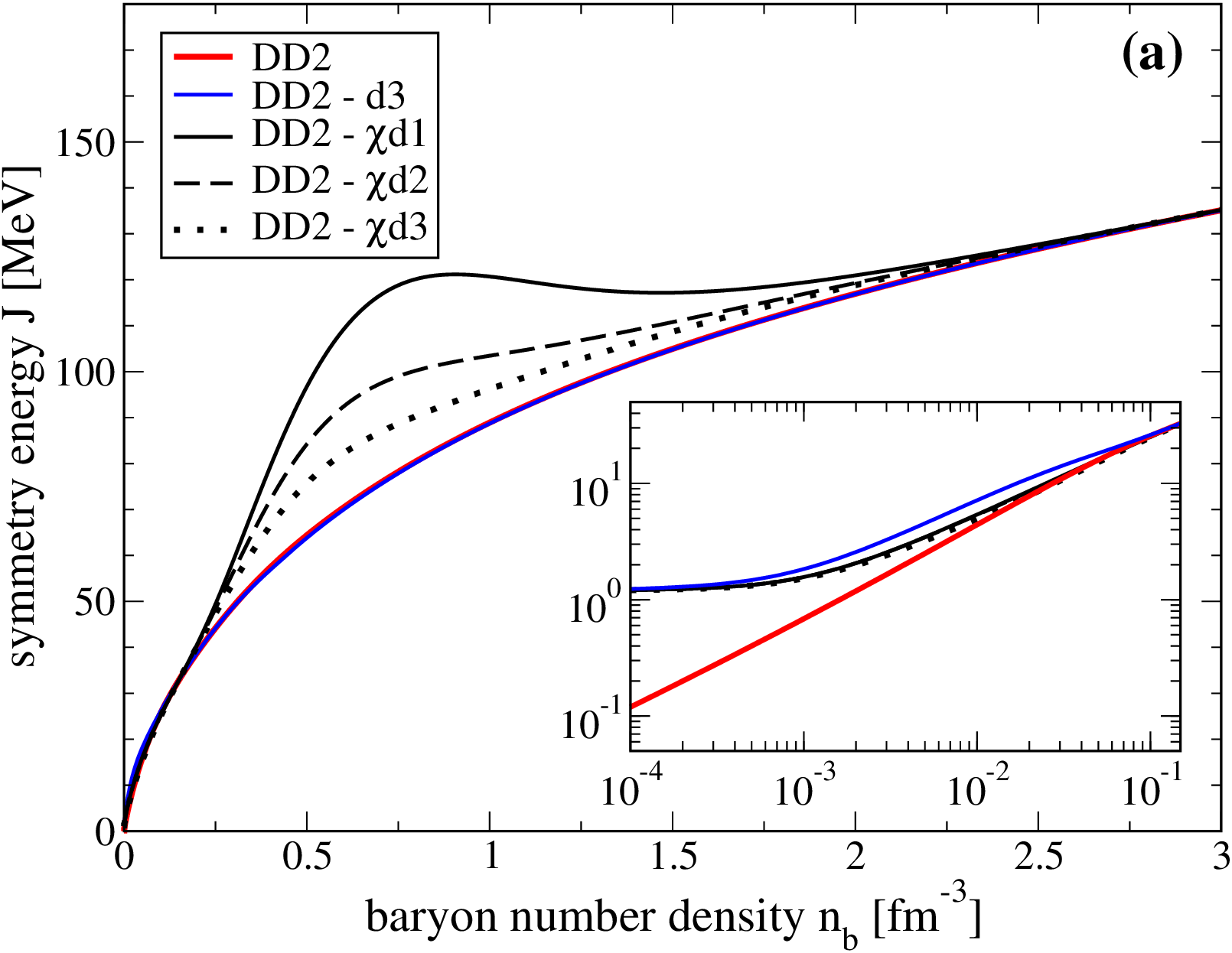} \qquad \includegraphics[width=.45\textwidth]{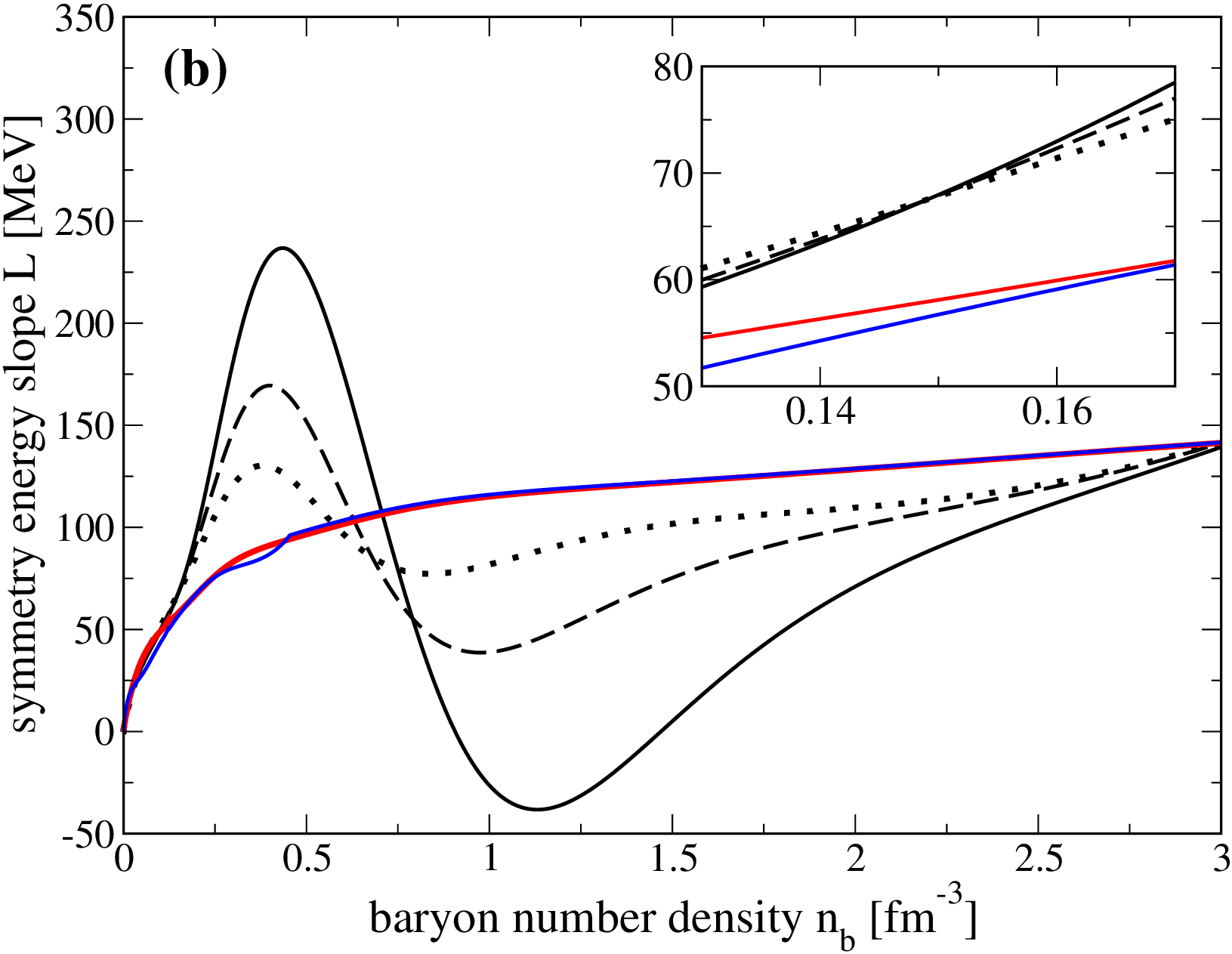} \caption{\label{fig:symmetry_energy_slope_param} 
Panel (a): Symmetry energy $J$ as function of the baryon density, as determined through Eq.\ \eqref{eq:symmetry_energy} for four selected parameterizations employed in this work. The inset shows a zoom at sub-saturation densities. 
Panel (b): Slope $L$ of the symmetry energy as function of the baryon density, as determined through Eq.\ \eqref{eq:slope}, for the same parameterizations as in panel (a). The inset shows a zoom  around the saturation density $n_{0}$. }
\end{figure*}

In the present work, the symmetry energy $J$ is calculated as the difference between the energies per nucleon in PNM and SNM
\begin{equation}
\label{eq:symmetry_energy}
J(n_{b}) = \left.\dfrac{E}{A}\right|_{\beta = 1}(n_{b}) - \left. \dfrac{E}{A}\right|_{\beta = 0}(n_{b}) \: .
\end{equation}
The quantity obtained %\red{in this way} 
through this equation is identical to the symmetry energy calculated from the usual definition
\begin{equation}
    J(n_{b}) = \frac{1}{2} \left. \frac{\partial^{2} (E/A)}{\partial \beta^{2}} \right|_{\beta = 0}
\end{equation}
using a second derivative of the energy per nucleon with respect to the asymmetry, if $E/A$ follows
a quadratic dependence on $\beta$.
The density dependence of $J$ is plotted in Fig.\ \ref{fig:symmetry_energy_slope_param}, panel (a). The inset of panel (a) shows a zoom at sub-saturation densities.  
Once again, the inset highlights the dissimilar behavior in the zero density limit, when the presence of the clusters is taken into account or not. It reflects the differences existing in the very dilute regime of the SNM EoS with respect to the deuteron-free case. If clustering is taken into account, the symmetry energy approaches indeed half of the deuteron binding energy in the zero-density limit in contrast to the simple description without explicit two-particle correlations. 
Some differences emerge among the parameterizations which account for the presence of the deuterons below $n_{0}$.  Apart from the constraint at saturation, no restrictions have been imposed on the density behavior of $J$. 

Concerning the behavior beyond saturation density, despite the presence of the deuterons, at least up to $0.45$~fm$^{-3}$, the blue curve remains close to the result of the standard DD2 parameterization. Huge differences are instead observed in the supra-saturation density region when the parameterizations with a reduced value of the deuteron-meson coupling scaling factor is considered. Furthermore, the size of the effect depends quite strongly on the mass fraction of the deuterons.
The differences vanish in the asymptotic limit, where the deuterons disappear. The black and blue curves approach the result of the standard DD2 parameterization as the contribution of the $\rho$ meson to the symmetry energy vanishes due to the suppression of its coupling, leaving the imbalance of the Fermi momenta of the nucleons as the main contribution to $J$.

Finally, it is interesting to look at the density derivative of the curves plotted in panel (a). The slope $L$ of the symmetry energy is numerically calculated here as
\begin{equation}
\label{eq:slope}
L (n_{b}) = 3 n_{b} \dfrac{d J}{d n_{b}}     
\end{equation}
and the result is shown in Fig.\ \ref{fig:symmetry_energy_slope_param}, panel (b). Except for the different behavior in the zero-density limit, the blue curve roughly coincides with the standard DD2 in the whole range of densities. A small kink is only observed for the DD2-d3 parameterization around 0.45 fm$^{-3}$. This kink is related to the disappearance of the deuterons in the SNM case. We recall that this feature was also responsible for the emergence of the discontinuity in the matter incompressibility discussed before. A huge double oscillation around the result obtained with the standard DD2 parameterization is observed for the parameterizations with a reduced scaling factor $\chi$. The magnitude of this oscillation depends again on the deuteron mass fraction predicted at supra-saturation densities, thus reflecting the result shown in panel (a) of Fig.\ \ref{fig:symmetry_energy_slope_param}. 
The inset of panel (b) displays the predictions for the symmetry energy slope around saturation. In spite of the extremely large differences in the high-density regime, reasonable values are obtained at saturation density with all the parameterizations considered in this work. These values, which are collected in Table\ \ref{tab:L0}, lie within the range usually assumed for the slope of the symmetry energy, see, e.g., \cite{Li:2021thg} and references therein. The three black curves, crossing each other at saturation, naturally provide the same value. Alternative scenarios  %\red{can be expected} 
manifest for the high-density behavior of the symmetry energy and its slope. The stiffness of the EoS in the supra-saturation density regime turns out to be %\red{determined} 
strongly affected by the value of the scaling factor. It is less dependent on the mere presence of the deuterons. Thus, as a general feature, one concludes that smaller values of the scaling factor correspond to higher stiffness values in the density region beyond saturation 
relevant for the applications of the model.
%\blu{Probably we should add something more here, to also conclude this section.} 
%\blu{Well, what else can we write here? }

\begin{table}[t]
\centering
\caption{\label{tab:L0} Values of the slope of the symmetry energy $L_{0}$, in MeV, at saturation density as derived according to Eq.\ \eqref{eq:slope}, for four selected parameterizations employed in this work. The result for the DD2 is also given for comparison.
} 
\begin{tabular}{*{6}{c}}
\toprule
& DD2 & DD2-d3 & DD2-$\chi$d1 & DD2-$\chi$d2 & DD2-$\chi$d3 \\
\midrule
$L_{0}$ [MeV]& $57.94$ & $56.49$ & $67.50$ & $67.50$ & $67.50$ \\
\bottomrule
\end{tabular}
\end{table}

%%%%%%%%%%%%%%%%%%%%%%%%%%%%%%%%%%%%%%%%%%%%%%%%%%%%%%%%%%%%%%%%%%%%%
\section{Conclusions and outlook}
\label{sec:conclusions}

In this paper, we have proposed and explored a novel approach to embed SRCs within the GRDF, a well-established phenomenological EDF based on nucleon and cluster degrees of freedom. In such a way, we aim to overcome the inconsistencies between recent experimental evidences, which brought to light the existence of SRCs, and the predictions of phenomenological models derived from mean-field approaches without explicit correlations at densities around saturation. Previous generalisations of these EDFs represent many-body correlation by cluster which dissolve, by construction, when the nuclear saturation density is approached from below.
Within an extended relativistic mean-field model with density dependent couplings, the idea of this work was to effectively account for the existence of SRCs through proper in-medium modifications of the cluster properties around saturation and above. They are considered as quasi-particles with density-dependent binding energies. 
Quasi-deuterons immersed in dense matter are used as surrogate for correlations in this first exploratory step. For the time being, the zero temperature case, where the deuteron fraction is determined by the density of a boson condensate, was addressed.

Suitable parameterizations of the cluster mass shift were derived, for the first time, for all baryon densities. The proposed mass shift functions comply with the available constraints and were employed to determine the density dependence of the quasi-deuteron mass fraction at arbitrary isospin asymmetries, thus for symmetric as well as for asymmetric nuclear matter. 
They were constrained by microscopic many-body calculations in the low-density limit, by specifying the actual deuteron fraction at saturation and by assuming a deuteron mass fraction behavior that respects the boson condensation condition at higher densities. 
The effective deuteron fraction around saturation was specified by extrapolating the experimental results on SRC pairs in nuclei
to infinite nuclear matter. Further constraints were moreover imposed at supra-saturation densities by the maximum allowed deuteron fraction, compatible with a non-negative value of the Dirac effective mass of the nucleons.

A proper description of well-constrained nuclear matter quantities at saturation required a refit of the nucleon-meson coupling strengths. An important role of the coupling scaling factor $\chi$ was revealed. It rules the coupling strength of the mesons with nucleons bound in the clusters. Such a scaling factor plays actually a primary role in the whole analysis developed in this work.

The natural choice was supposed to be that the nucleons inside the deuterons couple to the mesons with the same strength as the unbound nucleons. However, with this choice, the deuteron mass shift is not a monotonic function of the deuteron mass fraction for all baryon densities. As a result, the relation between the mass shift and the deuteron mass fraction can not be inverted uniquely in all cases. The same holds for any scaling factor $\chi$ larger than $1/\sqrt{2}$. Thus, as possible extreme values of two well distinct behaviors, two different values of the scaling factor, namely $\chi =1$ and $\chi = 1/\sqrt{2}$ were chosen, in the calculations performed in this work. The latter value is however significantly smaller than the universal scaling factor for the cluster-meson coupling strength.  %\red{It} 
A value smaller than $\chi=1$ was proposed in previous calculations of the EoS to take into account in-medium effects and to get a good description of the chemical equilibrium constants determined from recent experimental data.

As a general feature, our analysis shows that, for $\chi=1$, the only possible smooth solution for the density dependence of the deuteron mass fraction implies a sudden disappearance of the clusters at a density below the one corresponding to the emergence of a pole. In correspondence of this density, a discontinuity in the matter incompressibility emerges, analogous to the one observed at low density, owing to the disappearance of the pairing correlations and indicating the emergence of a second order phase transition. The analogy between the behavior of pairing and SRCs deserves however further investigation. For $\chi=1$, the density where the pole emerges is located around three times the saturation density. 
However, for more realistic values of the scaling factor, the pole is expected to appear at much higher densities, thus much beyond the range that is relevant in applications of the model. When the scaling factor value $\chi=1/\sqrt{2}$ is considered, alternative solutions exist, permitting smooth functions of the density dependence of the deuteron mass fraction for all densities. 
Three different parameterizations,  providing such a smooth behavior and characterized by different maximum deuteron fraction values at supra-saturation densities, were proposed.

Striking effects on some thermodynamic quantities are recognized, owing to the presence of the quasi-deuterons in the neighbourhood of saturation and at supra-saturation densities. In particular, a softening of the SNM EoS is systematically observed with respect to the standard DD2 parameterization, which does not include deuteron-like correlations. However, the stronger attraction, which is produced by the presence of the deuterons, might be counterbalanced by the repulsion driven by the modified coupling strengths. This delicate interplay is additionally tuned by the value of the scaling factor, which determines then alternative scenarios for the high-density behavior of the symmetry energy and its slope. In general, one concludes that smaller values of the scaling factor correspond to higher stiffness of the EoS in the supra-saturation density regime.

Last, but not least, it is worthwhile to recall that our analysis permits to also recover the correct low-density limit of the EoS. Indeed, at zero-density, both the energy per nucleon of SNM and the symmetry energy tend to be equal to one half of the deuteron binding energy in vacuum, in contrast to the predictions of standard mean-field models without cluster correlations.

The findings of the present study represent a first step to improve the description of nuclear matter and its EoS at supra-saturation densities in EDFs by considering correlations in an effective way. In a next step, the single-particle momentum distributions can be explored using proper wave functions of the quasi-deuteron in the medium. They have to be derived consistently with the interaction used in the model and will lead to prediction of the cluster mass shifts and fractions that can be compared to the suggested forms of the present work.
The many-body wave function of a cluster contains correlated nucleons with a specific momentum distribution. Then an imprint on the single-nucleon momentum distribution in nuclear matter is expected, such that a high-momentum tail develops even at zero temperature, as observed in the experimental study of SRCs by nucleon knockout with high-energy electrons.

The present approach can be generalized to finite temperatures, where a further change of the single-nucleon momentum distribution arises owing to the thermal change in the distribution functions. Also a momentum dependence of the mass shift and a more involved dependence on the isospin asymmetry might be considered in a future work, together with the effect of including heavier clusters and to investigate their relative importance.

As a perspective, we finally aim at investigating the effect of SRCs on neutron stars in the EDF framework, similarly to what was done in some prior studies, see, e.g., 
\cite{Lu:2021xvj}, %\cite{Cai:2015xga,Cai:2017rwj}) 
but our approach is to replace heuristic parameterizations of the momentum distributions with more microscopically founded descriptions.

%Shahrbaf:2022upc

More in general, we aim at achieving a more comprehensive description of correlations and clustering phenomena, which represents still a challenge from a theoretical point of view, despite the importance of these features in the widest scope of astrophysical applications and for general aspects of reactions dynamics in heavy-ion collisions.

%%%%%%%%%%%%%%%%%%%%%%%%%%%%%%%%%%%%%%%%%%%%%%%%%%%%%%%%%%%%%%%%%%%%%
\section*{Acknowledgments} 

The authors thank Maria Colonna and Gerd R\"{o}pke for their comments and suggestions on this work.
S. B. acknowledges support from the Alexander von Humboldt foundation.

%%%%%%%%%%%%%%%%%%%%%%%%%%%%%%%%%%%%%%%%%%%%%%%%%%%%%%%%%%%%%%%%%%%%%
%\clearpage
\appendix

%%%%%%%%%%%%%%%%%%%%%%%%%%%%%%%%%%%%%%%%%%%%%%%%%%%%%%%%%%%%%%%%%%%%%
\section{Mass fraction derivative of the deuteron mass shift}
\label{sec:ddeltamd_dxd}

The deuteron mass shift \eqref{eq:deltamd_high} is in general a function of the baryon density $n_{b}$, the asymmetry $\beta$, the deuteron fraction $X_{d}$, and the temperature T.
In this section, the derivative of %\red{$\Delta m_{d}$} 
$\Delta m_{d}^{\rm(high)}$ with respect to $X_{d}$ is derived
for constant $n_{b}$ and $\beta$ at $T=0$. In a first step, 
the derivative of the nucleon effective chemical potential 
with respect to $X_{d}$ is expressed as
\begin{equation}
\label{eq:dmuqdxd}
   \left. \frac{\partial \mu_{q}^{\ast}}{\partial X_{d}} \right|_{n_{b},\beta} =
  \frac{1}{\mu_{q}^{\ast}}
  \left( k_{q} \left. \frac{\partial k_{q}}{\partial X_{d}} \right|_{n_{b},\beta}
  + m_{q}^{\ast} \left. \frac{\partial m_{q}^{\ast}}{\partial X_{d}} \right|_{n_{b},\beta}\right)
\end{equation}
using Eq.\ \eqref{eq:mu_q_eff}.
The derivative of the Fermi momentum of a nucleon $q=n,p$ is found with help of the relation
\begin{equation}
\label{eq:dkndxd}
\left. \frac{\partial n_{q}^{(v)}}{\partial X_{d}} \right|_{n_{b},\beta}
= - \frac{n_{b}}{2} = \frac{3n_{q}^{(v)}}{k_{q}} 
\left. \frac{\partial k_{q}}{\partial X_{d}} \right|_{n_{b},\beta}
\end{equation}
for the vector density \eqref{eq:vec_dens}. Using Eqs.\ \eqref{eq:dmuqdxd}
and \eqref{eq:dkndxd},
the derivative of the scalar density of the nucleons can be written as
\begin{equation}
\label{eq:dnqsdxd}
\left. \frac{\partial n_{q}^{(s)}}{\partial X_{d}} \right|_{n_{b},\beta}
= - \frac{n_{b}}{2}\frac{m_{q}^{\ast}}{\mu_{q}^{\ast}}
  + f_{q}   \left. \frac{\partial m_{q}^{\ast}}{\partial X_{d}} \right|_{n_{b},\beta}
\end{equation}
with the factor
\begin{equation}
\label{eq:fq}
    f_{q} = 3 \left( \frac{n_{q}^{(s)}}{m_{q}^{\ast}}
    - \frac{n_{q}^{(v)}}{\mu_{q}^{\ast}}\right)
\end{equation}
after several steps of recasting the individual contributions.
Eq.\  \eqref{eq:dnqsdxd} contains again the
derivative of the effective mass that assumes the simple form
\begin{equation}
    \left. \frac{\partial m_{q}^{\ast}}{\partial X_{d}} \right|_{n_{b},\beta}
    = - C_{\sigma} \left. \frac{\partial n_{\sigma}}{\partial X_{d}} \right|_{n_{b},\beta}
\end{equation}
because $C_{\sigma}$ depends only on $n_{b}$. 
With the derivatives
\begin{equation}
    \left. \frac{\partial n_{d}^{(v)}}{\partial X_{d}} \right|_{n_{b},\beta}
    = \left. \frac{\partial n_{d}^{(s)}}{\partial X_{d}} \right|_{n_{b},\beta}
    = \frac{n_{b}}{2}
\end{equation}
of the deuteron densities, the derivative of the $\sigma$ meson source density 
\eqref{eq:nsigma} is found as
\begin{eqnarray}
  \left. \frac{\partial n_{\sigma}}{\partial X_{d}} \right|_{n_{b},\beta}
 & = &\frac{n_{b}}{1 + \left( f_{n}  + f_{p} \right) 
   C_{\sigma}}   \: \mathcal{U}_{d}
 \end{eqnarray}
with
\begin{eqnarray}
\label{eq:U_d}
  \mathcal{U}_{d} & = &
  \chi_{d\sigma} - \frac{m_{\rm nuc}^{\ast}}{2}
  \left(  \frac{1}{\mu_{n}^{\ast}}+ 
  \frac{1}{\mu_{p}^{\ast}} \right) \: 
\end{eqnarray}
\begin{comment}
\begin{eqnarray}
  \left. \frac{\partial n_{\sigma}}{\partial X_{d}} \right|_{n_{b},\beta}
 & = &\left[ 1 + \left( f_{n}  + f_{p} \right)
   C_{\sigma}\right]^{-1}
   \\ \nonumber & & 
 \left[ \chi_{d\sigma} n_{b}
   - \frac{n_{b}}{2}
   \left( \frac{m_{n}^{\ast}}{\mu_{n}^{\ast}}
  + \frac{m_{p}^{\ast}}{\mu_{p}^{\ast}} \right)
   \right]
\end{eqnarray}
\end{comment}
whereas 
\begin{equation}
\left. \frac{\partial n_{\omega}}{\partial X_{d}} \right|_{n_{b},\beta}
 = -n_{b}(1 - \chi_{d\omega})
\end{equation}
for the source density of the $\omega$ meson. Finally, the derivative of the mass shift
with respect to the mass fraction is obtained in the compact form
\begin{eqnarray}
\lefteqn{\left. \frac{\partial \Delta m_{d}^{\mathrm{(high)}}}{\partial X_{d}} \right|_{n_{b},\beta}} \\ \nonumber & = &
\left[ \frac{2 C_{\sigma}}{1 + (f_{n}+f_{p})C_{\sigma}} \mathcal{U}_{d}^2 
  - \frac{\pi^{2}}{2\mu_{n}^{\ast}k_{n}}
  - \frac{\pi^{2}}{2\mu_{p}^{\ast}k_{p}}
   \right] n_{b} 
\end{eqnarray}
with a contribution from the $\sigma$ meson and kinetic terms.

%%%%%%%%%%%%%%%%%%%%%%%%%%%%%%%%%%%%%%%%%%%%%%%%%%%%%%%%%%%%%%%%%%%%%
\section{Density derivative of deuteron mass shift}
\label{sec:ddeltamd_dnb}

%The deuteron mass shift \eqref{eq:deltamd_high} is in general a function of the 
%the baryon density $n_{b}$, the asymmetry $\beta$, and the deuteron fraction $X_{d}$ at zero %temperature.
In this section, the derivative of %\red{$\Delta m_{d}$} 
$\Delta m_{d}^{\rm (high)}$ with respect to $n_{b}$ is derived
for an arbitrary function $X_{d}(n_{b})$ and constant $\beta$ at $T=0$. 
It requires again several steps. First, the derivatives of the source densities have to be determined. For the $\omega$ meson one finds
\begin{equation}
\label{eq:dnodnb}
  \left. \frac{\partial n_{\omega}}{\partial n_{b}} \right|_{\beta}
  %=  1- \left(1-g_{d\omega}\right) X_{d}
  %-  \left(1 - g_{d\omega}\right) n_{b} \frac{\partial X_{d}}{\partial n_{b}}
    =  1- \left(1-\chi_{d\omega}\right) Y_{d}
\end{equation}
with the quantity
\begin{equation}
  Y_{d} 
  = \left. \frac{\partial (n_{b}X_{d})}{\partial n_{b}} \right|_{\beta}
  =  X_{d} + n_{b} \left. \frac{\partial X_{d}}{\partial n_{b}} \right|_{\beta}
  %= 2 \frac{\partial n_{d}^{(v)}}{\partial n_{b}}
\end{equation}
that contains the derivative of the deuteron mass fraction.
For the $\sigma$ meson, the calculation is more involved. Here,  
the relations
\begin{eqnarray}
  \left. \frac{\partial n_{\sigma}}{\partial n_{b}} \right|_{\beta}
  & = & \left( f_{n} + f_{p}\right)
  \left. \frac{\partial m_{\mathrm{nuc}}^{\ast}}{\partial n_{b}} \right|_{\beta}
   + \chi_{d\sigma} Y_{d}
   \\ \nonumber & &
  + \frac{m_{\mathrm{nuc}}^{\ast}}{2\mu_{n}^{\ast}}
  \left( 1 + \beta - Y_{d}\right)
  + \frac{m_{\mathrm{nuc}}^{\ast}}{2\mu_{p}^{\ast}}
  \left( 1 - \beta - Y_{d}\right)
\end{eqnarray}
with the factor $f_{q}$ defined in Eq.\ \eqref{eq:fq}
%\begin{equation}
%    f_{q} = 3 \left( \frac{n_{q}^{(s)}}{m_{\mathrm{nuc}}^{\ast}}
%    - \frac{n_{q}^{(v)}}{\mu_{\mathrm{nuc}}^{\ast}}\right)
%\end{equation}
and 
\begin{equation}
    \left. \frac{\partial m_{\rm nuc}^{\ast}}{\partial n_{b}} \right|_{\beta} =
  - C_{\sigma} \left. \frac{\partial n_{\sigma}}{\partial n_{b}} \right|_{\beta}
  - C_{\sigma}^{\prime} n_{\sigma}
%  =   - C_{\sigma} %\frac{\partial n_{\sigma}}{\partial n_{b}}
%  \left[ \frac{\partial n_{n}^{(s)}}{\partial n_{b}}
%  + \frac{\partial n_{p}^{(s)}}{\partial n_{b}}
%  + g_{d\sigma} \left( X_{d} + n_{b} \frac{\partial X_{d}}{\partial n_{b}}
%  \right) \right]
%  - C_{\sigma}^{\prime} n_{\sigma}
\end{equation}
for the derivative of the effective nucleon mass can be combined to obtain the form
\begin{eqnarray}
\label{eq:dnsdnb}
  \left. \frac{\partial n_{\sigma}}{\partial n_{b}} \right|_{\beta}
  & = & 
  \left[ 1 +\left( f_{n} + f_{p}\right)
  C_{\sigma} \right]^{-1}
  \\ \nonumber & &
  \left[ 
 -\left( f_{n} + f_{p}\right)  C_{\sigma}^{\prime} n_{\sigma}
   + \chi_{d\sigma} Y_{d}
  \right.  \\ \nonumber & & \left.
  + \frac{m_{\mathrm{nuc}}^{\ast}}{2\mu_{n}^{\ast}}
  \left( 1 + \beta - Y_{d}\right)
  + \frac{m_{\mathrm{nuc}}^{\ast}}{2\mu_{p}^{\ast}}
  \left( 1 - \beta - Y_{d}\right) \right] 
\end{eqnarray}
with an explicit dependence on $Y_{d}$. In the next step, the derivative of the mass shift assumes the form
\begin{eqnarray}
  \lefteqn{\left. \frac{\partial \Delta m_{d}^{(\mathrm{high})}}{\partial n_{b}} \right|_{\beta}}
  \\ \nonumber & = &
  \frac{k_{n}}{\mu_{n}^{\ast}}
  \left. \frac{\partial k_{n}}{\partial n_{b}} \right|_{\beta}
  + \frac{k_{p}}{\mu_{p}^{\ast}}
  \left. \frac{\partial k_{p}}{\partial n_{b}} \right|_{\beta}
%  \\ \nonumber & &
  + \left(  \frac{1}{\mu_{n}^{\ast}}
  + \frac{1}{\mu_{p}^{\ast}} \right)
  m_{\rm nuc}^{\ast} 
  \left. \frac{\partial m_{\rm nuc}^{\ast}}{\partial n_{b}} \right|_{\beta}
  \\ \nonumber & &
  + 2(1-\chi_{d\omega}) \left( C_{\omega} \
  \left. \frac{\partial n_{\omega}}{\partial n_{b}} \right|_{\beta}
  + C_{\omega}^{\prime} n_{\omega} \right)
  \\ \nonumber & &
  + 2 \chi_{d\sigma} \left( C_{\sigma} 
  \left. \frac{\partial n_{\sigma}}{\partial n_{b}} \right|_{\beta}
  + C_{\sigma}^{\prime} n_{\sigma} \right) 
\end{eqnarray}
with
\begin{equation}
 \left. \frac{\partial k_{n}}{\partial n_{b}} \right|_{\beta} = 
 \frac{k_{n}}{3n_{n}^{(v)}} \left( \frac{1+\beta-Y_{d}}{2} \right)
\end{equation}
and
\begin{equation}
 \left. \frac{\partial k_{p}}{\partial n_{b}} \right|_{\beta} = 
 \frac{k_{p}}{3n_{p}^{(v)}} \left( \frac{1-\beta-Y_{d}}{2} \right) \: .
\end{equation}
Using the expressions \eqref{eq:dnodnb} and \eqref{eq:dnsdnb}, the final result can be expressed in compact form as
\begin{equation}
\label{eq:ddmddnb_general}
    \left. \frac{\partial \Delta m_{d}^{(\mathrm{high})}}{\partial n_{b}}  \right|_{\beta}
    = \mathcal{W}_{d} - \mathcal{Z}_{d} Y_{d}
\end{equation}
with the auxiliary quantities
\begin{eqnarray}
    \mathcal{Z}_{d} & = &
    \frac{\pi^{2}}{2\mu_{n}^{\ast}k_{n}}
  + \frac{\pi^{2}}{2\mu_{p}^{\ast}k_{p}}
   + 2  \left(1-\chi_{d\omega}\right)^{2} C_{\omega} 
  \\ \nonumber & &
  - \frac{2C_{\sigma}}{1 +\left( f_{n} + f_{p}\right) 
  C_{\sigma}} \: \mathcal{U}_{d}^{2}
 % \left[  \chi_{d\sigma} - \left(  \frac{1}{\mu_{n}^{\ast}}
 % + \frac{1}{\mu_{p}^{\ast}} \right)
 % \frac{m_{\rm nuc}^{\ast}}{2} \right]^{2}
 \\
  \mathcal{W}_{d} & = &
  \frac{\pi^{2}}{2\mu_{n}^{\ast}k_{n}} \left( 1+\beta\right)
  + \frac{\pi^{2}}{2\mu_{p}^{\ast}k_{p}} \left( 1-\beta\right)
  \\ \nonumber & &
  + 2(1-\chi_{d\omega}) \left( C_{\omega} 
  +  C_{\omega}^{\prime} n_{\omega} \right)
  \\ \nonumber & &
  + 
  %\left[  \chi_{d\sigma} - \left(  \frac{1}{\mu_{n}^{\ast}}
  %+ \frac{1}{\mu_{p}^{\ast}} \right)
  %\frac{m_{\rm nuc}^{\ast}}{2} \right]
  \frac{2}{1 +\left( f_{n} + f_{p}\right)C_{\sigma}} \: \mathcal{U}_{d}
  \\ \nonumber & &
  \left[ C_{\sigma}^{\prime} n_{\sigma}
  + C_{\sigma} \frac{m_{\mathrm{nuc}}^{\ast}}{2}
  \left( %\frac{m_{\mathrm{nuc}}^{\ast}}{2\mu_{n}^{\ast}}
  \frac{1 + \beta}{\mu_{n}^{\ast}}
  + %\frac{m_{\mathrm{nuc}}^{\ast}}{2\mu_{p}^{\ast}}
  \frac{1 - \beta}{\mu_{p}^{\ast}} \right) \right]
\end{eqnarray}
and $\mathcal{U}_{d}$ as given in \eqref{eq:U_d}.
\section{Conversion of parameters 
%\red{in SNM with deuterons} 
at saturation}
\label{app:conv}

In order to find the coupling strengths $\Gamma_{\sigma,0}$ and 
$\Gamma_{\omega,0}$ as well as the deuteron mass shift and its density derivative at saturation, a step-by-step procedure can be followed. These quantities are determined as soon as 
the saturation density $n_{0}$, the binding energy per nucleon $B_{0}$, the effective nucleon mass $m^{\ast}_{\mathrm{nuc},0}$ and deuteron fraction $X_{d,0}$ of SNM are specified.

In a first step, the scalar and vector densities
\begin{equation}
n_{d, 0}^{(s)} = n_{d, 0}^{(v)} %= n_{d, 0} 
= n_0 \frac{X_{d, 0}}{2}
\end{equation}
of the deuteron
and the total vector density
\begin{equation}
n_{\mathrm{nuc}, 0}^{(v)} = n_0 \left( 1 - X_{d,0}\right)
%\frac{1 - X_{d, 0}}{2}
\end{equation}
of the nucleons %both nuclear species $q$ ($q=n,p$) 
are immediately obtained from $n_{0}$ and $X_{d,0}$ in SNM. Then the Fermi momentum
\begin{equation}
%\red{k_{\mathrm{nuc}, 0} = \left[ \frac{3\pi^{2}}{2}
%n_{\mathrm{nuc}, 0}^{(v)} 
%%n_0 \left(1 - X_{d, 0}\right)
%\right]^{1/3}}
k_{\mathrm{nuc}, 0} = \left[ \frac{6\pi^{2}}{g_{\mathrm{nuc}}}
n_{\mathrm{nuc}, 0}^{(v)} 
%n_0 \left(1 - X_{d, 0}\right)
\right]^{1/3}
\end{equation}
with degeneracy factor $g_{\mathrm{nuc}} = 4$ and the effective chemical potential
\begin{equation}
\label{eq:musnuc}
\mu_{\mathrm{nuc}, 0}^{\ast} = \sqrt{k_{\mathrm{nuc}, 0}^2 
+ \left( m^{\ast}_{\mathrm{nuc},0}\right)^{2}}
\end{equation}
allow to calculate the scalar density
\begin{eqnarray}
\lefteqn{n^{(s)}_{\mathrm{nuc},0}}
\\ \nonumber & = & \frac{g_{\mathrm{nuc}} m^{\ast}_{\mathrm{nuc},0}}{4\pi^{2}} \left[ 
k_{\mathrm{nuc}, 0}\mu_{\mathrm{nuc}, 0}^{\ast} - \left(m^{\ast}_{\mathrm{nuc},0}\right)^{2} \ln \frac{k_{\mathrm{nuc}, 0}+\mu_{\mathrm{nuc}, 0}^{\ast}}{m^{\ast}_{\mathrm{nuc},0}} \right]
\end{eqnarray}
using the effective nucleon mass $m_{\mathrm{nuc},0}^{\ast}$. Then the source densities
\begin{equation}
    n_{\sigma,0} =  n_{\mathrm{nuc}}^{(s)}
    %n_{n,0}^{(s)} + n_{p,0}^{(s)} 
    + 2 \chi n_{d,0}^{(s)} 
\end{equation}
and
\begin{equation}
    n_{\omega,0} = n_{\mathrm{nuc}}^{(v)}
    %n_{n,0}^{(v)} + n_{p,0}^{(v)} 
    + 2 \chi n_{d,0}^{(v)} 
%    \red{= n_{0}} 
%\red{= n_{0}[1-X_{d,0}(1-\chi)]}
\end{equation}
with the deuteron-meson coupling scaling factor $\chi$ are found and the pressure contribution 
\begin{equation}
p_{\mathrm{nuc},0} = \frac{1}{4} \left[ 
\mu_{\mathrm{nuc}, 0}^{\ast} n_{\mathrm{nuc}, 0}^{(v)} - 
m^{\ast}_{\mathrm{nuc},0} n_{\mathrm{nuc}, 0}^{(s)} \right]
\end{equation}
%\blu{\mbox{Prefactor} $1/4$!} \ora{I guess your result coincides with mine. Indeed, I was considering the scalar and vector density of each nuclear species, which are actually one half of $n_{\mathrm{nuc}, 0}^{(s)}$ and $n_{\mathrm{nuc}, 0}^{(v)}$ here considered.}
of the nucleons can be calculated immediately.

In the next step, the effective nucleon mass determines the scalar potential 
\begin{equation}
S_{\mathrm{nuc},0} = m_{\mathrm{nuc}}- m^{\ast}_{\mathrm{nuc},0}
\end{equation}
of the nucleons and thus the scalar coupling
\begin{equation}
C_{\sigma, 0} = \frac{S_{\mathrm{nuc},0}}{n_{\sigma, 0}} 
\end{equation}
and finally the coupling strength
\begin{equation}
\Gamma_{\sigma, 0} = m_{\sigma} \sqrt{C_{\sigma, 0}}
\end{equation}
of the $\sigma$ meson. The binding energy per nucleon $B_{0}$ gives
%the energy density
%\begin{equation}
%    \varepsilon = \left( m_{\mathrm{nuc}}- B_{0}\right) n_{0}
%\end{equation}
%at saturation 
the chemical potential
\begin{equation}
    \mu_{\mathrm{nuc},0} = m_{\mathrm{nuc}}- B_{0}
\end{equation}
at saturation and then the vector potential
\begin{equation}
    V_{\mathrm{nuc},0} = \mu_{\mathrm{nuc},0} 
    - \mu_{\mathrm{nuc},0}^{\ast} 
\end{equation}
of the nucleons. The latter quantity can be expressed in general as
\begin{equation}
\label{eq:Vnuc0}
    V_{\mathrm{nuc},0} = C_{\omega,0} n_{\omega,0} 
    + C_{\rho,0} n_{\rho,0} + U_{0}^{(r)}
\end{equation}
with the auxiliary quantity 
\begin{equation}
\label{eq:U0}
    U_{0}^{(r)} = V_{0}^{(r)} + W_{0}^{(r)}
\end{equation}
that also appears in the total pressure
\begin{eqnarray}
\label{eq:PO}
    \lefteqn{P_{0} = p_{\mathrm{nuc},0}}
    \\ \nonumber & &
    + \frac{1}{2} \left( C_{\omega,0} n_{\omega,0}^{2} + C_{\rho,0} n_{\rho,0}^{2}
    - C_{\sigma,0} n_{\sigma,0}^{2} \right) + U_{0}^{(r)} n_{0} \: .
\end{eqnarray}
For SNM, however, the $\rho$-meson contribution does not appear, since $n_{\rho}$ is identically zero.
The two equations \eqref{eq:Vnuc0} and \eqref{eq:PO} allow to solve for the $\omega$ coupling
\begin{eqnarray}
    C_{\omega,0} & = & %\frac{1}{n_{\omega,0}^{2}}
    \left( 2n_{\omega,0}n_{0}-n_{\omega,0}^{2}\right)^{-1}
    \\ \nonumber & &
    \left( 2p_{\mathrm{nuc},0} + 2V_{\mathrm{nuc},0} n_{0} 
    - C_{\sigma,0} n_{\sigma,0}^{2}\right) 
\end{eqnarray}
and further the coupling strength
\begin{equation}
\Gamma_{\omega, 0} = m_{\omega} \sqrt{C_{\omega, 0}} 
\end{equation}
using $P_{0}=0$.
With known $C_{\sigma,0}$ and $C_{\omega,0}$, their derivatives
$C_{\sigma,0}^{\prime}$ and $C_{\omega,0}^{\prime}$ can be determined
using the same functional density dependence of the couplings as in the %\red{DD2} 
reference parameterization. Thus also the rearrangement contribution
\begin{equation}
    V_{0}^{(r)}  
    = \frac{1}{2} \left( C_{\omega,0}^{\prime} n_{\omega,0}^{2} 
    + C_{\rho,0}^{\prime} n_{\rho,0}^{2} - C_{\sigma,0}^{\prime} n_{\sigma,0}^{2} \right)
\end{equation}
is given.

Finally, 
from Eqs.\ \eqref{eq:Vnuc0} and \eqref{eq:U0} one finds 
%\begin{equation}
%\red{
%W_{0}^{(r)} = C_{\omega,0} n_{\omega,0} 
%     + V_{0}^{(r)} - V_{\mathrm{nuc},0} 
%}
%\end{equation}
%\cya{
\begin{equation}
W_{0}^{(r)} = V_{\mathrm{nuc},0}  - C_{\omega,0} n_{\omega,0} - V_{0}^{(r)} 
\end{equation}
%}
and the deuteron mass shift derivative
\begin{equation}
    \left. \frac{d\Delta m_{d}}{d n_{b}} \right|_{n_{0}}
    = \frac{W_{0}^{(r)}}{n_{d,0}^{(s)}}
\end{equation}
at saturation.
The deuteron mass shift itself is determined as
\begin{eqnarray}
    \Delta m_{d,0} & = & B_{d}
    + 2 \left( \mu_{\mathrm{nuc},0}^{\ast}-m_{\mathrm{nuc},0}^{\ast}
    \right) 
    \\ \nonumber & &
    + 2  (1-\chi_{d\omega}) C_{\omega} n_{\omega}
    - 2 (1-\chi_{d\sigma}) C_{\sigma} n_{\sigma}
\end{eqnarray}
from the condensation condition with the binding energy of the deuteron in vacuum $B_{d}$.
%\blu{Attention: In neutron matter, the Fermi momentum, chemical potential. scalar density, effective mass etc.\ are different as compared to symmetric nuclear matter. Better define explicitly the corresponding quantities $k_{n}$, $\mu_{n}$, $n_{n}^{(s)}$, $m_{n}^{\ast}$. Of course, $n_{n,0}^{(v)}=n_{b,0}$ and $V_{n,0}=V_{\mathrm{nuc},0}$.}

The rescaling of the $\sigma$ and $\omega$ coupling strengths induces a modification of the energy per nucleon in PNM at saturation and thus of the symmetry energy. Within the parabolic approximation, the symmetry energy at saturation is indeed given by
\begin{equation}
\label{eq:J0}
J_{0} = \left.\dfrac{E}{A}\right|_{n_{0},\beta = 1} 
- \left. \dfrac{E}{A}\right|_{n_{0},\beta = 0} 
= \left.\dfrac{E}{A}\right|_{n_{0},\beta = 1} + B_0 \: .
\end{equation}
Then, constraining the value of $J_{0}$ implies a constraint on $E/A$ of PNM at $n_{0}$.
Taking into account Eq.\ \eqref{eq:fed}, the condition above writes
\begin{equation}
\mu^{\ast}_{n,0} + V_{n,0} - \left. \dfrac{P_{n}}{n_{0}} \right|_{n_{0},\beta=1} = m_{\mathrm{nuc}} + J_{0} - B_{0}
\end{equation}
with the effective chemical potential
\begin{equation}
 \mu_{n,0}^{\ast} = \sqrt{k_{n,0}^{2}+\left( m_{n,0}^{\ast}\right)^{2}}   
\end{equation}
of the neutron at the saturation density $n_{0}$.
The effective mass of the neutron 
$m_{n,0}^{\ast}=m_{\mathrm{nuc}}-\Gamma_{\sigma} n_{n,0}^{(s)}$ has to be determined self-consistently with the scalar density $n_{n,0}^{(s)}$, defined in Eq.\ \eqref{eq:nqs},
using the Fermi momentum $k_{n,0}= \left(3\pi^{2}n_{0}\right)^{1/3}$ of the neutron. Since there are no deuterons in PNM, the vector potential of the neutron is given by
\begin{eqnarray}
    V_{n,0} & = & C_{\omega,0} n_{\omega} + C_{\rho,0} n_{\rho}
    \\ \nonumber & &
 + \frac{1}{2} \left( C_{\omega,0}^{\prime} n_{\omega}^{2}
 + C_{\rho,0}^{\prime} n_{\rho}^{2} 
 - C_{\sigma,0}^{\prime} n_{\sigma}^{2}\right)
\end{eqnarray}
and the pressure assumes the simple form
\begin{eqnarray}
 P_{n}(n_{0}) & = & \frac{1}{4} \left( \mu_{n,0}^{\ast} n_{n,0}^{(v)} - m_{n,0}^{\ast} n_{n,0}^{(s)}\right)
 \\ \nonumber & &
 + \frac{1}{2} \left[ D_{\omega,0} n_{\omega}^{2} + D_{\rho,0} n_{\rho}^{2} - D_{\sigma,0} n_{\sigma}^{2} \right]
\end{eqnarray}
with $n_{\omega}=n_{\rho}=n_{n,0}^{(v)}=n_{0}$ and $n_{\sigma}=n_{n,0}^{(s)}$. The rescaled couplings 
$C_{\omega,0}$, $C_{\sigma,0}$, $C_{\omega,0}^{\prime}$, $C_{\sigma,0}^{\prime}$, $D_{\omega,0}$, and $D_{\sigma,0}$, 
c.f., Eq.\ \eqref{eq:Ddef}, at saturation are already known and thus $C_{\rho,0}$ can be deduced from
\begin{eqnarray}
 C_{\rho,0}  & = & 
 \frac{2}{n_{0}}\left[ m_{\mathrm{nuc}} + J_{0} - B_{0}
 - \frac{3}{4} \mu^{\ast}_{n,0}
  \right. \\ \nonumber & & \left.
    - \frac{1}{4} m_{n,0}^{\ast} \frac{n_{n,0}^{(s)}}{n_{0}}
  - \frac{1}{2} C_{\omega,0} n_{\omega}
   - \frac{1}{2n_{0}} C_{\sigma,0}  n_{\sigma}^{2} \right]
\end{eqnarray}
and, finally, 
\begin{equation}
\Gamma_{\rho, 0} = m_{\rho} \sqrt{C_{\rho,0}}  
\end{equation}
for the coupling of the $\rho$ meson at saturation.

%%%%%%%%%%%%%%%%%%%%%%%%%%%%%%%%%%%%%%%%%%%%%%%%%%%%%%%%%%%%%%%%%%%%%
\section{Analytical expressions for the mass shift parameters}
\label{sec:cd}

The following analytical expressions permit to calculate the parameters $c$ and $d$ appearing in Eq.\ \eqref{eq:deltamd_param}:
%\gre{
\begin{eqnarray}
c &=& \left[1 - \tanh(e) \right]^{-1} 
\\ \nonumber & & \left[\Delta m_{d, 0} - \dfrac{a}{1+b} - f\tanh(g) \right] \\
\eta &=& \left[1 - \tanh(e) \right]^{-1}
\\ \nonumber & & 
\left\{ 
\dfrac{\partial \Delta m_{d}}{\partial n_{b}}\right|_{n_{0}}n_{0} - \dfrac{a}{(1 + b)^2} + \dfrac{ce}{\cosh^2(e)} 
\\ \nonumber 
 & &  \left. - f \gamma \tanh(g)  -\dfrac{fg}{\cosh^2(g)} - c[1 - \tanh(e)]\right\}
\end{eqnarray}
%}
where the values for $a$, $b$, $e$, $f$ and $g$ are determined as explained in Section \ref{sec:unified_mass_shift}, while $\Delta m_{d, 0}$ and $\left.\dfrac{\partial \Delta m_{d}}{\partial n_{b}}\right|_{n_{0}}$ indicate the mass shift and its density slope at saturation density $n_{0}$.

\bibliographystyle{apsrev4-1}
\bibliography{deuteron}

%%%%%%%%%%%%%%%%%%%%%%%%%%%%%%%%%%%%%%%%%%%%%%%%%%%%%%%%%%%%%%%%%%%%%
\end{document}